CENTRO DE INVESTIGACIÓN Y DE ESTUDIOS AVANZADOS DEL INSTITUTO POLITÉCNICO NACIONAL, UNIDAD IRAPUATO

LABORATORIO NACIONAL DE GENÓMICA PARA LA BIODIVERSDAD

# Functional and evolutionary genomics of the *Streptomyces* metabolism

A thesis presented by

**Pablo Cruz Morales**

To obtain the degree of Doctor in Science
Specialized in Plant Biotechnology

Thesis director:

Dr. Francisco Barona Gómez

My PhD studies were funded by the National Council of Science and Technology of México (Conacyt PhD Fellowship 28830) to which I am deeply grateful.

This work has mainly been done at the National Laboratory of Genomics for Biodiversity (LANGEBIO) at Irapuato Guanajuato, Mexico.

# Abstract

This thesis is focused in the study of the evolution of the metabolic repertoire of *Streptomyces,* which are renowned as proficient producers of bioactive Natural Products (NPs). The main goal of my work was to contribute into the understanding of the evolutionary mechanisms behind the evolution of NP biosynthetic pathways. Specifically, the development of a bioinformatic method that helps into the discovery of new NP biosynthetic pathways from actinobacterial genome sequences with emphasis on members of the genus *Streptomyces*. I developed this method using a comparative and functional genomics perspective. My studies indicate that central metabolic enzymes were expanded in a genus-specific manner in *Actinobacteria*, and that they have been extensively recruited for the biosynthesis of NPs.

Based in these observations, I developed EvoMining, a bioinformatic pipeline for the identification of novel biosynthetic pathways in microbial genomes. Using EvoMining several new NP biosynthetic pathways have been predicted in different members of the phylum *Actinobacteria*, including the model organism *S. lividans* 66. To test this approach, the genome sequence of this model strain was obtained, and its analysis led to the discovery of an unprecedented system for peptide bond formation, as well as a biosynthetic pathway for an arsenic-containing metabolite.

Moreover, this work also led to the identification of expansions on a conserved metabolic node in the glycolytic pathway of *Streptomyces.* These expansions occurred before the radiation of *Streptomyces* and are concomitant with the evolution of their capability to produce NPs. Experimental analyses indicate that this node evolved to mediate the interplay between central an NP metabolism.

# Resumen

Esta tesis está enfocada al estudio de la evolución del repertorio metabólico de *Streptomyces*, el género de bacterias productoras de Productos Naturales (NPs) bioactivos más prolífico. La meta general de este trabajo fue contribuir en el entendimiento de los mecanismos evolutivos detrás del surgimiento de las vías biosintéticas de NPs. Específicamente, a través del desarrollo de un nuevo método bioinformático que facilita el descubrimiento de vías biosintéticas de NPs a partir de secuencias genómicas de actinobacterias. Dicho método fue desarrollado desde una perspectiva de genómica funcional y comparativa. Los resultados de mi trabajo indican que las enzimas del metabolismo central se expandieron de una manera género-especifica en la familia de las *Actinobacterias*, y posteriormente han sido reclutadas frecuentemente para la biosíntesis de NPs.

A partir de esta observación y mediante el uso de la teoría evolutiva desarrollé una nueva estrategia de minería genómica que quedó plasmada en una herramienta bioinformática llamada EvoMining. Usando esta herramienta se han predicho muchas vías biosintéticas en diferentes miembros del phyllum *Actinobacteria*, incluyendo al organismo modelo *S. lividans* 66. Para validar la eficacia de EvoMining se obtuvo el genoma de este organismo y su análisis condujo al descubrimiento de un nuevo sistema para la formación de enlaces peptídicos así como la vía biosintética para un metabolito que contiene arsénico.

De manera inesperada este trabajo condujo también a la identificación de expansiones en un nodo metabólico conservado en la vía glucolítica de *Streptomyces*. Estas expansiones ocurrieron antes de la radiación del género y son coinciden con la evolución de su capacidad para producir NPs. El análisis experimental del nodo glucolítico indica que evolucionó para mediar la interacción en el metabolismo central y el de NPs.

# Others contributions to this work

Hilda Ramos Aboites worked HPLC and mass spectrometry analysis, her work was fundamental for the discovery of the livipeptin biosynthetic pathway.

Luis Yanez Guerra contribuited with his studies on the function of SLI1096, his work was crucial for the prediction of the arsonolividin biosynthetic pathway.

Christian Martinez Guerrero worked on the improvement of EvoMining, his work allowed the implementation of a more efficient and user friendly interface for the bioinformatic pipeline.

Fernanda Iruegas Bocardo worked in the *Streptomyces lividans* 66 genome sequencing project during her master studies, she isolated genomic DNA and helped in the preliminary analysis of the sequence.

Paulina Valdez Camacho helped in the cloning of glycolytic genes from *S. coelicolor* during her summer internship.

Alejandro Aragón Raygoza performed phenotypic analyses of *S. coelicolor* mutants of glycolytic genes.

Dr. Sandra Perez Miranda provided useful protocols and discussions on metal-related phenotypes in *S. lividans* 66.

Dr. Erik Vijgenboom from Leiden University provided RNAseq data, insights and protocols for the study of the metal homeostasis in *S. lividans* 66.

Professor Gilles van Wezel provided PacBio and Illumina data for the *S. lividans* 66 genome sequencing project, he also supervise me during my internship at the Molecular Biotechnology group in Leiden University where part of the project was developed.

# Contents

# 1 Introduction

## 1.1 Natural products and the discovery of antibiotics

### 1.1.1 What is a Natural Product?

The main subject of my thesis is microbial metabolism and its evolution, with emphasis on the chemical diversity present in natural products metabolism which is also known as secondary metabolism. Since the nomenclature of this class of molecules has undergone many changes, as described in the excellent review and discussion of the book by Richard Firn, Nature Chemicals (published in 2010) is important to establish a definition of the natural products term (NP). This is important given the pervasiveness and utility of NPs both in scientific contexts and in society in general. The definition that I consider most appropriate, and is referred to throughout this thesis is the following:

"A diverse group of low molecular weight chemical compounds produced by microorganisms, whose production is not essential at all times, but it is essential for adaptation and survival of the producing organism."

This definition implies that chemical compounds or metabolites are not conserved and may even be specific to a given strain within a specie. This also implies that these compounds are produced only under certain conditions which tend to differ from laboratory conditions such as axenic culture.

NPs are extraordinarily diverse and it can be assumed that they are not incorporated into other metabolic pathways as precursors, i. e. NPs are endpoints of the metabolic network. This may be related to the fact that they are not universally distributed, on the contrary, their distribution is practically species-specific. This diversity, therefore, is the natural consequence of their role in adaptation to specific environmental conditions and lifestyle of each organism (Firn, 2010; Firn and Jones, 2009). In contrast, central metabolism involves catabolic and anabolic reactions which produce an increase in biomass. In other words, it includes all the necessary functions for the production of energy and reducing power to be used in the synthesis of complex macromolecules, structural and storage elements. Since the basic physiological functions are conserved in most organisms, central metabolism may be seen as a tightly regulated and integrated network of universal biochemical conversions.

In many organisms NPs biosynthesis is activated in the later stages of development, so it is often assumed that the production of these molecules is associated with nutrient limitation on the environment. This assumption has led to the idea that a natural product is a molecule produced when the cell has stopped growing or is growing slowly (Stone and Williams, 1992). Whereas central metabolism, active during the production of biomass in the exponential phase of a liquid culture, has been called primary metabolism, the metabolism active in the late stationary phase in liquid cultures has been called secondary. Is very common that the terms Natural Product and secondary metabolite are used equivalently which can be confusing, to clarify this issue, these definitions will be revisited later in the light of the Darwinian theory of evolution.

NPs are a consistent source of drugs, over 20,000 bioactive NPs of microbial origin are known. Over 40% of the new compounds reported in recent years are NPs derived from microorganisms, 60% of compounds with anti-cancer activity and 70% of clinical antibiotics used today are NPs or derivatives from them (Demain, 2009). Studies conducted after the explosion of genomic data have revealed that the discovery of potential NPs and their diversity has been underestimated, therefore new strategies for their identification and exploitation for human use are much needed (Van Lanen and Shen, 2006). In this thesis I addressed this challenge with an evolutionary view, however, at this point, a short review is necessary to introduce the chemical and biological basis, of the genetic and evolutionary diversity behind NPs metabolism.

### 1.1.2 The origin of antibiotics

The study of NP microbial metabolism especially that concerning the biosynthesis of molecules of therapeutic and industrial use, has its origins in the late nineteenth century. Between 1871 and 1877 Joseph Lister, John Tyndall, Louis Pasteur and Jules Joubert reported the inhibitory effect caused by fungal growth in some bacterial species. In 1897 Ernest Duchesne in his thesis *Contribution à l'étude de la concurrence vitale chez les micro-organismes: antagonisme entre les moisissures et les microbes*, reported the inhibitory effect of *Penicillium glaucum* on the bacilli that causes typhoid fever (probably *Salmonella tiphy*) during *in vivo* studies on horses. These experiments were designed after he noticed that in the stalls, the horse saddles were stored in dark, damp places to allow the growth of molds that helped to heal the wounds caused by the saddles (Glick and Pasternak, 2009). Even though the agent responsible for this effect was not specifically described, is clear that at the beginning of the twentieth century the inhibitory effect of some organisms over others was a common observation. Furthermore, the use of this effect for human therapy had been already proposed.

It did not take long for these observations to become a great discovery when Alexander Fleming observed the formation of inhibition zones on plates inoculated with *Staphylococcus* in which a contaminating mold (later identified as *Penicillium notatum*) had grown. This fluke led Fleming investigations to the isolation of an antimicrobial agent active against human pathogens which was named penicillin (Figure 1). Fleming also showed that penicillin was not toxic to human leukocytes, and suggested its use as a therapeutic agent (Fleming, 1929).

A decade later, in 1939, Rene Dubos isolated a bacteria capable of inhibiting the growth of other Gram-positive bacteria from soil samples. This bacterium was named as *Bacillus brevis*. Dubos, in collaboration with Rollin Hotchins, defined chemical nature of the active substance as two polypeptides, a lysozyme and gramicidin (Figure 1). This discovery revitalized research on penicillin, which at that time still did not rouse enough interest for its therapeutic application. Finally, Howard

Florey and Ernst Chain managed to produce and purify large quantities of penicillin, and demonstrate their use as a chemotherapeutic agent (Chain et al 1940) their work made the therapeutic use of penicillin practical. During WWII, both gramicidin and penicillin were widely and successfully used, starting the era of antibiotics.

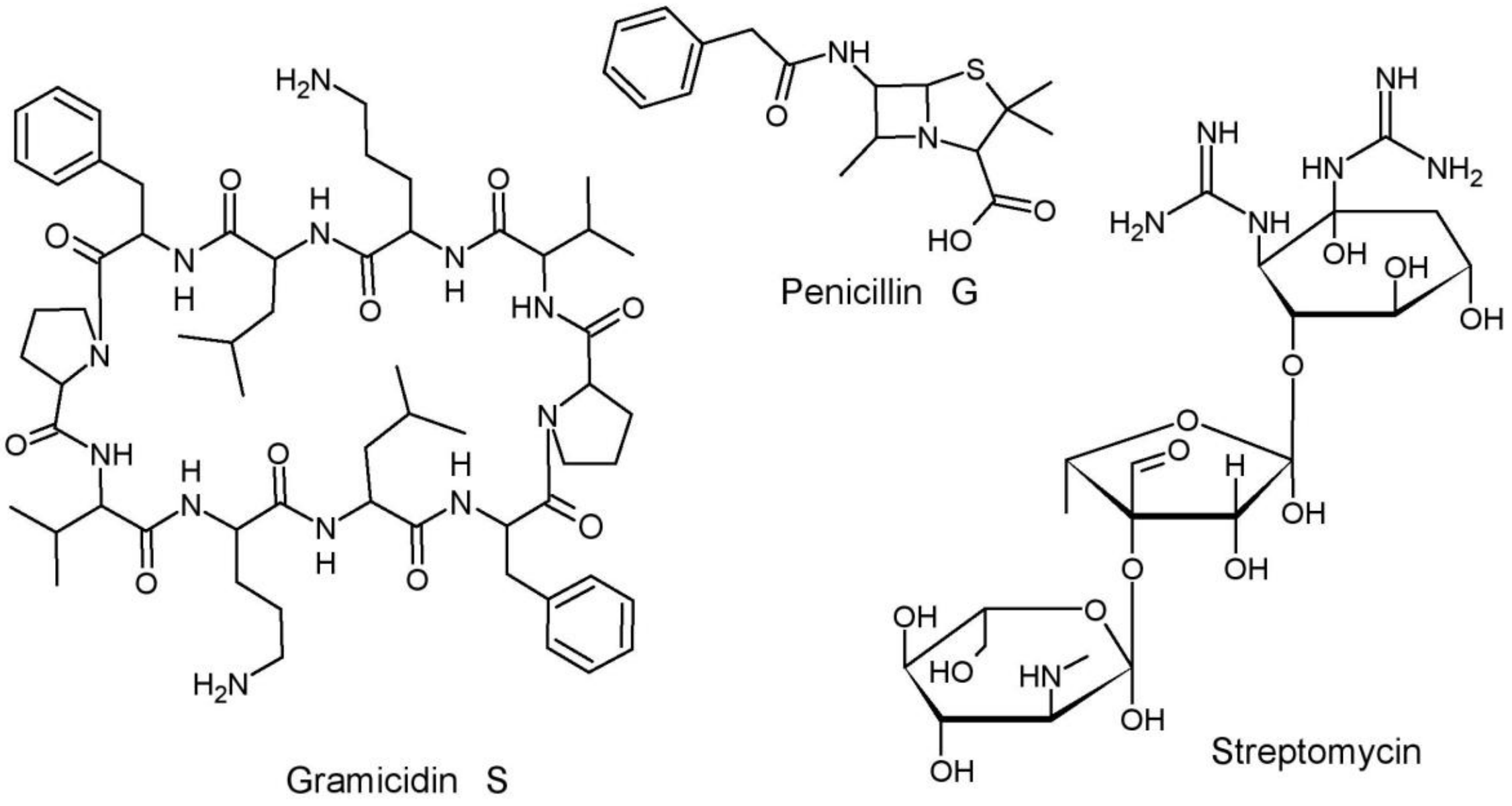

**Figure 1. The first NPs of microbial origin, discovered between 1930 and 1950, with antibiotic activity.** Gramicidin S is a cyclic variant of the Dubos gramicidin.

### 1.1.3 The golden age of antibiotic discovery: “The remedies are in our backyards”

In the 40's the potential of microorganisms as the sources of useful molecules to control the pathogens that threatened human health was well established. Between 1941 and 1943 the use of penicillin became massive, this served to promote the idea that there could be more similar molecules awaiting to be discovered. The crystallization of this idea came as a large scientific effort involving exhaustive screenings that led to the isolation of microorganisms capable of producing such molecules. An example of this efforts is the International *Streptomyces* Project, (ISP), which resulted in a large number of culture media designed expressly to promote the production of NPs in laboratory conditions (Shirling and Gottlieb, 1966).

In 1944 Albert Schatz and Selman Waskman identified a compound from *Streptomyces griseus*, a member of the *Actinobacteria* phylum, which was isolated from the soil obtained from a farm. The molecule was named streptomycin (Figure 1). This molecule was able to inhibit Gram-positive bacteria, including *Mycobacterium tuberculosis* (Schatz et al, 1944), the causative agent of tuberculosis. This antibiotic went quickly to mass production and saved millions of people from death despite the side effects caused by the high toxicity of the molecule.

The search for antibiotics intensified during the following years and one after another new therapeutic agents with different ranges of action against different pathogens appeared. Among the sources of these compounds the members of the genus *Streptomyces* standout, therefore it became evident that these bacteria possessed extraordinary chemical capabilities. In 1949, in an interview with the Time magazine, the by then famous microbiologist Selman Waskman stated, "The remedies are in our own backyards" alluding to the discovery of the tuberculosis cure from a soil sample taken from a farm. At that time, when the era of antibiotics was just beginning and the biosynthetic potential of microorganisms was barely known, this observation was both fascinating and controversial. The decades of the 40s and 50s were so prolific for the seekers of antibiotics, that this period is now known as the golden age of antibiotic discovery.

Such optimism, however, could not come without its equivalent concerns. Since the early days a problem was derived from the clinical use of antibiotics: the resistance to these compounds by pathogenic bacteria. After only one year of penicillin use, the first resistant strains of *Staphylococcus aureus* appeared, subsequently the resistance became highly prevalent, prompting replacement of penicillin with other more efficient antibiotics (Walsh, 2003). This phenomenon was ubiquitous in all cases in which an antibiotic were used to control a pathogen, so that the need for new molecules with antibiotic activity became constant. In the decades that followed the golden age of antibiotic discovery to date, a large number of molecules for therapeutic use are the result of iterative semi-synthetic modification of NPs discovered in the golden age, leading to second and third generation antibiotics (Clardy et al, 2006). Nowadays it is considered that resistance to each new generation or class of antibiotics inevitably arise in periods ranging between one year and a decade, long enough for the arising and spreading of the necessary mutations among the population of the targeted pathogens that become resistant. Thus, the lifetime of a molecule begins to run out at the moment when it reaches the market and is used clinically (Walsh, 2003).

## 1.2 The *Streptomyces* lifestyle and the role of their natural products

### 1.2.1 The *Streptomyces* developmental program

The members of the genus *Streptomyces* are Gram-positive bacteria, like other members of the family of *Actinobacteria* they have high G+ C content in their DNA, with an average of 72%. These microorganisms have been isolated from a wide range of habitats, mainly from soils of every kind and marine sediments. They are also often found as symbionts of insects such as ants, beetles and wasps (Seipke et al, 2012), even a handful of known pathogenic *Streptomyces* are known to infect plants, fungi and animals (Seipke et al, 2012). However, most *Streptomyces* are saprophytic free-living bacteria that degrade organic material from their habitats, which are often oligotrophic and populated by a large number and variety of organisms (Hodgson, 2000). Thus, bacteria of the genus *Streptomyces* had manage to get their resources through a complex development program that includes the formation of vegetative mycelium that grows embedded in its substrate, normally solid. When nutrients become scarce in the niche, the vegetative mycelium undergoes programmed cell death and triggers the development of aerial hyphae, its biomass is formed from the remains of the colony, aerial hyphae eventually led to the formation of spores, (Schwedock et al, 1997). Spores are more resistant than the mycelium and have a hydrophobic surface that allows them to be washed away for dispersion (Hodgson, 2000). If a spore finds a new niche suitable for development, it germinates and starts the cycle again (Figure 2).

The colony may be more vulnerable during maturation of the aerial mycelium and sporulation, which could explain the fact that this step is tightly synchronized with the production of NPs (van Wezel & MacDowell, 2011), it is believed that the production of these molecules in many instances is intended to protect the colony. Such molecules are precisely those which have been exploited as antibiotics.

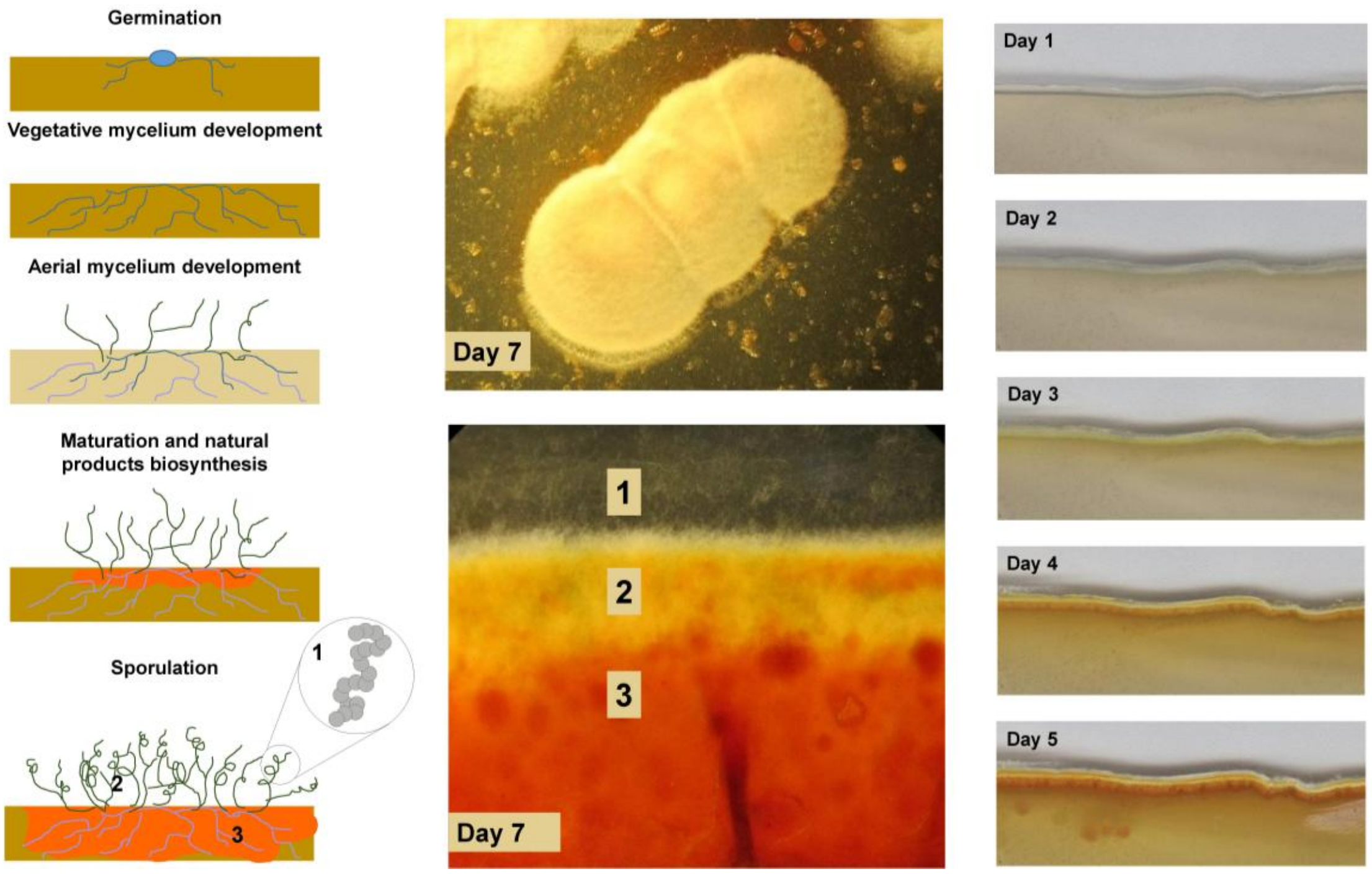

**Figure 2. Typical growth of a member of the genus *Streptomyces*. (*S. parvus* CH2, a strain isolated from the Coahuila desert in México).** In the first column the developmental process is schematically shown, in the second column two pictures are shown: aerial and transversal perspectives of a fully developed colony. The number 1 indicates sporulating aerial mycelia, number 2 indicates the aerial mycelia maturation process, and number 3 indicates the remnant vegetative mycelia and the copious production of an unknown pigmented NP. The last column shows a series of pictures of the growth of *S. parvus* CH2 during 5 days.

## 1.2.2 Natural products… in *Streptomyces* nature

The antibiotic activity of many of the molecules produced during the development of aerial mycelium and spores appears to be evidence of their potential role in competition for resources. This idea has led to the belief that NPs with antibiotic activity act as defense mechanisms for colonization of ecological niches. Even, at the extreme of anthropocentrism, they have been involved in an alleged arms race between the inhabitants of the soil (Riley, 1998). The truth is that the role of NPs produced by *Streptomyces* and other bacteria is still vastly unknown and the scenario seems more complicated than simply inhibiting the growth of competitors (Figure 3).

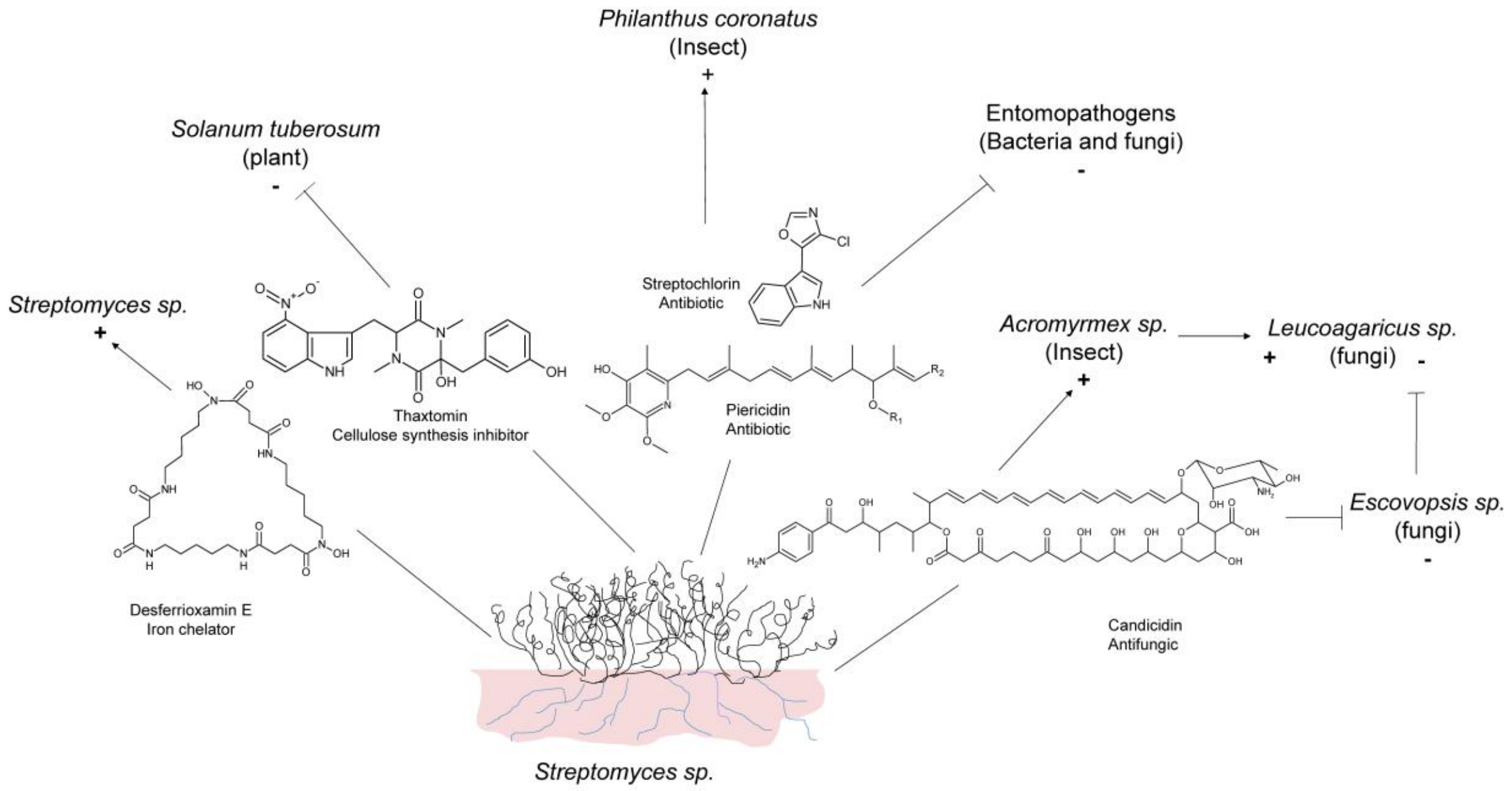

**Figure 3. The *Streptomyces* NPs and their role in nature.** The NP mediated interactions between *Streptomyces* and other organisms are represented with lines. Inhibition is represented with bar topped lines, while positive interactions are shown with arrows. The + and – symbols are used to represent the effect of the interaction over the organisms

A simple observation casts doubt on a general role as antibiotics for the NPs: To inhibit the growth of a pathogenic bacteria using an antibiotic NP of clinical use, the quantities of the compound are much higher than the amounts that a colony can produce in their natural environment. Purity is also an important issue: while the clinically used antibiotics are administered in very pure versions, the *Streptomyces* colonies appear to produce cocktails of different molecules (Yim et al, 2007). Perhaps the most interesting observation about the role of NPs in the *Streptomyces* lifestyle is that many antibiotics in high (therapeutic) concentrations can inhibit the growth of a microorganism, while in lower (natural) concentrations the same compounds can alter its growth and developmental programs, that is, antibiotics can act as signaling and regulatory molecules of biological processes (Yim et al, 2007; Fajardo and Martinez 2008), rather than only as chemical weapons e. g. The SapB lanthipeptide produced by *S. coelicolor* has antibacterial activity against other species but is also important for aerial mycelium development of *S. coelicolor* itself (Kodani et al, 2004). Unfortunately, since most of the research on natural products focus on its inhibitory role, its role as regulatory molecules that mediate biological processes and species interactions is not entirely clear.

It is clear, however, that *Streptomyces* NPs are molecules that allow these species to adapt to their environments. Clear examples of these functions are metal chelators, molecules that trap insoluble metals and allow their assimilation (Barona-Gómez et al, 2006; Kallifidas et al, 2010) and carotenoids (Takano et al, 2005) that prevent damage caused by solar radiation. In some cases, *Streptomyces* can produce molecules that allow them to invade a host, e.g. *Streptomyces scabies* a potato plants pathogen produces thaxtomin, a molecule chemically similar to many antibiotics which function is to allow tissue invasion through the inhibition of the cellulose biosynthesis in the plant (Loria et al, 2008).

Some of the most striking examples of species interactions include antifungals, even of clinical use. Remarkably, some compounds with such activity have no antibacterial activity, and therefore could not be used to compete with other bacteria, so apparently the selection of these molecules is exclusively due to their antifungal role (Barke et al, 2010; Haeder et al, 2009): The leaf-cutter ants of the genus *Acromyrmex* which have a mutualistic relation with fungus of the genus *Leucoagaricus* on which the ants cultivate fungal gardens in exchange for food, also maintain a symbiotic relationship with *Pseudonacoradia*, *Amycolatopsis*, (also mycelium and NP producing *Actinobacteria*) and *Streptomyces* genera members (Seipke et al, 2012). These ants suffer from the attack of a pathogenic fungus from the genus *Escovopsis* which invade their fungal gardens threatening their survival. It has been shown that the actinobacterial symbionts of ants live in specialized structures in their bodies (Currie et al, 2006) and are responsible for producing NPs that can inhibit the fungal garden invasion (Currie et a1, 1999). Among these compounds candicidin stands out, this compound is an antifungal used in human therapy against fungal infections. It has been demonstrated that candicidin is also effective in combating the invasive fungus while is innocuous to the mutualistic one. This is a function that does not directly benefit the organism producing the metabolite, as would be the case of direct inhibition of a competitor, but is also evident that ants play an important role in the survival and dissemination of their actinomycete symbionts.

Another case is that of the predatory wasp *Philanthus coronatus*, which possess specialized organs on their antennae were its symbiont *Streptomyces philantii* can grow. The nesting habits of these wasps includes oviposition in holes on the ground, where the larvae is nourished on the bodies of paralyzed honeybees that the mother deposited in the nest. The larvae form cocoons where they develop during the winter. Due to the warm, moist and nutrient-rich conditions of the nests, they are ideal for the growth of harmful bacteria and fungi. To prevent the infection of the larvae, the female wasp impregnate the nest with a paste, mainly constituted by the mycelium of *Streptomyces philantii* using its specialized organs. The larvae incorporate this secretion to their cocoons, and thus is inoculated with the symbiont (Kaltenpoth et al, 2005). It has been shown that these *Streptomyces* symbionts provide a NPs cocktail that can inhibit a broad range of entomopathogenic organisms including fungi and other bacteria (Kroiss et al, 2010). These two examples are very useful to show how complex relationships that have been built over millions of years, and include the development of habits and specialized structures for symbiosis have emerged around NPs. Moreover, these examples have shown that before Fleming, Waskman or any human being, other living beings have been exploiting the chemical potential of *Streptomyces*.

### 1.2.3 *Streptomyces* central metabolism

Since members of the genus *Streptomyces* are generally saprophytic, the main sources of nutrients are the most nature abundant polysaccharides, such as chitin, cellulose and xylan (Swiatek et al, 2012). These compounds form part of the membranes and structural elements of plants, insects, fungi and bacteria, including the cell walls of *Streptomyces*. In order to use these nutrients, *Streptomyces* possess a wide range of extracellular enzymes. e. g. *S. coelicolor* has 60 proteases, 13 chitinases, 8 cellulases and 3 amylases (Bentley et al, 2002; Chater et al, 2010). These enzymes degrade complex macromolecules to units that can be transported and metabolized. The monomer unit of chitin, N -acetyl -glucosamine (GlcNAc), is in fact the preferred *S. coelicolor* substrate (Swiatek et al, 2012) it has been shown that the main transport system coupled to the transfer of phosphate groups (phosphotransfer system or PTS) is used to preferentially transport this compound (Nothaft et al, 2003). The study of *Streptomyces* central metabolism has revealed that the availability of GlcNAc is a defining checkpoint for the metabolic fate of these organisms *Streptomyces* transport GlcNAc monomers through the PTS, this molecule enters the cytosol and is immediately phosphorylated, being phosphoenolpyruvate (PEP) the phosphate group donor for this reaction, this molecule, in turn, is acetylated and deaminated in consecutive steps, then is ready to enter glycolysis as fructose-6-phosphate (F6P), which is also the entry point for other carbohydrates commonly used as carbon sources (Figure 4).

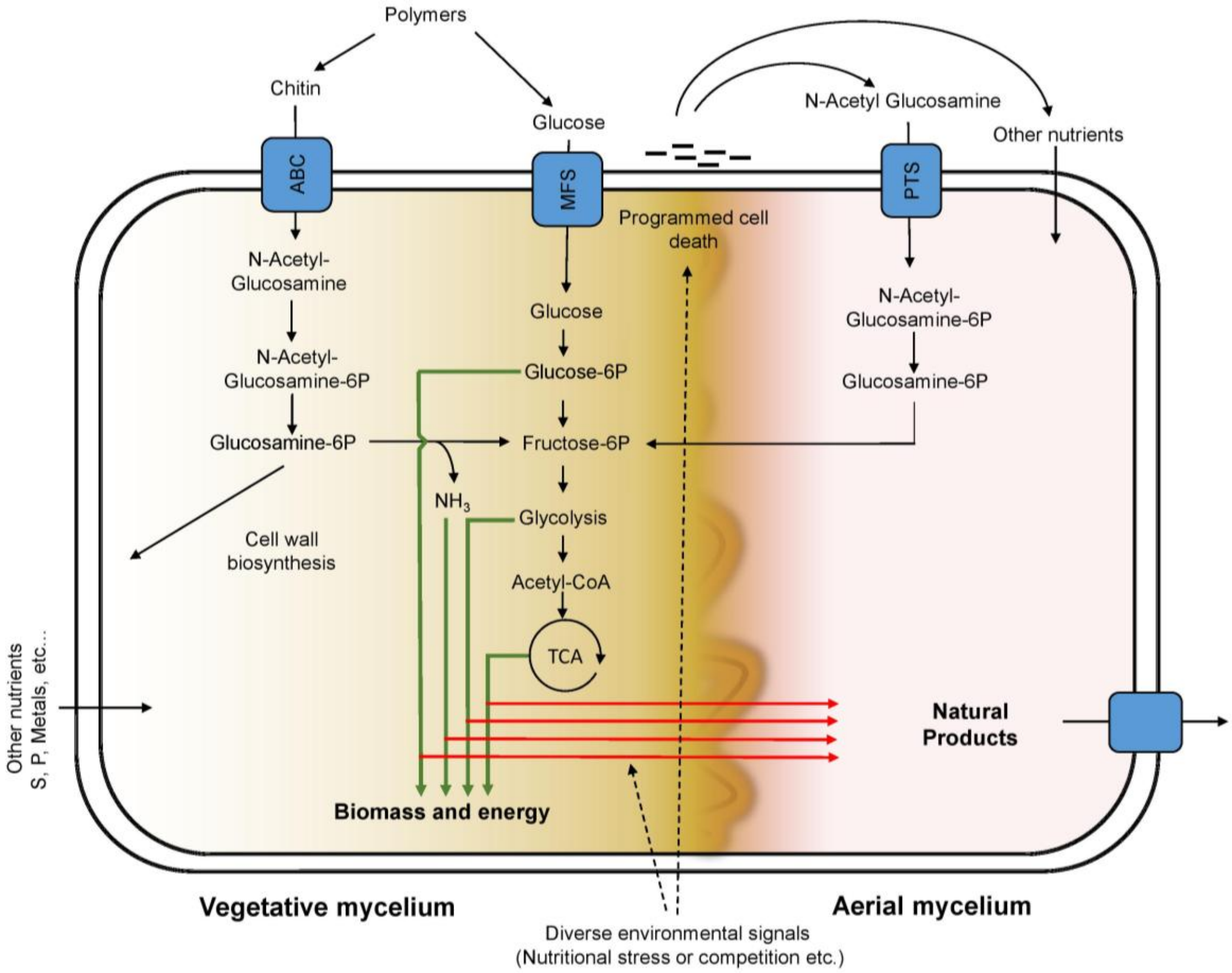

**Figure 4. Links between *Streptomyces* metabolism and development**. The right side of the figure represents a *Streptomyces* cell as part of the vegetative mycelia, in which most of the metabolic processes are devoted to biomass production, GlcNAc is obtained from polymers available in the environment such as chitin. The left side of the figure represents a *Streptomyces* cell as aerial mycelium, is in this stage in which the biosynthesis of NPs occurs, GlcNAc is acquired directly from the environment due to the lysis of the vegetative mycelia, which is in turn triggered by diverse environmental stresses.

In nutrient-rich conditions, i. e. in not-limiting concentrations of phosphorus, nitrogen, carbon, sulfur and other trace elements, GlcNAc comes from complex polymers, primarily chitin and the metabolic processes are in general directed toward the production of biomass, in other words, towards the formation of vegetative mycelia, while the development of aerial hyphae, sporulation and the synthesis of NPs, are repressed (Rigali et al, 2008; Rigali et al, 2006). In the absence of adequate nutritional conditions *Streptomyces* undergoes programmed cell death of vegetative mycelium. It has been shown, that a high concentration of GlcNAc, probably because of the action of autolytic enzymes that degrade the polymers forming part of the membrane, combined with the absence of other nutritional elements, triggers the processes leading to cell differentiation, aerial hyphae development, synthesis of NPs and finally sporulation (Rigali et al, 2008; Rigali et al, 2006)  This GlcNAc-mediated signaling mechanism, could explain how *Streptomyces* are able to "decide" its metabolic fate depending on environmental conditions: In case of abundance, growth is favored, and in case of famine, the sessile structures are destroyed, NPs are produced to protect the resources of the colony and the spores, that are the more resistant cell form, are formed, in that way, the next generation of *Streptomyces* can escape from a limited environment and possibly find a suitable place to restart their life cycle.

Besides the control of the concentration of GlcNAc to induce NP metabolism (Rigali et al, 2008), the concentration of phosphate regulates the deployment of the *Streptomyces* development program. A *Streptomyces* growing in phosphate-limited medium, but rich in other nutrients, is generally able to produce biomass, but phosphate limitation triggers NP metabolism, this induction results in early and abundant production of NPs (Liras et al, 1990; Chouayekh and Virolle 2002), this phenotype is often exploited to exacerbate their production.

To survive in complex media, the microorganisms have developed nutrient utilization systems that operate in such way that the optimal substrate for the organism is selected, discriminating alternative substrates which may be more abundant but have lower energy efficiency (Görke and Stülke, 2008). This mechanism is conserved in all organisms and is known as carbon catabolite repression (CCR), CCR is normally mediated by the PTS system, however, since in *Streptomyces* glucose transport occurs through a member of the major facilitator superfamily (MFS), (van Wezel et al, 2005) is in fact the glucose kinase, the enzyme that phosphorylates glucose after entry through MFS, who is responsible for the regulation of this process (Angell et al, 1992; Kwakman and Postma, 1994). Glk mediated CCR system is primarily responsible for the utilization of carbon sources, induction of central metabolic enzymes and production of NP precursors. Additionally, a glk-independent CCR system has been found in *S. coelicolor*, this system affects the urea cycle and the production of certain signaling molecules (Gubbens et al, 2013). In general the topology and nature of metabolic reactions used to produce the building blocks necessary for growth in *Streptomyces* are not very different from other organisms. *Streptomyces* are generally obligate aerobic organisms that can tolerate the temporary absence of oxygen (van Keulen et al, 2007). Their metabolic repertoire includes oxidative phosphorylation, glycolysis, the pentose pathway, the TCA cycle, which in many members of the genus includes glyoxylate pathway as well as the biosynthetic pathways for the production of all proteinogenic amino acids, nucleic acids and most cofactors.

# 1. 3 The biosynthesis of natural products

## 1.3.1 Natural product biosynthetic gene clusters

The extraordinary biosynthetic capacity of the members of the genus *Streptomyces* and other family members of the actinomycetes has been highlighted through this thesis. The exploitation of this chemical diversity for human benefit has also been acknowledged, as well as how this chemical repertoire is used by their owners to survive in complex and changing environments. It is now necessary to look at the genetic, chemical metabolic and enzymatic basis of this diversity.

The first genetic studies on the biosynthesis of natural products can be traced to 1979 when Hopwood and Rudd determined the location of the genes responsible for the biosynthesis of actinorhodin, a pigment with antibiotic activity produced copiously by *Streptomyces coelicolor* which was named after this compound (*coelicolor* means sky-colored in Latin). Through the use of mutants lacking the pigment, co-synthesis (two mutants growing together could produce the compound), genetic mapping and allelic crosses, Rudd and Hopwood determined the order in which the products of the genes should act to produce actinorhodin and unambiguously established that these genes would be grouped in a discrete region on the chromosome (Figure 5). Given this gene organization, they proposed on their original article, that the cloning of what they called a cluster of genes, perhaps in the form of operon, was feasible given the advances in the study of plasmids and the transformation of *Streptomyces* via protoplasts. It took five years for this goal to be achieved. In 1984 Malpartida and Hopwood cloned and heterologously expressed an antibiotic biosynthetic gene cluster for the very first time. In their 1984 article, Hopwood and Malpartida went further again and envisioned that cloning antibiotic biosynthetic cluster will have a high impact on biotechnology. These pioneer statements were confirmed multiple times: All NPs biosynthetic pathways that have been studied in bacteria maintain a genetic organization similar to that described by Rudd and Hopwood for actinorhodin, suggesting that evolution has favored an adjacent location for the genes involved in NP biosynthetic pathways. One of the implications of this genetic organization is that the regulation of these genes is finely mediated, which is compatible with the complex program of development of *Streptomyces* and the concerted action of the enzymes involved, the presence of suitable precursors and even an accelerated production of these compounds.

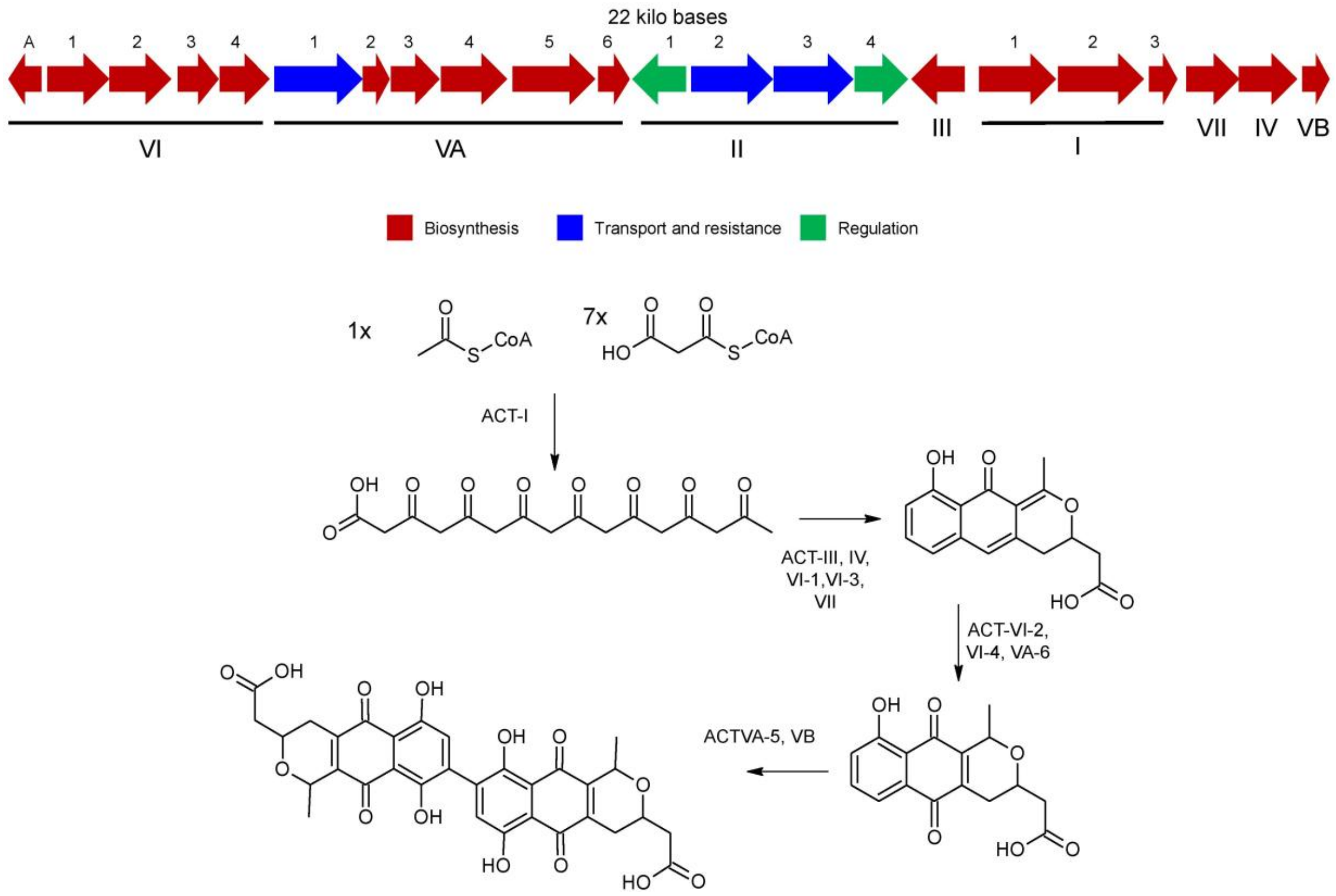

**Figure 5. The actinorhodin biosynthetic gene cluster (ACT).** All the genes needed for actinorhodin biosynthesis, regulation, transport and resistance are located within a continuous DNA region. This genetic organization is common to all bacterial NP biosynthetic pathways.

### 1.3.2 The biosynthetic logic of the natural product megasynthases

Six years after the cloning of the actinorhodin biosynthetic cluster, the cloning (Tuan et al, 1990) and sequencing of the genes that direct the biosynthesis of the polyketide erythromycin was achieved. Erythromycin is a macrolide compound produced by *Saccaropolyspora erythraea* an *Actinobacteria* that also possess a broad metabolic repertoire. This development represented a key step in deciphering the biosynthetic secret responsible for a large part of the known diversity of NPs. In subsequent work, developed independently and published almost in parallel between 1990 and 1991, Donadio and Cortes identified three large genes encoding 28 domains that together constitute the 200 KDa megasynthase responsible for the biosynthesis of the erythromycin aglycone (a cyclic molecule derived from polyenol which is subsequently glycosylated to form the final product). The nature and organization of the domains resulted highly similar to that of the fatty acid synthases (FAS), and it was determined that the complex acts as an assembly line in which acyl units extend the precursor polyketide molecule and are finally lactonized (Donadio et al, 1991). This biosynthetic logic proved to be the general strategy through which most of the polyketide NPs are synthesized. A similar domain architecture was defined for the non-ribosomal peptide synthases, another class of megasynthases involved in the biosynthesis of several antibiotics including emblematic compounds such as gramicidin (Stachelhaus et al, 1995) and penicillin (Smith et al, 1990). These two groups of enzymes, polyketide synthases (PKSs) and nonribosomal peptide synthetases (NRPS) are the most representative of the NP metabolism and account for the biosynthesis of the majority of known compounds for the past 80 years. For this reason, the elucidation of its biosynthetic logic paved the way to solve the biosynthetic pathways of a large number of compounds.

After two decades of characterization of polyketide and non-ribosomal peptide biosynthetic systems the detailed biosynthetic mechanism of these enzymes is deeply known nowadays (Figure 6). The substrates for the megasynthases are activated with a handle molecule, in the case of polyketide synthases, the extension and starter units are thioesters of coenzyme A, while amino acids that feed the non-ribosomal peptide synthases are loaded by adenylation domains (A) which require ATP. These domains as well as acyl transferase domains (AT) in the PKSs, maintain substrate specificities for each monomer unit that enters the assembly line, while the successive condensation of the monomers is conducted in the condensation domains (C) of the NRPS or ketosynthase domains (KS) in the PKSs.

In both cases the monomers and subsequent intermediates are carried by phosphopantetheinyl molecules, to which they are attached as thioesters. Additionally, during the assembly process, certain modifications may be made prior to the release and / or cyclization of the product, in the PKSs, reductions (KR), dehydration (DH) and enolization processes (ER) can be performed with their corresponding domains. In the case of NRPSs these modifications may be epimerizations (E) (Hertweck, 2009; Walsh and Fischbach 2010). In other words a subunit of these megasynthases capable of condensing two monomers (acyl-CoAs or amino acids in a polyketide or a peptide, respectively), must contain at least two recognition and activation domains (AT / A) two phosphopantetheinyl carrying domains (ACP / PCP) and a condensation domain (KS / C).

The key to the versatility of these enzyme complexes is its modularity: Just a change in a module of monomeric substrate recognition is needed to change the chemical structure of the final product of the pathway. Adding modification modules adds extra diversity to the product. Furthermore, the compatibility of these systems has been demonstrated by the discovery of PKS-NRPS hybrid synthases that further expand the chemical repertoire derived from these modular systems. Currently, a large number of domains for the recognition of different amino acid and acyl groups is known (Ansari et al, 2004). These building blocks can be the same as those used for the analogous megasyntahses in central metabolism: acetyl / malonyl-CoA to PKSs and FAS, and proteinogenic L-amino acid to for ribosomes and NRPSs. However, the advantage of the PKSs and NRPSs is that their substrate activation domains may also recognize unique precursors: PKSs can incorporate acyl groups such as butyryl- acetoacetyl-, succinyl-, cyclohexenyl-, etc., while the NRPSs are able to incorporate D amino acids and all kinds of non-proteinogenic amino acids produced *ex-professo* for the biosynthesis of NPs, or which are intermediates in other central metabolic pathways such as ornithine (Figure 6).

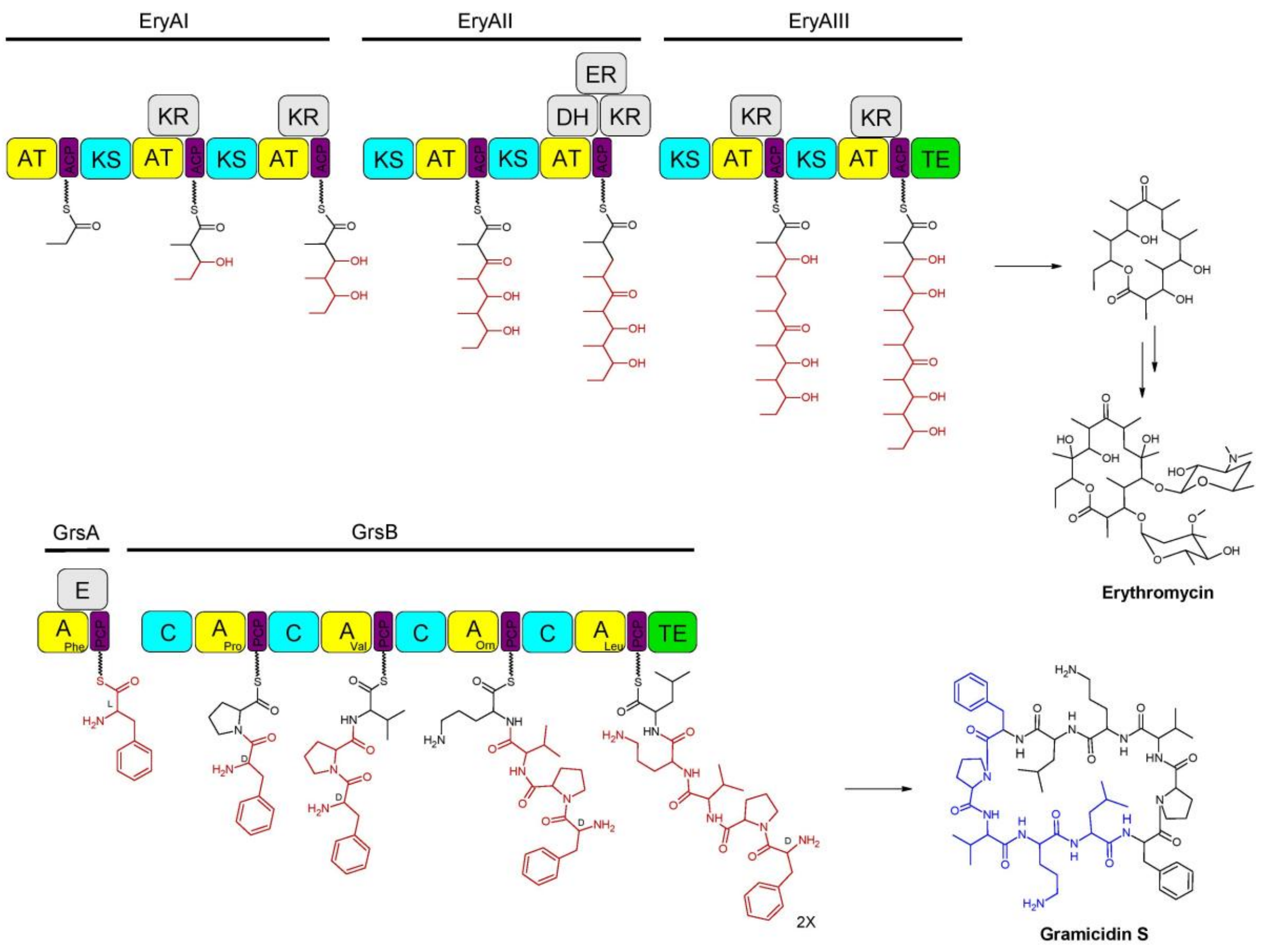

**Figure 6. Erythromycin and Gramicidin S synthases, as models of the biosynthetic logic of polyketide and non-ribosomal peptide assembly.** Equivalent domains of both systems are colored in identical colors, while the accessory domains are marked in gray. The precursors that are incorporated in each step are marked in black and previously incorporated precursors are marked in red.

### 1.3.3. Natural products biosynthetic pathways beyond megasynthases

A typical NP biosynthetic gene cluster also includes other enzymes besides megasynthases such as genes encoding enzymes responsible for the modification and incorporation of precursors generally from central metabolism, e. g. the calcium-dependent antibiotic (CDA) produced by *S. coelicolor* contains a methyl-glutamate residue, which is obtained through the action of a methyl transferase which incorporates the methyl group to a molecule of alpha-ketoglutarate, the amination of this molecule results in methyl-glutamate which is incorporated CDA peptide backbone via an NRPS (Mahlert et al, 2007). Once the biosynthetic process to produce a natural product is completed, it is likely that the final product is able to exert its function, such as a metal chelation (Barona -Gomez et al, 2006) or antibiotic activity, since these functions must occur outside the cell, these molecules should be exported, otherwise, in the case of antibiotics, the accumulation of a toxic molecule will have negative consequences in the producer organism, to prevent this and allow the action of the molecules against their possible targets, NPs biosynthetic gene clusters also include genes encoding transporters, in many cases these systems are efflux pumps, these transmembrane proteins allow active transport of pathway end products in coordination with the biosynthetic enzymes, thereby preventing accumulation of the metabolite, this provides a natural mechanism of resistance and simultaneously release of the molecules to the environment (Méndez and Salas, 2001). The biosynthetic gene clusters may also include enzymes that modify the final product to avoid suicide, e. g. an acetyl transferase responsible for resistance to the phosphinothricin tripeptide, which biosynthesis is discussed below, (Thompson et al, 1987) or enzymes that modify the antibiotic molecular targets, e. g. The avilamycin biosynthetic gene cluster, that produces an oligosaccharide with antibiotic activity which targets the 23s ribosomal subunits, includes two genes coding for rRNA methylases, which modify the ribosome, preventing the producer organism *S. viridochromogenes* T57 to commit suicide (Treede et al, 2003).

Other biosynthetic gene clusters may include genes homologous to the target molecule, but with modifications that make it immune to their action, e. g. in the case of pentalenolactone, a sesquiterpene produced by *S. avermitilis* and *S. arenae* that specifically inhibits the activity of the glyceraldehyde 3-phosphate dehydrogenase enzyme (3PGDH), the biosynthetic gene cluster includes a 3PGDH paralog that is insensitive to the inhibition by pentalenolactone (Tetzlaff et al, 2006), when the toxic compound is produced the sensitive version of 3PGDH is replaced by the resistant version (Frohlich et al, 1989).

Finally NPs biosynthetic gene clusters include genes coding for regulatory proteins that coordinate the expression of the genes in the cluster and respond to physiological and environmental signals. The regulation of the biosynthesis of NPs is a complex phenomenon that goes beyond the interest of this thesis, so it will not be further addressed, however, the in-depth review that Van Wezel and MacDowall (2011) have wrote about the topic will be very useful for the reader interested in this fascinating subject.

### 1.3.4 Other natural product biosynthetic pathways

Despite that a substantial fraction of known NPs are synthesized by PKS and NRPS enzymes, a large and growing number of NPs are produced by other enzyme systems including extraordinary chemical conversions. This section will address some of these alternative biosynthetic systems, which I have selected for their relevance to this work, as will be seen later.

#### 1.3.4.1 Ribosome-assembled natural products

A large number of NPs are peptides, many of these, are produced by NRPSs. These biosynthetic systems provide the advantage of being able to incorporate modified amino acids that the ribosome could not recognize, and to which tRNAs are not available. Even so, a large number of NPs include peptide fragments which are restricted to proteinogenic amino acids encoded by small genes that are ribosomally translated (Velasquez and Van der Donk, 2011). The consequent limitation of chemical diversity derived from the use of only 20 chemical precursors is sorted by a large number of enzymes capable of post-translationally modify these amino acids forming chemical structures with the distinctive properties of this family of NPs, known as lantipeptides. Examples of these enzymatic conversions can be seen in the structure of thiostrepton (Figure 7), a compound with antibiotic activity produced by *Streptomyces laurentii* (Kelly et al, 2009).

The biosynthetic gene clusters of such compounds include a small gene encoding the peptide precursor, this gene include a region that is known as the leader peptide, which is removed leaving only the necessary amino acids for the synthesis of the NP. The peptide precursor is subsequently modified by a series of enzymes which may result in non-proteinogenic amino acid equivalent residues. A cyclization reaction provides a specific three-dimensional structure with rigidity which allows them to perform their function. In many biosynthetic systems of this class, this reactions are catalyzed by lanthipeptide dehydratases and cyclases, which proceed through the dehydration and condensation of serine and threonine residues in the peptide precursor, producing meso-lanthionine or 3-methyl-lanthionine residues. The relevance of this conversion is that the cyclization of the molecule occurs by joining contiguous residues in the peptide sequence (Knerr and van der Donk, 2012).

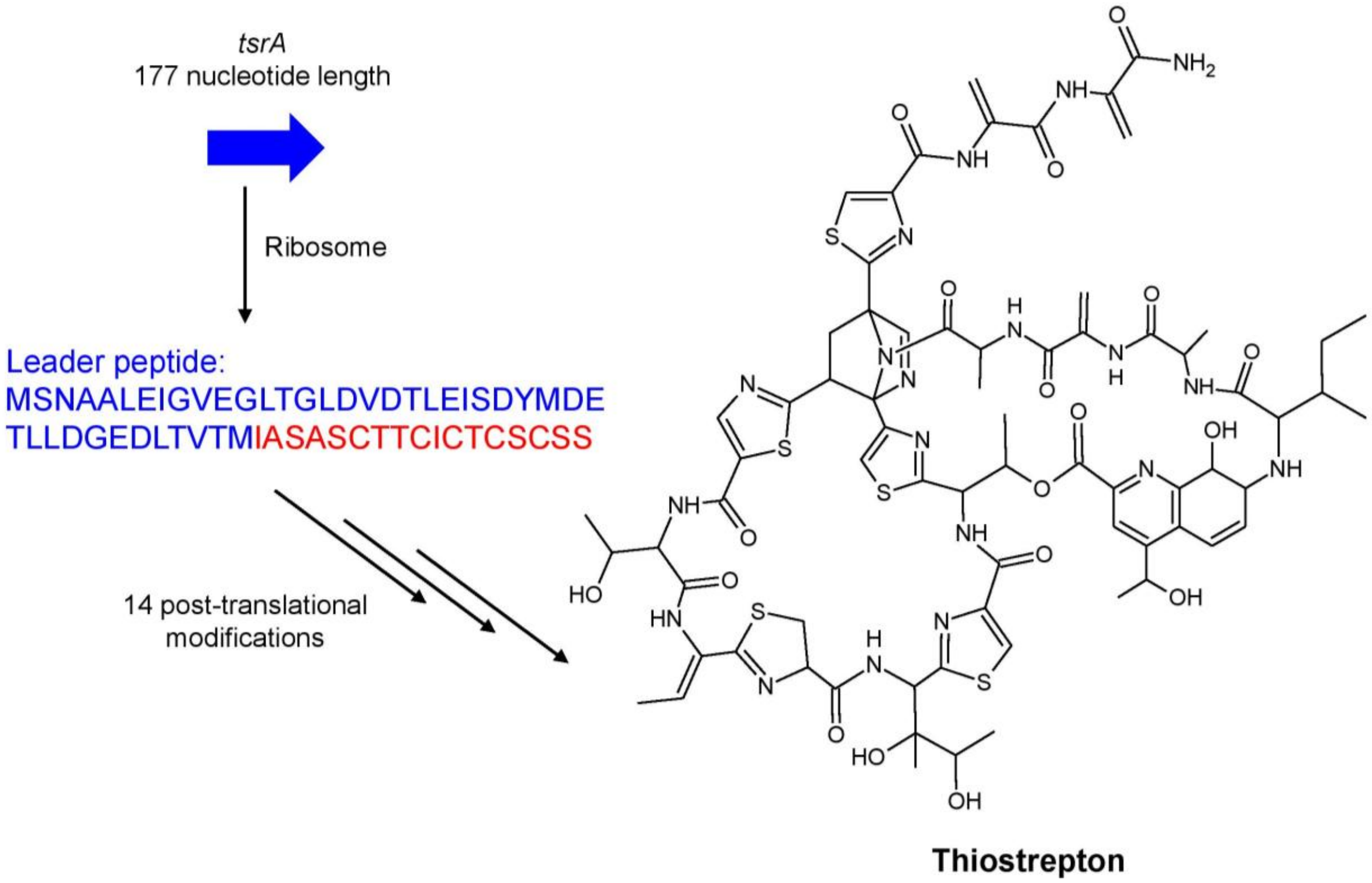

**Figure 7. Thiostrepton, a ribosomally assembled peptide natural product**. The thiostrepton precursor, TsrA, is a post-translationally modified peptide synthesized by the ribosome.

### 1.3.4.2 tRNA-dependent non-ribosomal peptides

In the last 5 years, NPs biosynthetic enzymes that exploit the versatility of the tRNAs for the condensation of amino acids have been discovered. These enzymes which have no evolutionary or structural relationship with the ribosome or NRPS systems, use proteinogenic amino acids activated as aminoacyl tRNAs (Figure 8) for the formation of amide bonds by a condensation reaction. The first case (Garg et al, 2008) was identified in the biosynthetic pathway of valanimycin (Figure 8), in which a homologue of MprF, an enzyme that transfers an amino acid residue to the phosphatidyl glycerol of cell walls in some organisms, is responsible for the condensation of an ester derivative of valine and a serine residue activated by a tRNA. This molecule in turn serves as a precursor for the formation of the characteristic azoxy group of valanimycin. The second example was discovered in the biosynthetic pathway for albonoursin, an antibiotic containing a diketopiperazine, a six-membered ring which includes two nitrogen atoms and two ketone groups in *para* position (Figure 8). This ring is produced by the head to tail condensation of two amino acids, leucine and phenylalanine. This reaction is catalyzed by a small protein that condenses both amino acids as aminoacyl-tRNAs (Gondry et al, 2009). The most recent case reported to date was discovered in the biosynthetic pathway pacidamycin, a five amino acid peptide that is bound to a nucleoside moiety. The synthesis of a tetrapeptide is performed by a NRPS, then, the fifth residue, an alanine, is added in a post-NRPS reaction by a FemX homologue, an enzyme that peptidoglycan modifying enzyme. Again, this amino acid is incorporated in its activated form as a tRNA cognate (Zhang et al, 2011).

In these three cases, the transfer of the amino acids is enhanced by the presence of tRNA, which acts as an electron sink allowing the nucleophilic attack of the ester bond by the other amino acid, leading to the condensation of both substrates in a substrate-catalyzed reaction. This is a reaction mechanism shared by different families of biosynthetic enzymes, including ribosomes, NRPS, membrane modifying enzymes and the new tRNA-dependent enzymes related to the biosynthesis of NPs. In fact, only one more enzyme family is known to be able to use tRNA-activated amino acids for the formation of peptide bonds: The Leucyl/Phenylalanil-tRNA protein transferases (LFT) that are involved in the N-rule proteolytic pathway. These enzymes add leucine or phenylalanine residues at the N- terminal end of other folded proteins which marks them for degradation via the proteasome (Leibowitz and Soffer, 1969, Fung et al, 2011). Given this function, it has been speculated that this enzyme family could be involved in the biosynthesis of NPs (Zhang et al, 2011), although to date not a single biosynthetic gene cluster that includes a LFT homologue has been identified.

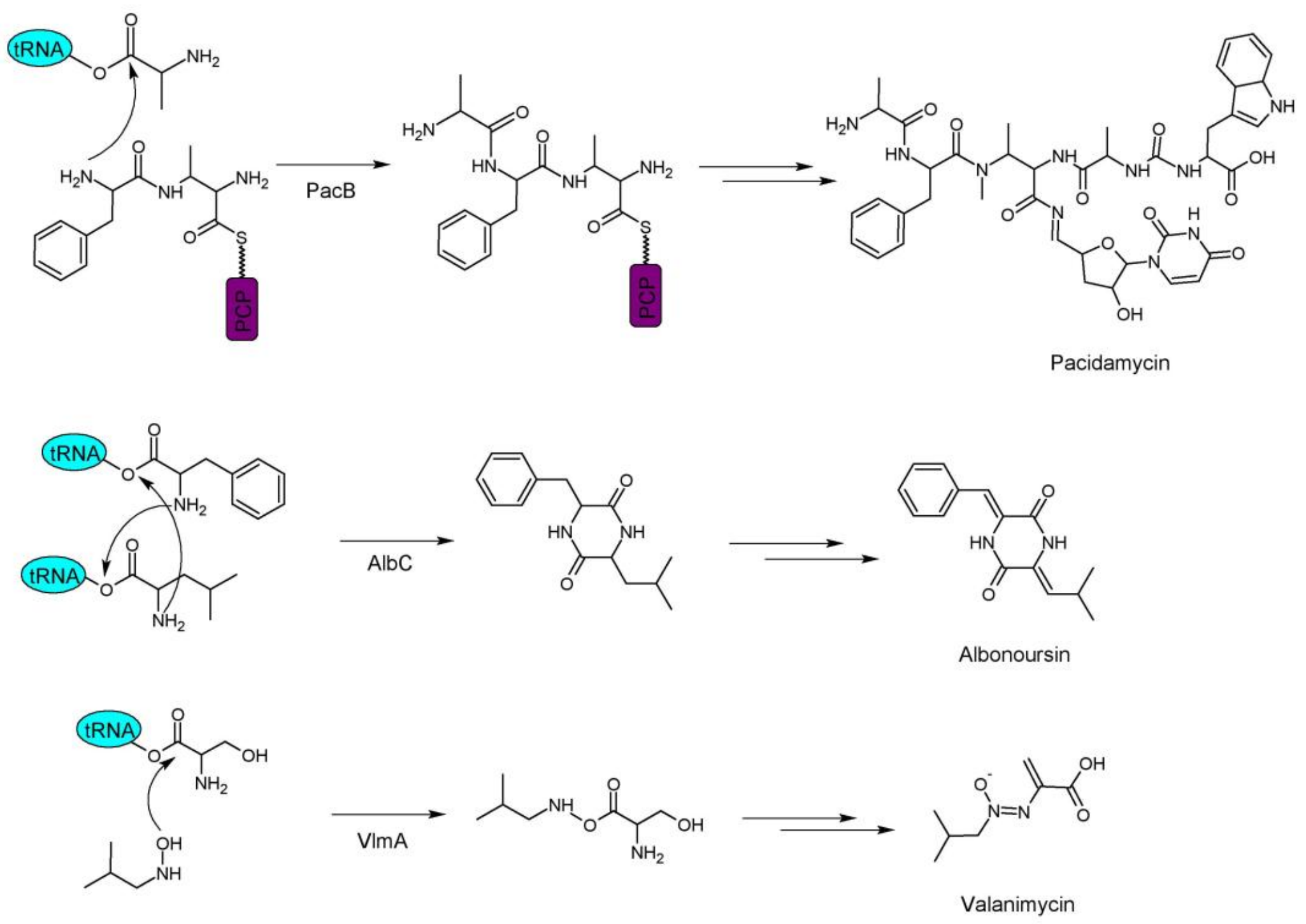

**Figure 8. tRNA-dependent enzymes in natural product biosynthesis.** PacB, AlbC and VlmA share a common reaction mechanism that is based in a nucleophilic substitution for the formation of peptide bonds. The figure shows the substrates and products of these enzymes as well as the the final products of their corresponding pathways, pacidamycin, albonoursin and valanimycin

### 1.3.4.3 Carbon-phosphorous bond containing natural products

Some NPs mimic molecules that are part of the central metabolism, however, these molecules can cause inhibition when recognized by their target enzymes. An example of such NP is the phosphinothricin tripeptide, a compound that contains two alanine residues and a glutamate analogue called phosphinothricin which includes a carbon-phosphorus bond (CP) (Figure 9). Its structural analogy to glutamate allows it to be recognized by the glutamate synthase but also prevents its modification by this enzyme, thereby inhibiting its activity and affecting the growth of the organism that has incorporate it on its metabolism. Since phosphinothricin is not able to cross the membranes by itself, the molecule is bound to other amino acids via a NRPS, which facilitates its active transport as a peptide. Once inside a cell the tripeptide is degraded releasing the toxic molecule (Schinko, 2009).

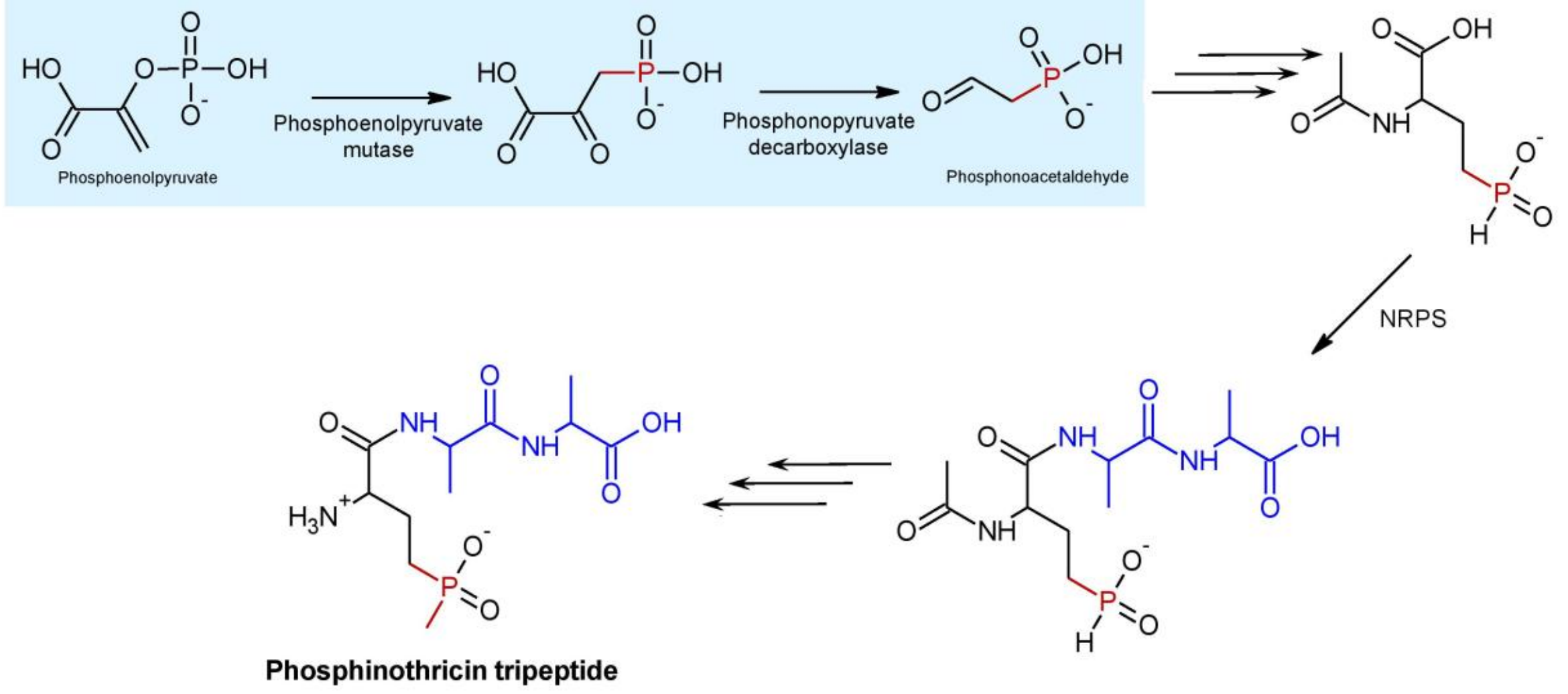

**Figure 9. Phosphinothricin biosynthesis.** The two common reactions for phosphonate and phosphoinate biosynthesis are highlighted in a blue box. The next biosynthetic steps towards the formation of the P-containing amino acid that enters the NRPS assembly line had been summarized, two more biosynthetic steps are required to form the phosphinothricin tripeptide a phosphinate natural product.

The carbon-phosphorus bond on phosphinothricin is formed from phosphoenolpyruvate (PEP), a high energy metabolite formed in the glycolysis. This molecule is isomerized by a specialized phosphoglyceromutase that exchanges the ester bond (P-O) to a carbon-phosphorous bond (C-P) leaving a carboxylic group where there was an enol. Because of the high activation energy of this reaction, this conversion is not favorable unless is coupled to the action of a decarboxylase which releases the newly formed carboxylic group as $CO_2$. This strategy makes this reaction highly favorable, so the balance is irreversibly brought into phosphonoacetaldehyde formation (van der Donk and Metcalf, 2009). Other functional groups are added to this molecule and after a series of consecutive reactions phosphinothricin (Phn) is produced. Phn enters an NRPS assembly line that produces the tripeptide Ala-Ala-Phn (Schwartz et al 2004; Blodgett et al, 2005) (Figure 9).

The first two steps of this biosynthetic pathway are essential for the formation of C-P bond. In fact the biosynthesis of all NPs including C-P bonds involves one mutase, and in most cases a decarboxylase.

The only exception is fosfomycin biosynthesis, an inhibitor of cell wall biosynthesis. This compound can be synthesized by two different routes, the first discovered in *Streptomyces fradieae*, includes the mutase-decarboxylase pair (Woodyer et al, 2006), whereas the second version of the biosynthetic pathway found in the genome of *Pseudomonas syringae* lacks the decarboxylase. In this case the mutase reaction is coupled to the reaction of a homologue of the citrate synthase, as in the mutase-decarboxylase system, the citrate synthase reaction switches the reaction equilibrium reaction towards the C-P bond formation (Kim et al, 2012). Other known phosphonate biosynthetic pathways include rhizocticin, a tripeptide that inhibits threonine synthase and is produced by *Bacillus subtilis* (Borisova et al, 2010), and the compound FR900098, an antimalarial agent that inhibits the synthesis of isoprenoids and is produced by *Streptomyces rubellomurinus* (Eliot et al, 2008).

Other classes of NPs such as oligosaccharides, aminoglycosides, isoprenoids, nucleosides, hydroxamates and phenazines, are biosynthesized by enzymes that together with the previously described enzymes form a remarkable repertoire of biocatalysts which are responsible for the production of a seemingly infinite number of chemical structures. To review all these enzymatic functions in one document is a formidable task that is beyond the aim of this thesis. The biosynthetic systems that I have reviewed were selected due to its ubiquity and its relevance in the context of my research, however, the reader interested in this fascinating subject can find more information on the excellent reviews written by Knerr and van der Donk (2012) on ribosomal peptides, Metcalf and van der Donk (2009) on C-P containing NPs, Hertweck (2009) on polyketides, Park et al (2013) on aminoglycosides, and the general review on NP biosynthesis by Zhang and Walsh (2010), among others.

## 1.4 Cryptic biosynthetic gene clusters.

### 1.4.1 The discovery of cryptic clusters

The first complete genome sequence of a *Streptomyces* was that of *S. coelicolor* (Bentley et al, 2002). Its analysis revealed the presence of at least 20 NP biosynthetic gene clusters, a number of putative compounds larger than the number of compounds known by then in that organism. This same scenario was later found in the genomes of *S. avermitilis* (Ikeda et al, 2003), the producer of avermectin, an insecticide and anthelmintic compound, and in *S. griseus* (Ohnishi et al, 2008), the streptomycin producer organism. In the last decade, with the technological advances in genome sequencing, the genomes of dozens of species of the genus *Streptomyces*, and other antibiotic-producing bacteria have been sequenced and deposited in public databases (Table S2). The analysis of a large number of these genomes, especially the NPs producing species belonging mainly to the *Actinobacteria* phyllum (Donadio et al, 2007, Nett et al, 2009; Doroghazi and Metcalf, 2013), has revealed the presence of a huge number of NP biosynthetic gene clusters and increased the number of known and chemically characterized metabolites. Although many of these systems could produce variations to known NPs an huge fraction of the predicted systems are unknown, therefore these biosynthetic gene clusters have been called "cryptic clusters" (Zerikly and Challis, 2009). This observation implies that the unique biosynthetic ability of these organisms, despite having been exploited extensively in the past, has been underestimated. This new scenario has revitalized research in NPs produced by *Streptomyces*, which had declined after the golden age of discovery of antibiotics, mainly due to cost and little success in discovering new interesting molecules using conventional techniques (Nett et al, 2009).

Perhaps the main reason why many cryptic biosynthetic gene clusters were not discovered before the genome sequencing became affordable for many NP research groups, is the lack of suitable conditions for their expression in laboratory conditions. Due to the adaptive nature of NPs, it is expected that the absence of factors associated with the environment in which the strain was isolated (including axenic culture) prevents expression of the entire metabolic repertoire of these organisms. The challenge of the lack of expression of biosynthetic clusters has been addressed with the use of genetic engineering: "Superhost" strains for heterologous expression of NP biosynthetic gene clusters have been obtained from *S. coelicolor* (Gomez-Escribano and Bibb, 2012) and *S. avermitilis* (Komatsu et al, 2013), in these strains the native NP biosynthetic gene clusters have been removed, so that the precursors from central metabolism can be used by the heterologously introduced biosynthetic pathways, thus enabling the production of the metabolites that are difficult to obtain on their original physiological contexts.

### 1.4.2 Analytical methods for cryptic natural product identification

Once a cryptic biosynthetic gene cluster has been predicted, different analytical methods can be followed to identify their products. Some strategies are based on the prediction of the identity of the precursors or the structural characteristics of the product from the pathway, using chemical and biosynthetic models based on the sequence (i.e. domain organization) of the enzymes in each cluster (Figure 10). Gregory Challis one of the pioneers in this area, has reviewed these methods (Challis, 2008), including the following:

The **genomisotropic approach**. The producer organism is fed with the predicted precursors, which are labelled. The extracts from the fermentation broth are screened to identify the product of the cluster.

The ***in vitro* reconstitution approach**. The enzymes from the biosynthetic cluster are purified and incubated with the predicted substrate. The products of the reaction are determined (Zhao et, al 2008).

When the substrate specificity or product properties predictions are not confident enough, direct approaches are followed.

The **Knockout/comparative metabolic profiling approach**. This approach involves the inactivation of a hypothetically essential gene in the cryptic cluster. Metabolites absent in the mutant and present in the wild type are likely to be the product of the cryptic cluster. Prediction of putative physico-chemical properties of the products from a cryptic cluster can be used to narrow the search and identification of the product in fermentation broths of the producer organism (Lautru *et. al*, 2005).

The **heterologous expression/comparative profiling approach.** In this case the whole biosynthetic gene cluster is cloned in a suitable vector and expressed in a suitable host. The metabolites present in the transformants and absent in the non-transformed host are likely to be the product of the cryptic cluster.

**Mixed approaches**. The recent development of new mass spectrometry techniques of in parallel with the use of well-established bioinformatic methods has produced new approaches that guide the search for cryptic metabolites with real data from the metabolic repertoire of the organism under study (Kersten et al, 2011). Structural information derived from the systematic analysis of in the metabolome of an organism, in combination with the known biosynthetic logic of these compounds allows for the identification of the genes responsible for the biosynthesis of the observed metabolic products.

## 1.5 *Streptomyces* as model organisms for the study of the evolution of metabolism

### 1.5.1 *S. coelicolor* y *S. lividans*, model species for genetic studies

After the discovery of many antibiotics from *Streptomyces*, the study of the biosynthesis of these metabolites naturally derived in the entry of these bacteria to the genetics era by the hand of Professor David Hopwood. For his bacterial genetics studies, Hopwood choose *S. coelicolor* because of its copious production of actinorhodin, which provided the colonies with an intense blue color, easy to distinguish with the naked eye, this trait was an ideal genetic marker (Figure 10). This strain quickly became a favorite for the first *Streptomyces* studies, including the final classification of *Streptomyces* as bacteria (at that time they were suspected to be fungi), the detailed study of the morphological changes and developmental program, and the mapping of the genes involved in antibiotic biosynthesis (Hopwood, 1999). Another *Streptomyces* model strain, *S. lividans* 66, a close relative of *S. coelicolor*, was introduced in the early 60s by the Russian researcher Natalia Lomovskaya (Hopwood, 1999). Both bacteria belong to the *violaceoruber* subgroup of *Streptomyces*, which are characterized by the production of blue (actinorhodin) and red pigments (prodigiosines) (Figure 9). *S. lividans* 66 was adopted immediately by the new community of *Streptomyces* geneticists e. g., *S. lividans* 66 was the model strain to study phages, plasmids, and other mobile genetic elements (Hopwood et al, 1983). Moreover, it was on S. *lividans* 66 that the first linear chromosome in the genus was identified (Lin et al, 1994), this soon became a paradigm (all known *Streptomyces* chromosomes are linear molecules). Finally the first DNA thiolation system was also identified in *S. lividans* 66, this system has subsequently been identified in many other bacteria (Zhou et al, 2005).

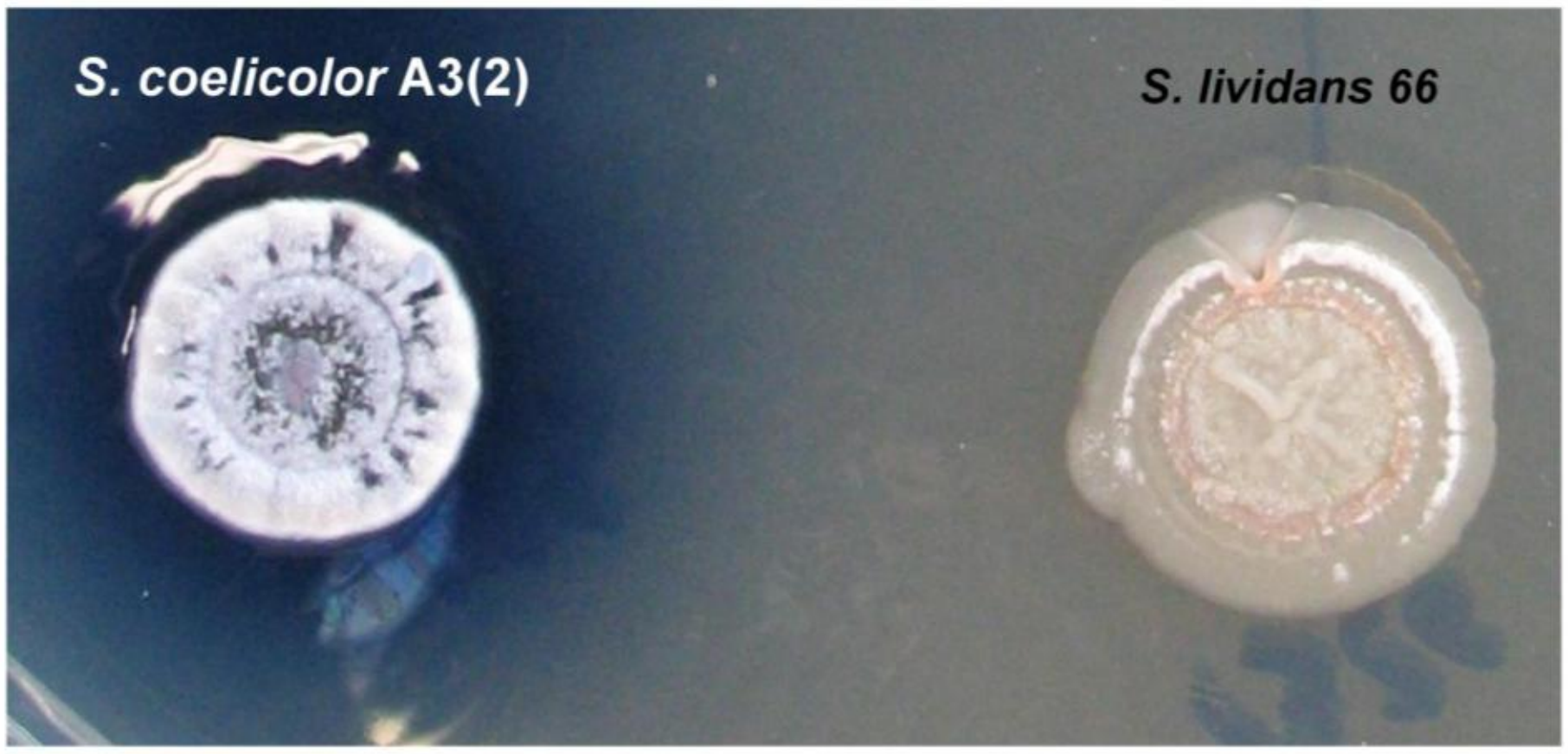

**Figure 10. *Streptomyces coelicolor* (sky-colored in Latin) and *S. lividans* (pale in Latin).** Their phenotypic differences among which the abundant production of actinorhodin is evident at plain sight.

*S. lividans* 66 also became the favorite strain for *Streptomyces* genetic manipulation, mainly because it is able to accept methylated DNA which makes it ideal for gene cloning. This feature, together with its low endogenous proteolytic activity are the main reason why this strain is widely used as a host for expression of antibiotic biosynthetic pathways and proteins (Anné et al, 2012). This has resulted in a very important role for *S. lividans* 66 and its derivatives e. g. strain TK24, in industry.

### 1.5.2 *Streptomyces* chromosome organization

The members of the genus *Streptomyces* have large genomes, with an average of 8.3 MB. (Table S1). The central region of the *Streptomyces* chromosome encodes for most of the essential functions of the organism, such as the central metabolism. This region shares a high degree of synteny with the chromosomes of other actinomycetes. In contrast, the chromosome ends or "arms" in conjunction with plasmids, are highly variable regions which are subject to constant genetic flows. These regions are rich in of horizontally transferred genes, gene loss and gene replacement events. These chromosomal “arms” are where the species-specific or contingency functions are encoded, including most of the NPs biosynthetic pathways (Bentley et al, 2002; Choulet et al, 2006). This arrangement may be a positively selected genetic trait in *Streptomyces* and other microorganisms with broad metabolic repertoires. One possible reason for this conserved genomic architecture is that the chromosomes can be selected as an evolutionary platform that provides protection for essential functions, while allowing the recruitment of new genetic material. The genetic mechanisms involved in these processes are unknown, but should include gene duplication and horizontal gene transfer, the gained genetic material could then be the starting point for the emergence of new functions needed for survival in the complex environments where these species thrive.

### 1.5.3 *Streptomyces* mobile genetic elements

Since the early genetic characterization of *Streptomyces*, the presence of plasmids was detected. The first discoveries about these mobile elements were related to their transmissibility and the phenotypes associated with their transfer known as pocks, now is known that pocks are associated with the plasmid mobilization through the mycelium (Hsu and Chen, 2010). Given that *S. lividans* and *S. coelicolor* are the most widely used model species for the study of natural product biosynthesis and that within the context of this thesis both species are analyzed from the perspective of comparative genomics, it is necessary to review what is known about their mobile genetic elements.

The presence of two plasmids called SLP2 and SLP3 was detected in *S. lividans* 66 (Hopwood et al, 1983), it has been shown that SLP2 is a linear molecule with terminal proteins that are similar to telomeres (Lin et al, 1994),this finding was later extended to all *Streptomyces* chromosomes. SLP2 has been purified using pulsed field electrophoresis and its 50 Kb have been completely sequenced (Chen et al, 1993, Huang et al, 2003).

To date, however, little is known about SLP3, although it is generally assumed that it is a molecule that remains inserted into the chromosome of *S. lividans* 66, and that can be mobilized and amplified (Eichenseer & Altenbuchner, 1994). Additionally, SLP3 has been associated with resistance to mercury, the genetic determinants of this trait have been found in the plasmids of other *Streptomyces* strains isolated from highly contaminated sites (Sedlmeier & Altenbuchner, 1992; Ravel et al, 1998). However, the location, genetic content and functions contained in SLP3 remains a mystery.

Moreover, in *S. coelicolor* two plasmids, SCP1 and SCP2, have also been detected and sequenced. SCP1 is a large molecule of 350 Kb, indeed, the presence of giant plasmids, later became a common trait among *Streptomyces*. During the characterization of the plasmid SCP1 it was found that the $SCP1^+$ strains inhibited $SCP1^-$ strains, this phenotype was associated with a diffusible compound with antibiotic activity against other organisms named methylenomycin (Vivian, 1971). Subsequently it was demonstrated that the sensitivity of the $SCP1^-$ strains was due to the fact that the genetic determinants of methylenomycin resistance were associated with the biosynthetic genes of the metabolite, therefore, the loss of the ability to produce methylenomycin came with the loss of the resistance to the compound (Bibb et al, 1980). Methylenomycin was the first example of an antibiotic encoded in a mobile genetic element, this discovery brought with it the expectation of finding more biosynthetic pathways among *Streptomyces* plasmids (Hopwood, 1999). However, for more than two decades, methylenomycin was the only example with such characteristic.

Recently, the genome of *Streptomyces clavuligerus* (Medema et al, 2010) revealed the presence of a 1.8 mega base pairs "mega plasmid" (20% of the entire genome). In this plasmid the presence of 25 NPs biosynthetic gene clusters was predicted using bioinformatic approaches. This high number of putative NP biosynthetic pathways, contrasts with the lack of central metabolism-related enzymes encoded within the genetic element, whether *S. clavuligerus* can survive without its mega plasmid is not clear, the genetic content of the element indicates that this is an extra- chromosomal element rather than a second chromosome. Regarding this possibility, the use of the term "Chromid" has been recently proposed for these elements, by this definition, the main difference between the chromosomes and chromids is that the genetic content of chromids is less conserved and their replication systems are similar to those from plasmids. According to Medema et al (2010), it is possible that the *S. clavuligerus* large extra-chromosomal element arose after a recombination event of a small mobile element that dragged down one arm of the chromosome.

A more recent example of plasmid-borne NP biosynthetic gene clusters has been found in *Streptomyces aureofaciens* CCM3239 which genome contains a linear mobile element of about 240 Kb. This plasmid encodes the biosynthetic pathway for auricin, a natural product with antibiotic activity and two more cryptic biosynthetic gene clusters (Novakova et al, 2013). Despite the relatively low number of examples of this kind, given the adaptive nature of NPs, the possibility that they can be transferred via mobile genetic elements is attractive and has been previously proposed (Kinashi, 2011).

# 2 Background

## 2.1 Theories to explain the complexity of modern metabolism

Once the *Streptomyces* ability of for the production of a large number of compounds to help them adapt to the environment has been highlighted, and after a short overview of the main biosynthetic systems responsible for the production of metabolic repertoire, this section will focus in the evolutionary aspects of metabolism. A first step to study the evolution of the NP metabolism is to recognize that the concepts of "primary" and "secondary" metabolism, which have been associated with exponential and stationary phases of growth, can be confusing given the lack of theoretical foundations for the division of metabolism in such categories. This is evident after considering the fact that the exponential and stationary phases are laboratory artifacts since many soil organisms such as *Streptomyces* do not grow and develop properly in liquid cultures. Therefore to study metabolism, is important to take into account that the same evolutionary rules apply to all metabolic pathways and the difference between them is that they evolve with different restrictions due to the force of natural selection (Firn and Jones, 2009). At this point is important to take a look at the different models that have been proposed to explain the evolutionary process that are behind the diversification of modern metabolism.

In 1945, Horowitz proposed a retro-evolution scenario for anabolism. Organisms exploiting substrates that were available in the environment, when faced whit the challenge imposed by a drop in their availability, had a selective advantage when they evolved the ability to produce their own precursors. According with this model, the first enzymes to appear in a metabolic pathway would be the most downstream, and therefore the pathways evolved backwards. In 1990, Cordon expanded these concepts for catabolism, describing a forward evolution mechanism (Cordon 1990). In this scenario, degradation of the substrates for efficient acquisition of energy will provide a selective advantage. The most upstream enzymes within these pathways would be the first enzymatic activities to appear.

Ycas (1974) and Jensen (1976) proposed that promiscuous catalytic activities provided a selective advantage and were recruited to perform new functions as the network evolved. The early recruitment of promiscuous enzymes was fortuitous and increased the opportunities for acquisition of novel functions at the level of substrate specificity. The recruited enzymes specialized to increase fitness and produced key metabolites from intermediates, even if these were available in remote reactions from the evolving pathways. This resulted in a mosaic of homologous enzymes scattered in metabolic pathways. This hypothesis is known as the patchwork hypothesis, and it assumes that there is already an active enzymatic core with multifunctional and/or specialized enzymes from which new enzymes evolve (Hughes, 1994; Caetano-Anolles *et. al*, 2009).

Finally, the shell hypothesis postulated by Morowitz in 1999 provides a conceptual framework that fit all the previous ideas. This scenario assumes that the current metabolic complexity has evolved through the sequential recruitment of elements present in the earlier stages of metabolism. In other words, the new layers of chemical complexity arose from the catalysts present in the previous layers. Following this idea, the evolutionary history of metabolism can be seen as an onion on which layer after layer of metabolic complexity are built on the top of each other. The latest layers of metabolism contain the most diverse and recently evolved enzyme functions and pathways, while the internal layers contain ancestral and universal metabolic reactions (Figure 11). From this perspective, the metabolism of NPs is the latest layer of metabolic complexity and can be assume that has recently emerged from the enzyme repertoire present in the central metabolism (Vining, 1992).

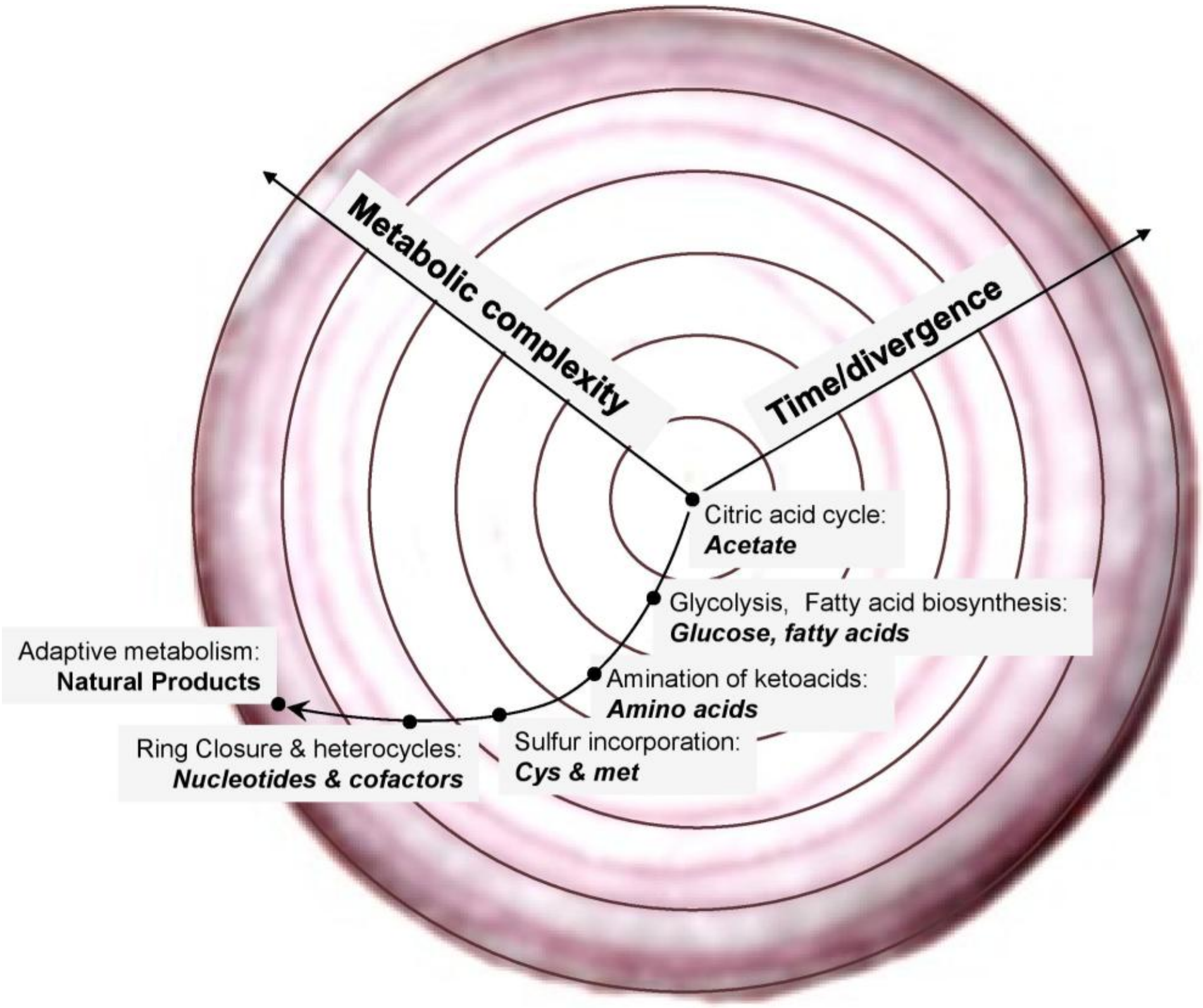

**Figure 11.** The evolution of metabolism seen as an onion. This scheme is based on the shell hypothesis by Morowitz (1999). According with this hypothesis, the current metabolic complexity. The recent layers of the metabolism evolved after the recruitment of the catalysts in the previous layers. Therefore, NPs metabolism is the latest layer of metabolism and evolved from central metabolism.

## 2.2 Genetic mechanisms for the expansion of the metabolic repertoire

Within this scenario, gene duplication (Ohno, 1970) is often assumed as the main mechanism for the functional divergence and therefore the expansion of the metabolic complexity. However, this same effect can be achieved through horizontal acquisition of homologous genes that perform the same enzymatic function (Iwasaki and Takagi, 2009). In fact, the paradigm of gene duplication as the main source of adaptive innovation has been recently delimited by Treangen and Rocha (2011) who have shown that in prokaryotes the main mechanism of protein family expansions, and therefore of adaptive innovation is horizontal gene transfer. Once the need for the release of functional constraints affecting a gene that evolves towards a new function, either by the appearance of a paralog or xenolog has been established, it is evident that the presence of an extra copy of a gene homologue is not strictly necessary for the evolution of new functions. Two additional alternatives can provide the scenario for the expansion of the metabolic repertoire:

On one hand, functional redundancy can be achieved through the acquisition of a gene encoding an enzyme analogue, i. e. non-homologous redundancy. A good example of these cases are the type I and II dehydroquinate dehydratases, which catalyze the conversion of identical substrates and products by two different chemical strategies and which have very different protein folds (Gourley et al, 1999). Systematic analysis of such cases revealed that analogous enzymes are often found in a variety of metabolic reactions (Galperin et al, 1998; Omelchenko et al, 2010). On other hand, the presence of alternative and redundant pathways that functionally converge or produce identical or equivalent metabolic products also have the effect of releasing the selection pressures. This in turn dampens the effect of mutations on the function of genes involved in redundant or alternative pathways, contributing to the evolution of new functions. The presence of metabolic shortcuts and alternative routes is indeed abundant and therefore its contribution to the evolution of new enzymatic functions should not be neglected (Ramsay et al, 2009; Firn, 2010).

## 2.3 Links between central and natural products metabolism

### 2.3.1 Physiological Links

About 8000 genes were annotated in the genome sequence of *Streptomyces coelicolor* (Bentley et al, 2002), it has been suggested that this large number of genes may well reflect its extensive metabolic repertoire and derives from the presence of new enzymes families, as well as a high number of expansions in known protein families. Many of them appear to be involved in regulation, transportation and extracellular degradation of nutrients (Bentley et al, 2002). It has even been speculated that 35% of the genome of another related specie, *Streptomyces avermitilis*, had arisen by duplication (Ikeda et al, 2003). Today this statement is no longer supported, since it has been shown that gene duplication in bacteria is less common than by horizontal gene transfer (Treangen and Rocha 2011).

The truth is that many of the expansions of known gene families may be associated with essential functions of the central metabolism, e. g., *S. coelicolor* has three phosphofructokinase (PFK) homologues, pfkA3 (SCO1214), pfkA (SCO2119) and pfkA2 (SCO5426) (Bentley et al, 2002; Alves et al 1997). Removal of one of them, *pfk*A2 (the only PFK homologue detectable *in vitro* activity) leads to overproduction of actinorhodin and undecyl-prodigiosin, while the deletion of the other two homologues does not seem to cause any effect in the same conditions (Borodina et al 2008). Similarly, in *S. lividans*, deletion of the two *zwf* homologous genes, coding for glucose-6-phosphate dehydrogenases, involved in the first step of the pentoses phosphate pathway induces the over-production of actinorhodin (Buttler et al, 2002). This increase in the synthesis of NPs has been attributed to an incrase of the metabolic flux through the pentoses phosphate pathway or glycolysis, which in turn increases the amount of available precursors for polyketides biosynthesis.

In a more recent study, two homologues of the malic enzyme from *S. coelicolor* were characterized, one of them is $NAD^{+-}$dependent and another $NADP^{+}$-depdendent. The elimination of each enzyme has no effect on growth, however, the lack the $NAD^{+-}$dependent enzyme leads to a drop in the production of actinorhodin, revealing the role of anaplerotic reactions in the provision of precursors for NPs biosynthesis (Rodriguez et al, 2012). Taken together, these observations suggest a potential link between the expansion of families of central metabolic enzymes and a mechanism that mediates the interaction between central and NPs metabolism through the deviation of precursors for their biosynthesis (Figure 4).

### 2.3.2 Evolutionary links

The availability of the genomes of antibiotic producer strains, primarily members of the genus *Streptomyces* has helped to establish the genetic basis for the biosynthesis of a large number of NPs while generating large catalog of specialized enzymes for this class of metabolism (Van lanen and Shen, 2006, Bode and Muller, 2005; Nett et al, 2009; Run and Challis, 2009; Table S1)

Within this repertoire, the modular enzyme systems such as NRPSs and PKSs have been extensively studied studied, mainly in terms of their structure, chemical and catalytic mechanisms, as well as their evolution, (Koglin and Walsh, 2009; Hertweck 2009). In an exceptional study it has been established that the PKSs are evolutionarily related to fatty acid synthases (Jenke, Kodama et al, 2005), with which they share biosynthetic strategy comparable to an assembly line. Meanwhile, the origin of the NRPSs remains unclear but might relate to the enzymes that generate amide bonds in central metabolism.

While the role of PKSs and NRPSs in the diversity of microbial metabolism has been extensively studied and efficiently exploited through the mining of genomic sequences to identify novel NPs (Jenke- Kodama and Dittman, 2009; Ziemert et al, 2012), little has been done to understand the contribution of central metabolism in the evolution and diversification of the NP metabolism. This is surprising, since many, and in principle, all of the NP biosynthetic enzymes must have evolved from ancestors present in central metabolism. In fact, in many studies focused on the biosynthetic pathway of a in particular natural product homologues of central metabolic enzymes have been identified, although the potential of these observations has been generally underestimated, in a few cases these expansion and recruitment events have been recognized as relevant evolutionary events, these are the case of the homologues of enolase (ENO / PHPH) and aconitase (ACN / PMI involved in phosphinothricin biosynthesis (Heinzelmann et al, 2001; Blodgett et al, 2007) and the homologues the dehydroquinate dehydratase involved in the biosynthesis of acarbose and validamycin (Asamizu et al, 2012, Figure 12). In these examples, the chemical conversions performed by NP specialized enzymes occur through mechanisms that are shared with their ancestor in central metabolism, it can be deduced that the only changes selected during the evolutionary process are those that enabled the acquisition of a new substrate specificity.

This natural strategy for the evolution of new enzymatic functions results in divergent groups of enzymes that share partial reaction mechanisms, intermediates or transition states and three-dimensional structure, these groups are known as mechanistic diverse superfamily of enzymes (Gerlt and Babbitt, 2001). An extended review of these concepts and their potential for protein design can be found in the master thesis "Functional Migration between different folds sharing common reaction mechanisms" (Cruz-Morales, 2009).

With these evolutionary basis, it is easy to speculate that many of the specialized enzymes of central metabolism arose after the acquisition of new substrate specificities while conserving their original reaction mechanisms, intermediates and transition states. Therefore, the new enzymes and their ancestors would be part of mechanistic diverse superfamilies. The phylogenetic reconstruction of these superfamilies, should reflect the divergence and recent emergence of these NP specialized sub-families (Figure 3).

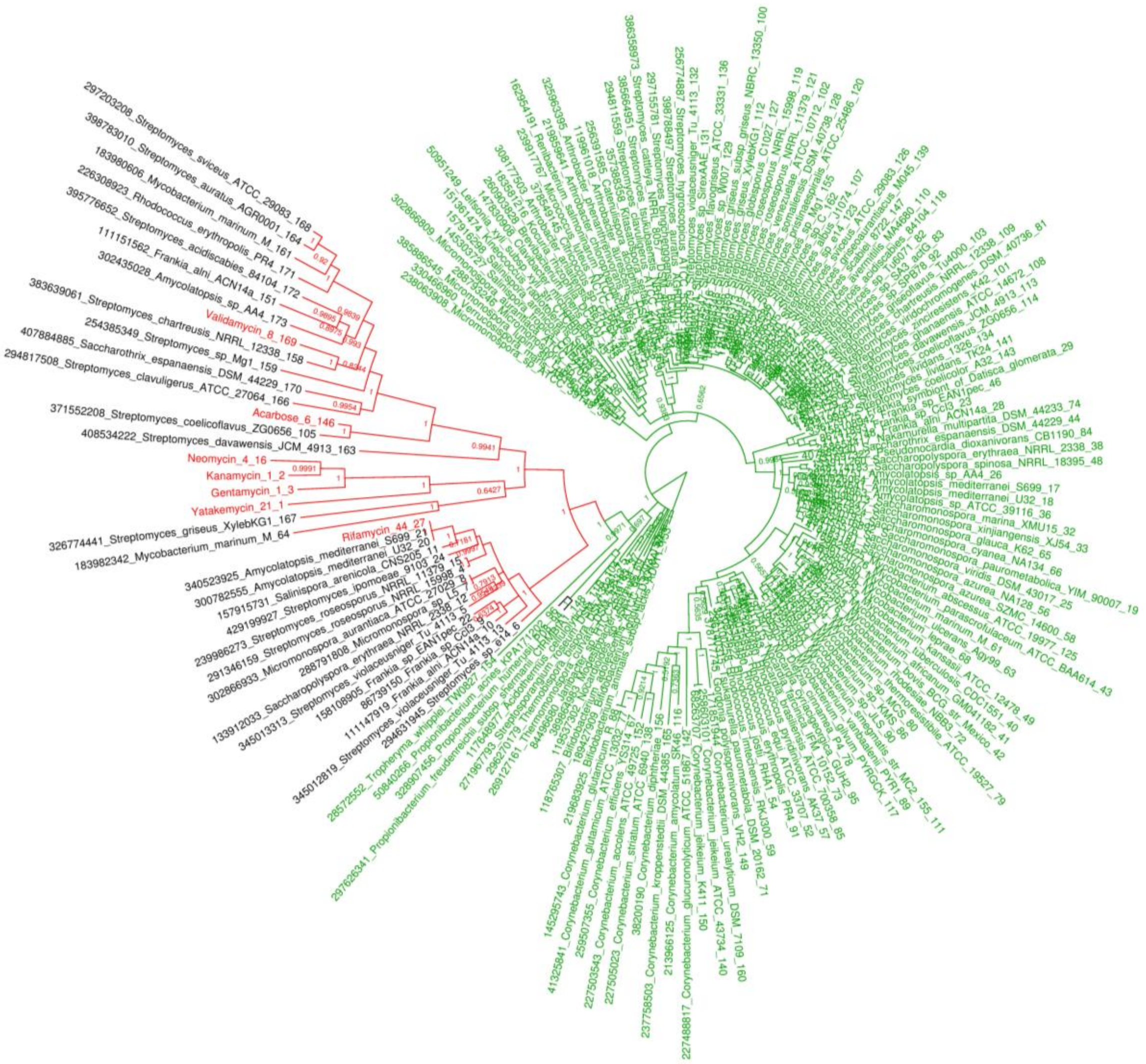

**Figure 12. Phylogenetic reconstruction of the dehydroquinate dehydrogenase enzyme family as an example of the evolutionary relationship between central and NP metabolic enzymes.** The amino acid sequences of the enzyme homologues were obtained from a collection of actinomycetes genomes (Table S2). The clades marked in green are connected to the central metabolism, each organism has a homologue in this clade while the more divergent clade shown in red contains additional homologues (expansions) present only in some genomes, this clade also includes enzymes involved in biosynthesis of NPs whose names are also highlighted in red.

## 2.4 Natural products genome mining

The current approaches for the discovery and identification of NPs are mainly based on the search for cryptic biosynthetic gene clusters. The most common method involves the identification of the genes encoding the enzyme complex assembly of NPs as PKSs and NRPSs, and other enzymes typically related with NPs (e.g. lantipeptide synthases, terpene synthases, etc.). The protein sequences of these enzymes are identified in the genomes of NPs producer organisms using bioinformatic tools based on sequence homology, such as Blast (Altschul et al, 1990).

Several years of deep study of these modular systems have provided todays NP researcher with a plethora of structural, mechanistic, genetic, and chemical knowledge from PKSs and NRPSs Hetweck, 2009, Walsh, 2004, Walsh and Fischbach 2010; Koglin and Walsh, 2009; Run and Challis, 2009, Jenke-Kodama et al, 2009). The available information is so abundant that it is possible to infer the substrates used by a PKS or NRPS, and thus, to predict the structures and chemical properties of their many products and intermediates from DNA sequences in a few minutes. This remarkable progress is mainly based on the exploitation of databases that include the sequences of PKSs and NRPSs responsible for the biosynthesis of known NPs. These databases allow the search of sequence signatures of the conserved domains dedicated to recruitment, incorporation and / or modification of precursors, within the megasynthase assembly lines described in the section 1.3.2 in query sequences (Ansari et al, 2004, Figure 6). Recognition of these domains in combination with the knowledge of the chemical structures previously identified NPs, produces highly efficient predictions which are useful for subsequent identification of the new metabolic product (Barona-Gomez et al, 2004; Lautru et al, 2005; Song et al, 2006).

The incorporation of these mining strategies in specialized bioinformatic pipelines, and the implementation of these on free-access web interfaces since almost ten years ago (Ansari et al, 2004; Anand, 2010; http://www.nii.res.in/NRPS-pks.html, Bachman and Ravel 2009; http://nrps.igs.umaryland.edu/nrps/; Weber et al, 2009; Starcevic et al, 2008; Ziemert et al, 2012; http://napdos.ucsd.edu/) had made these tools available for everyone. These bioinformatic tools have indeed been successfully used for the discovery of a large number of new metabolic pathways (they have been collectively cited more than 500 times). Recently, the necessary level of knowledge for the development of bioinformatic tools for genome mining has reached NPs biosynthetic systems others than PKSs and NRPSs such as lantipeptides and glycosidic compounds (Velazquez and van der Donk, 2011; Medema et al, 2011).

This progress has enabled the emergence of more advanced and user-friendly platforms. The best currently available system for automated NPs genome mining is AntiSMASH, or antibiotics and secondary metabolite analysis shell (Medema et al, 2011; http://antismash.secondarymetabolites.org/). AntiSMASH is a web server, which allows the introduction of unannotated genomic sequences from which it generates a biosynthetic gene cluster catalog, its location in the genome, the kind of compound which produces, and even in some cases, possible chemical structures derived from the analysis of the PKSs and NRPSs domains identified. AntiSMASH, like its predecessors, is based on the identification of homologous biosynthetic enzymes involved in the assembly of siderophores, polyketides, non-ribosomal and ribosomal peptides, aminoglycosides, etc., therefore, it heavily relies on a limited universe of enzymes families that are mined and led directly to the identification of the biosynthetic clusters, e. g., to identify biosynthetic gene clusters for nonribosomal peptides, the simplest way is to locate NRPSs homologous sequences in a genome. This has the disadvantage that the searches inevitably lead to the identification of known biosynthetic systems.

NPs genome mining is a well-established field that is also in constant advance, nowadays genomic sequencing, identification of new biosynthetic systems, expression, biosynthetic mechanism and structural elucidation of the products of interest is feasible using practical amounts of time and resources (Figure 13), however, the main challenge that genome mining and in general the microbial NPs research community faces, is the discovery of new classes of compounds, with novel chemical properties and biosynthetic mechanisms that expand the metabolic repertoire that could be exploited for human benefit in the future. In this thesis, this challenge has been approached from an evolutionary perspective.

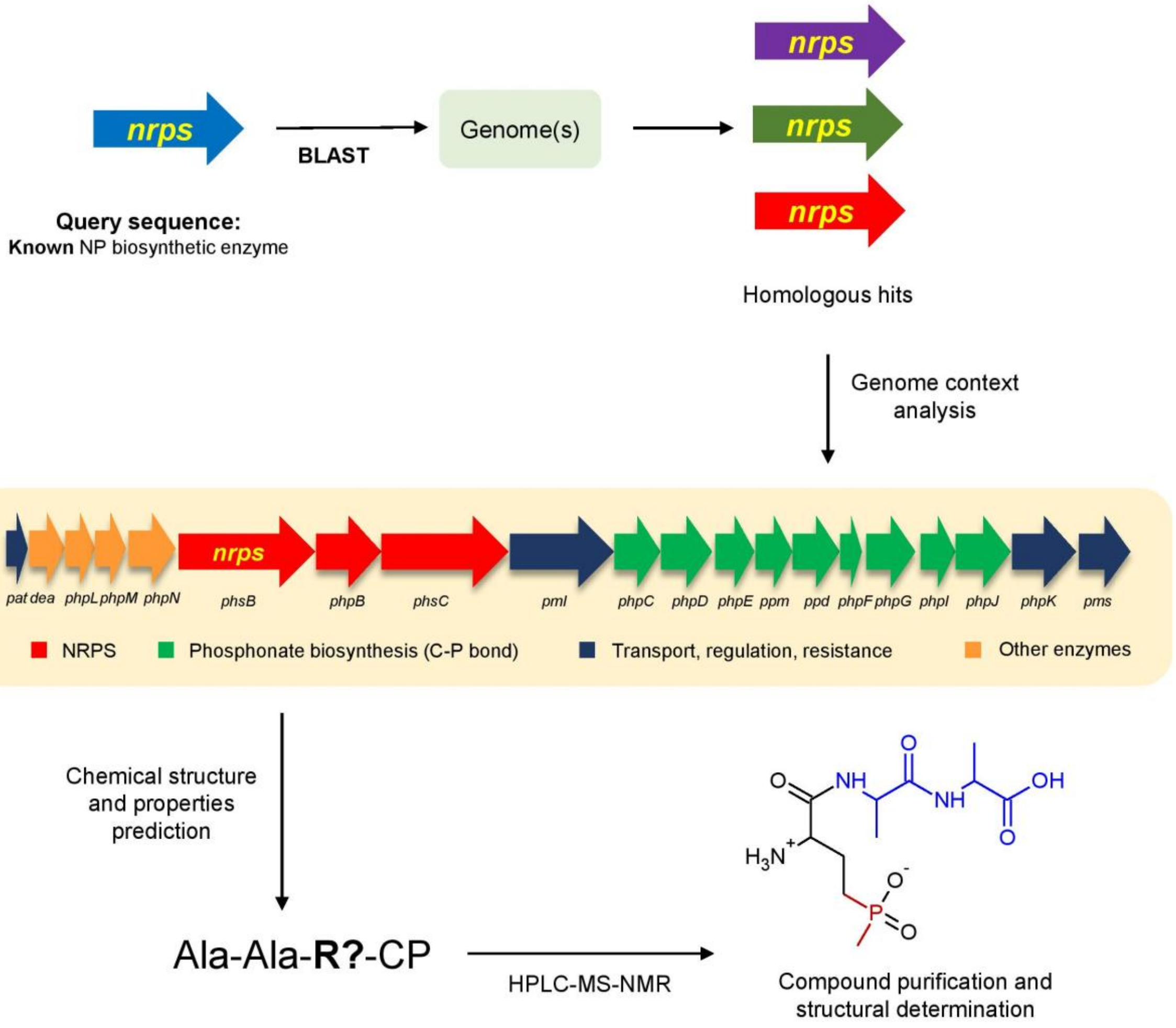

**Figure 13. Modern Strategy for discovery of NPs**. The first part of the process involves the identification of biosynthetic gene clusters, once identified, chemical predictions made from the genetic information is used for determination of the final product of the pathway. The figure shows the cluster of genes responsible for the biosynthesis of phosphinotricin, as an example of this proccess.

# 3 Hypothesis

The broad metabolic repertoire encoded in the *Streptomyces* genomes evolved following the shell model, therefore, NP biosynthetic pathways evolved from central metabolism after enzyme family expansion and recruitment.

The expansion and recruitment events among catalysts in the actinobacterial central metabolism can be detected using bioinformatic methods such as the phylogenetic analysis and comparative genomics. Such analysis can be exploited for the discovery of novel NP biosynthetic pathways.

# 4 Objectives

## 4.1 General objective

Identify and characterize the enzyme family expansion events within the central metabolism of *Actinobacteria* as well as the recruitment of members of such families for natural product biosynthesis, using the phylogenetic reconstruction of expanded enzyme families with recruitments, predict novel metabolic pathways and experimentally validate such predictions.

## 4.2 Specific objectives

1. Establish an enzyme expansion index in qualitative (absolute) and quantitative (relative to the biosynthesis of NPs) terms in the *Streptomyces* genus.
2. Develop strategies and bioinformatic tools for the identification on new natural product biosynthetic pathways in actinobacterial genomes based on evolutionary concepts such as the expansion of enzyme superfamilies and the shell hypothesis for the evolution of metabolism.
3. Identify recruitment events among enzyme families from central metabolism for natural product biosynthesis where a clear link between genotype and phenotype can be established.

4. Select and analyze recruitment events that can be used as proof of concept for the novel genome mining strategies developed in specific objective two.
5. Characterize such model recruitment events in order to identify novel NPs and their biosynthetic pathways.

# 5 Materials and Methods

## 5.1 *Streptomyces lividans* 66 genome sequencing and assembly strategy

Genomic DNA from *S. lividans* 66, obtained from the John Innes Centre culture collection (equivalent to strain 1326) was purified using standard protocols (Kieser et al, 2000) and sequenced using next generation sequencing platforms. Pyrosequencing (454, Life Sciences) was performed at the Langebio core sequencing facility (Irapuato, Mexico); ligation-based approaches (Illumina) were performed at Baseclear (Leiden, The Netherlands) and single molecule real-time sequencing (Pacific Biosystems RS) at Service XS (Leiden, The Netherlands). An initial set of contigs was obtained from Illumina and 454 reads using Celera Assembler 7.0 for Hybrid *de novo* assembly (Myers et al, 2000). The resulting contigs were sorted using *S. coelicolor* M145 and *S. lividans* TK24 genomes as references with R2CAT (Husemann and Stoye, 2010). This assembly was improved with IMAGE 2 (Tsai et al, 2010) by iterative mapping and extension using paired-end Illumina reads. The second assembly and Illumina mate pair reads were used to obtain genome scaffolds with SSPACE (Boetzer et al, 2011). A second round of sorting with R2cat and the genome of *S. lividans* TK24 as reference yielded a single scaffold. PacBio RS (long) reads were corrected with the PacBiotoCA pipeline (Koren et al, 2012) using Illumina reads. These corrected reads were used to manually fill gaps in the main chromosomal scaffold. Oligonucleotides used for closing gaps by PCR are included in table S5.

## 5.2 Genome annotation, mining and comparative analysis

The *S. lividans* 66 genome sequence was annotated automatically with RAST (Aziz et al, 2008) inspected and improved manually. Comparative genomic analysis between the chromosomes of *S. coelicolor,* and *S. lividans* TK24 and 66 strains, was performed using sequence comparisons with BlastN (Altschul et al, 1990), and visualized in Artemis Comparative Tools (Carver et al, 2005) a score cutoff of 200 was used to visually determine the presence or absence of genes trough the genomes. Jspecies (Richter and Roselló, 2009) was used to obtain the blast-calculated average nucleotide identity (ANIb) between genomes.

## 5.3 Transcriptional analysis of the *S. lividans 66* response to metal stress

RNA-seq libraries from *S. lividans* 66 produced by Dwarakanath et al (2012) were used for transcriptional analysis. The annotated genome of *S. lividans* 66 was used as reference after manually removing ribosomal genes. Differentially expressed genes were identified with CLC genomics work bench (CLC-Bio, Denmark). A cutoff of FDR $\leq$ 0.05 was used to filter the genes with significant changes. To obtain the coverage-plots the RNAseq reads were mapped against the 66 genome using BWA (Li & Durbin, 2009), and the alignments were processed with SAMtools (Li et al, 2009) for graphic representation in Artemis (Carver et al, 2005).

## 5.4 Comparative metal tolerance phenotypic analysis

Fresh spores were titrated, droplets containing 10E0, 10E2, 10E4 and 10E6 spores of *S. coelicolor* M145, *S. lividans* 66, and *S. lividans* Tk24 were plated in Soy Flour Media (SFM) and R5 media containing different metal salt concentrations. The plates were incubated for 3 days at 30 ºC. M145 and Tk24 strains were obtained from the John Innes Centre collection. The media and handling procedures of the strains used were done according to Kieser et al (2000).

## 5.5 EvoMining Database integration

**Natural Products gene clusters database.** NP biosynthetic pathways which biosynthetic gene clusters, products and biosynthetic logic have been described were obtained from the available literature until January 2013. A total of 129 biosynthetic pathways were included in the database. The amino acid sequences of the biosynthetic enzymes other than the modular assembly complexes such as PKSs and NRPSs, regulatory and transport proteins from the NP biosynthetic gene clusters were included in the database. The accession numbers of the gene clusters and references of the characterization of these pathways are shown in table S3.

**Genome database.** Amino acid sequences from complete and draft genomes of 140 members of the *Actinobacteria* family were obtained from the GenBank until January 2013. A Complete list of the strains, accession numbers, genome length and number of proteins is available on table S2.

**Central metabolic pathways database.** Amino acid sequences from the proteins involved in 11 central metabolic pathways (tables 5-1 and S1) were collected from the GenBank using the genome scale metabolic reconstructions of *Streptomyces coelicolor* (Borodina et al, 2005), *Corynebacterium glutamicum* (Kjeldsen and Nielsen, 2008), and *Mycobacterium tuberculosis* (Jamshidi and Palsson, 2007).

**Table 5-1.** Queries used for Enzyme Expansion analysis on the central metabolism of *Actinobacteria* from *S. coelicolor* (Borodina et al, 2005); *C. glutamicum* (Kjeldsen and Nielsen, 2008), and *M. tuberculosis* (Jamshidi and Palsson, 2007).

| Model Pathway | Main products | Steps[1] | Complexes[2] | Queries |
|---|---|---|---|---|
| Glycolysis | Pyruvate, Phosphoenol-Pyruvate and ATP | 10 | 0 | 49 |
| Pentose phosphate pathway | Fructose, E-4 Phosphate, ribose 5-P | 8 | 0 | 38 |
| Citric acid cycle | Acetyl CoA, oxaloacetate, α-Ketoglutarate | 8 | 3 | 94 |
| Amino acids from AKG | Glu, Gln, Pro, Arg | 13 | 1 | 50 |
| Amino acids from THR and PYR | Ala,Ile, Leu, Val | 14 | 2 | 42 |
| Amino acids from Oxaloacetate | Asp, Asn, Thr, Met, Lys | 18 | 0 | 64 |
| Amino acids from 3PGA | Gly, Ser, Cys | 6 | 0 | 25 |
| Amino acids from R5P | His | 10 | 1 | 37 |
| Amino acids from E4P and PEP | Tyr, Phe, Trp | 17 | 2 | 56 |
| Purines | ADP,GTP, dADP, d GTP | 21 | 5 | 88 |
| Pyrimidines | UTP, CTP, dCTP, dUTTP | 18 | 3 | 66 |
| **Total: 11** | **37** | **143** | **17** | **609** |
| | **Non-redundant queries** | | | **499** |

[1]: Enzyme-catalyzed reaction in a metabolic pathway

[2]: Protein complexes participating in each pathway, each protein of the complex was used as query

## 5.6 Expanded enzyme families identification on actinobacterial central metabolism

The central metabolic enzymes on the actinobacterial genomes database were retrieved using BlastP (Alstchul, 1990) using the 499 amino acid sequences in the central metabolic pathway database as queries. The hits were filtered with an e-value cutoff of 0.0001 and a score cutoff of 100. The outputs from these searches were processed to obtain the number of homologues for each metabolic reaction (HN) in central metabolic pathways in every genome analyzed. Media (μ) and standard deviation (σ) were also calculated. An enzymatic family expansion was accounted if $\boldsymbol{HN => \mu + \sigma}$. A graphical representation of the expansion profiles is shown in supplementary table S4.

## 5.7 Enzyme recruitments identification.

Enzyme recruitment events were defined as cases where members of expanded central enzyme families were found in the NP gene cluster database. To identify enzyme recruitments, the amino acid sequences of the natural product gene clusters database were used as queries for blastP searches against expanded enzyme families identified in the previous step using an e-value cutoff of 0.0001 and a score cutoff of 100.

## 5.8 Phylogenetic analysis of expanded enzyme families with recruitments

Members of the recruited enzyme families were retrieved from the actinobacterial genome and the NP gene clusters databases using BlastP. The collected sequences were aligned using Muscle (Edgar, 2004). The alignments were inspected and curated manually using JalView (Waterhouse et al, 2009; http://www.jalview.org). The phylogenetic reconstructions of the expanded/recruited enzyme families were constructed with MrBayes (Huelsenbeck and ronquist, 2001) with the following parameters: aamodelpr=fixed(wag), samplefreq=100, burninfrac=0.25. For one million generations. Graphic representations of the phylogenetic trees were created with FigTree (http://tree.bio.ed.ac.uk/software/figtree/) and inspected manually. The process was executed semi-automatically using scripts written in Perl.

## 5.9 Natural product pathway prediction

The phylogenetic reconstructions obtained were analyzed in order to identify enzyme family members linked to novel biosynthetic pathways. The criterion of selection is described in detail in chapter II, this criterion was confirmed after individual inspection of the topology on the phylogenetic reconstructions of the 33 recruited enzyme families identified herein. Artemis (Carver et al, 2012) was used for the graphic display of the gene context of the homologues. The annotation of the genomes in the genome database (DB2) was obtained from the GenBank, (a complete list of the accession numbers of the genomes is available in supplementary table S2).

## 5.10 *S. viridochromogenes* PTT enolase molecular modelling and comparison

A molecular model of PhpH was constructed with Modeller (Sali and Blundell, 1993) using as template the crystal structure of the dimeric yeast enolase in complex with magnesium, 2-Phosphoglycerate (2-PGA) and Phosphoenolpyruvate (PEP) obtained from the protein data Bank PDB (PDB ID=2ONE) (Zhang et al, 1997), which shares a 33% identity with the Carboxy-phosphoenolpyruvate synthase (phpH or PTT enolase) from *S. viridochromogenes* (Blodgett et al, 2007). Query and template sequences were aligned with ClustalW (Larkin et al, 2007). A model of the product of the PTT enolase, carboxy-phosphoenol pyruvate was built with VegaZZ (Pedretti et al, 2004), and located in an analog position with respect to PEP in the active site of the PTT enolase, using superimpositions of the model and template in Pymol (The PyMOL Molecular Graphics System, Version 0.99 Schrödinger, LLC; http://www.pymol.org/)

## 5.11 Gap closing and sequence annotation of the putative phosphonate natural product biosynthetic gene cluster in *S. sviceus*

The gaps and misassembles found in the region 8 Kilobases downstream and 24 Kilobases upstream of the PTT enolase (ZP_06914376.1) of the *S. sviceus* draft genome sequence (GI NZ_ABJJ00000000) were closed using PCR product sequencing (table 5-2), for gap 3 which was too long for a single PCR sequencing reaction 3 iterative rounds of sequencing and primer synthesis were required until the gap was closed. Two misassembles were also corrected by PCR sequencing (sequences 5 and 6). The corrected sequence spanning a total of 34 Kbs was annotated manually using BlastP against the NR database and available literature.

**Table 5-2.** Primers used for gap closing on the *S. sviceus* genome sequence

| **Fragment** | **Total sequenced bases** | **Primer direction-Sequence (5'-3')** |
|---|---|---|
| 1 | 148 | F-TGCCGCCCAGTTCGAGCAGA; R-ATCCGAACGCACACCGCTG |
| 2 | 566 | F-CCAGCGTTCTGGCCAGGGCT;R-CACGATCGCGACCGACGACT |
| 3 | 2726 | FA-AAGGCGCCCTGCTTGATGAA; RA-CAAACTCCAGGCCTTCTACG<br>B-GAAGTTGATGCGGAACGCCA; RB-GCCGAGAACATCCTGCACGTG<br>FC-GCTGATGGGTTTGTCGTCGC; RC-GGTGGCGTGATGGTCACAGC<br>RD-CGTGTGCACCACCGGCAAGTC; |
| 4 | 538 | F-ATTCCGGTTGTTGGCGTGCC; R-TAGTTGTTGATGCTCCACAC |
| 5 | 484 | F-GTCGTCGAAGTCATGGGCGT; R-CATGGTCTTCGACACCCTGG |
| 6 | 535 | F-GAGTGGTCGGCATGGGCCGG; R-GTGACCTCGTGATCCGGGAC |

## 5.12 Construction of knock-out mutants

### 5.12.1 Mutants in *S. coelicolor*

All the *S. coelicolor* gene knock-out mutants reported in this thesis were done using in-frame PCR-targeted gene replacement of their coding sequences with an apramycin resistance cassette (*acc(3)IV*) using the Redirect system (Gust et al, 2003). The plasmid pIJ773 was used as template to obtain the mutagenic cassette containing the apramycin$^R$ marker by PCR amplification using the primers reported in table 5-3. The mutagenic cassettes were used to disrupt the coding sequences of the genes of interest from the cosmid clones (Redenbach et al, 1996) reported in table 5-4. The gene disruptions were performed on DH5α *E. coli* transformant cells carrying the plasmid pIJ790 (λ-RED recombination system) and the corresponding cosmids (Gust et al, 2003). The mutagenized cosmids were selected, purified, and introduced in *S. coelicolor* M145 via conjugation using the methylation-deficient *E. coli* strain ET12567 carrying the plasmid pUZ8002 (Gust et al, 2003). Double cross-over ex-conjugants were selected for its resistance to apramycin and kanamycin sensitivity.

**Table 5-3.** Primers used for in frame PCR-targeted gene replacement and confirmation of gene knock outs in *S. coelicolor*

| Primer | Sequence | Use |
|---|---|---|
| **P1_SCO3096** | tcacagtgaagcgtcacattgaagaaggagatgctcgtgattccggggatccgtcgacc | Gene disruption |
| **P2_SCO3096** | ggggacgtacgtacgacgctgccttctaccgggccgtcatgtaggctggagctgcttc | Gene disruption |
| **P1_SCO7638** | gccgggtcacggcaccgacaaagggaaggaactcccatgattccggggatccgtcgacc | Gene disruption |
| **P2_SCO7638** | cacggcgggccggcacggcggcgggccggtcgcgactcatgtaggctggagctgcttc | Gene disruption |
| **P1_SCO2014** | aattcgctacccggcagtaagtcataggctcgactcatgattccggggatccgtcgacc | Gene disruption |
| **P2_SCO2014** | gccctcatggacgcccacgtcggccccaagtactgatcatgtaggctggagctgcttc | Gene disruption |
| **P1_SCO5423** | tattccgcgccgaaacaaaccgataggatggcacccatgattccggggatccgtcgacc | Gene disruption |
| **P2_SCO5423** | gggggccctcggcgcgtacgacggggagggggggcggtcatgtaggctggagctgcttc | Gene disruption |
| **P1_SCO4209** | cgcgggggatcagggccttggattacgctcggaagcatg-attccggggatccgtcgacc | Gene disruption |
| **P2_SCO4209** | ggagaaaccgcaggtagggggcctgttcgtgcgcgctta-tgtaggctggagctgcttc | Gene disruption |
| **P1_SCO2576** | ggaggcgacgcacaccgacgaggactggcgatgagcgccattccggggatccgtcgacc | Gene disruption |
| **P2_SCO2576** | tttagcctgcgacctgccggaaagtgaaatcgccgttcatgtaggctggagctgcttc | Gene disruption |
| **ScrSCO3096_F** | acgacgccgacctgatctac | confirmation |
| **ScrSCO3096_R** | atcctggtcgcggtggagaa | confirmation |
| **ScrSCO7638_F** | gatgcggtcctcgtgcttga | confirmation |
| **ScrSCO7638_R** | gcgcgttgctgtcactgctt | confirmation |
| **ScrSCO2014_F** | cgcacaattcaacacttgtc | confirmation |
| **ScrSCO2014_R** | cgaacaaccaaccgactgtc | confirmation |
| **ScrSCO5423_F** | ggtaagtgaatcggggaatc | confirmation |
| **ScrSCO5423_R** | cttcaccaacgtgaaggtgg | confirmation |
| **ScrSCO4209_F** | aacttagaaacggctctctc | confirmation |
| **ScrSCO4209_R** | gaaccgccgagcaagaaacg | confirmation |
| **ScrSCO4209_F** | gtggaaggactgccccgagc | confirmation |
| **ScrSCO4209_R** | gacccgagccgcatggtcctg | confirmation |
| **R1** | cttggtgtatccaacggcgt | confirmation |
| **F2** | tgatcttcctgcatccgcca | confirmation |
| **P1** | attccggggatccgtcgacc | Resistance cassette amplification |
| **P2** | tgtaggctggagctgcttc | Resistance cassette amplification |

The genotype of the clones was confirmed by PCR, two PCR products were obtained for each mutant to confirm the proper insertion of the cassette. The primers for genotype confirmation are reported in table 5-3. Scr_F primers were used in combination with the R1 primer while Scr_R primers were used in combination with the F2 primer to amplify PCR products including fragments for the cassette and the neighboring upstream and downstream regions of the disrupted allele. The strains and plasmids of the redirect system were obtained from the John Innes Centre (Norwich, UK).

**Table 5-4**.*S. coelicolor* cosmid clones used for gene disruption

| Disrupted gene | Cosmid | Region in the *S. coelicolor* chromosome | Coding sequences contained |
|---|---|---|---|
| **SCO2014** | 7H2 | 2125895-2168549 | SCO2001-SCO2023 |
| **SCO3096** | E25 | 3352895-3398366 | SCO3060-SCO3106 |
| **SCO3096** | E41 | 3386902-3422929 | SCO3092-SCO3122 |
| **SCO5423** | 8F4 | 5862037-5895909 | SCO5400-SCO5424 |
| **SCO7638** | 10F4 | 8455566-8491805 | SCO7629-SCO7666 |
| **SCO2576** | stC123 | 2760885-2802160 | SCO2560-SCO2590 |
| **SCO4209** | 2stD46 | 4597090-4632885 | SCO4188-SCO4228 |

### 5.12.2 Mutants in *S. lividans* 66

*S. lividans* 66 mutant SLI0883-5: *acc(3)IV* was obtained by double crossover recombination of a mutagenic cassette cloned in the pWHM3 plasmid (Vara et al, 1989) following the strategy previously described by van Wezel et al (2005). pWHM3 is a high copy number shuttle vector which can replicate in *E. coli* (ampicillin resistance) and in several streptomycetes (thiostrepton resistance), however is instable and can be curated after a few rounds of non-selective growth in *Streptomyces.*

The mutagenic cassette was constructed using PCR products containing the 1.5 KB upstream (product 1) and downstream (product 2) of the start and stop codons respectively. The restriction sites EcoRI-XbaI (product 1) and XbaI-HindIII (product 2) were introduced in the PCR primers. The PCR products were digested, purified and ligated in a digested pWHM3 (EcoRI-HindIII) in a single reaction (triple ligation). The products of the ligation reaction were introduced in *E. coli* DH5α competent cells via thermal shock. The positive clones were selected in ampicillin and further confirmed by PCR and digestion. The apramycin resistance cassette was obtained from the plasmid

pIJ773 (Gust et al, 2003) by digestion with XbaI, the purified cassette was ligated into the pWHM3 plasmid carrying product1 and 2, which was previously digested with XbaI, the ligation products were introduced in *E. coli* DH5α cells, via thermal shock and selected with apramycin. The construction was confirmed by PCR and digestion.

The mutagenic cassette cloned into pWHM3 was introduced into wild type *S. lividans* 66 via protoplast transformation, according with the standard protocols reported by Kieser et al (2000). Apramycin resistance was selected among transformant clones, several colonies were picked and grown in apramycin for one more generation and then selected in thiostrepton, those clones with apramycin resistance (integration of the cassette and los of the wild type allele by double cross-over) and thiostrepton sensitivity (loss of the pWHM3 cassette delivery vector ) were selected. The mutants were confirmed by PCR. Table 5-5 includes the PCR primers used for the whole process.

**Table 5-5.** Primers used for in frame gene replacement and confirmation of gene knock outs in *S. lividans* 66

| Primer | Sequence | Use |
|---|---|---|
| **P1_NeoNRP** | ctgaggatccgtccccgcacctgggctcc | Gene disruption |
| **P2_NeoNRP** | gaagttatccatcacctctagagtgtgcgaacgggttcccgga | Gene disruption |
| **P3_NeoNRP** | gaagttatcgcgcatctctagagtgtgggtcaggcaccgcctc | Gene disruption |
| **P4_NeoNRP** | ctga-aagctt-ggcctggtgagctccgagc | Gene disruption |
| **NeoNRP_F** | gtcacgtcactaagtggcccgg | Confirmation |
| **NeoNRP_R2** | gtgcggcgaggagttgtattgc | Confirmation |

## 5.13 Cloning and chromosomal integration of *S. coelicolor* genes using pAV11b

The *S. coelicolor* glycolytic genes SCO7638, SCO3096, SCO2014, SCO5423, SCO4209 and SCO2576 were cloned into the pAV11b plasmid, a successfully tested plasmid, which however, was not fully sequenced (Khaleel et al, 2011). Therefore the plasmid was sequenced using PCR fragment sequencing in collaboration with Karina Verdel and Lianet Noda, both students of the Evolution of Metabolic Diversity Group at Langebio, the sequence was assembled and annotated manually (The sequence is provided as supplementary material S9). Once the integrity of pAV11b was confirmed,

PCR primers were designed for the amplification of the genes from the start to the stop codon (table 5-6). The oligonucleotides were phosphorylated prior to the PCR reaction using the T4 polynucleotide kinase, the PCR was performed with PFU-DNA polymerase, the products were purified and ligated to EcoRV digested and de-phosphorylated pAV11b vector (blunt end). The proper orientation of the positive clones (start codon downstream the promoter of pAV11b) was confirmed by PCR (Table 5-6) and digestion, the process was carried out in *E. coli* DH5α and the selection of the clones was done with hygromycin in Luria-Bertani Medium with half the normal concentration of NaCl (5 gr).

The glycolytic genes cloned into pAV11b were introduced in *S. coelicolor* via conjugation, for that purpose, the plasmid clones were introduced by electroporation into the non-DNA methylating strain of *E. coli* ET12567 strain carrying the plasmid pUZ8002, the selection of the trasnformant strains was performed with kanamycin, hygromycin and chloramphenicol. Conjugation of *E. coli* ET12567/pUZ8002, carrying the pAV11b plasmid clones with *S. coelicolor* was performed in SFM media and the selection of ex-conjugants was done with hygromycin. The integration of the constructs was confirmed by PCR with pAV11b specific primers adjacent to the cloning site.

**Table 5-6.** Primers used for cloning on the pAV11b plasmid and confirmation of the proper integration of the construct in *S. coelicolor.*

| Primer | Sequence | Use |
|---|---|---|
| **SCO7638_F** | atgtccgcaacagcagtc | Cloning |
| **SCO7638_R** | tcagacccggcgcagcgc | Cloning |
| **SCO3096_F** | gtgccgtccatcgacgtc | Cloning |
| **SCO3096_R** | tcagcccttgaagcgggg | Cloning |
| **SCO2014_F** | atgcgccgagcaaagatcgtc | Cloning |
| **SCO2014_R** | tcacttgggaatgtcgtcctc | Cloning |
| **SCO5423_F** | atgcgccgttcgaaaatcgtc | Cloning |
| **SCO5423_R** | tcagccgcgccgtgtctcgcc | Cloning |
| **SCO4209_F** | atggccgacgcaccgtacaag | Cloning |
| **SCO4209_R** | ttacttcttcttgccctggtt | Cloning |
| **SCO2576_F** | atgagcgccaccggtgaggtg | Cloning |
| **SCO2576_R** | tcagacgtcgtcgccgatcac | Cloning |
| **pAV11b_For** | gatagtggtaggatccctatc | Confirmation |
| **pAV11b_Rev** | catcaaggtcaaggcgtaggtc | Confirmation |

## 5.14 Transcriptional analysis of glycolytic genes

The transcriptional analysis of specific genes in *S. coelicolor* and *S. lividans* 66 under different conditions was performed by RT-PCR, total RNA was isolated from mycelium grown on liquid minimal media for *S. lividans* 66 with and without Magnesium Chloride and Potassium Chloride at 300 mM, and in solid R2YE media for *S. coelicolor* using the Nucleospin II RNA Kit (Macherey-Nagel) using the protocol provided by the manufacturer. In both strains HrdB (a house-keeping sigma factor) was used as control. The RT-PCR reactions were done using the OneStep RT-PCR kit (Quiagen) using the protocol provided by the manufacturer. The PCR primers used are presented in table 5-7. In all cases the PCR reactions were carried out for 30 cycles with a melting temperature of 55 ºC

**Table 5-7**. Primers used for transcriptional analysis of glycolytic genes in *S. coelicolor*

| Primer | Sequence |
|---|---|
| **rtSCO4209_F** | acacgtccgtccagaagc |
| **rtSCO4209_R** | cttcttgccctggttcttca |
| **rtSCO2576_F** | gagaccggcgtgaaccag |
| **rtSCO2576_R** | cagcagcagttggtcagg |
| **rtSCO2014_F** | cgactcgtacgaccagatca |
| **rtSCO2014_R** | aggtcgtcctcgtccttctt |
| **rtSCO5423_F** | cggtgacgagttcaccatc |
| **rtSCO5423_R** | tcgtgatcatcgactccatc |
| **rtSCO3096_F** | ctgatcgaccaggccatgtt |
| **rtSCO3096_R** | gtaggagccgtccttgtaga |
| **rtSCO7638_F** | tgccgacaatccgctggact |
| **rtSCO7638_R** | agtgcctcgtcggtgcagaa |
| **rtSLI0883_F** | atgatcggatcgccacttac |
| **rtSLI0883_R** | acgaagtgatggtcgtttcc |
| **rtSLI0884_F** | cgtcaacgaactgacctacga |
| **rtSLI0884_R** | tggtgagaggcaaggtgtact |
| **rtHrdB_F** | cgcagcctcaaccagatcct |
| **rtHrdB_R** | ggagttcgccagcttgtcct |

## 5.15 Phenotypic characterization

For the phenotypic characterization of the *S. coelicolor* and *lividans* wild type and mutant strains fresh spores were harvested and titrated, similar number of spores were plated on minimal media with different carbon sources, using ammonium sulfate as the nitrogen source, as well as R2YE or R5 medium, ISP2, DNA and SFM supplemented with different salts or compounds when necessary. The spores were plated either as droplets or as loans in the surface of the plates, For phenotypic array experiments, SF-N2 and SF-P2 plates (Biolog) were used, the colloid matrix was 0.2 %, gellan gum according with the manufacturer instructions and a similar number of spores was inoculated in each well using a multichannel micropipette. Liquid cultures were performed in beakers with cotton caps and steel springs to avoid pellet formation, incubation was always carried out at 30 ºC, and liquid cultures were agitated always at 200 RPM.

## 5.16 General PCR conditions

In general, *Streptomyces* PCRs were performed in the same conditions, primers were used at a final concentration of 40 pMol per 50 µl reaction, all the reaction mixtures included 10% DMSO, *Streptomyces* colony PCRs were performed before the development of aerial mycelium, typically PCR primers designed in this work worked well at a melting temperature of 55 ºC. The typical PCR protocol included 95 ºC/5 minutes-95º/30 seconds-55ºC/30 seconds-72ºC/30 seconds per each 0.5 kb of product length.

## 5.17 Materials and services

The oligonucleotide synthesis were performed by Sigma, and the DNA synthesis services of the Biotechnology Institute-UNAM, PCR product sequencing were performed at the DNA synthesis and sequencing services of the Biotechnology Institute-UNAM and at the Langebio sequencing services. Most molecular biology enzymes used herein (Polymerases and restriction enzymes and ligases, T4 polynucleotide kinase, alkaline phosphatase, etc.) were purchased from Fermentas and New England Biolabs and used indistinctively. PCR product and plasmid DNA purification was performed with kits purchased from Quiagen. In general the solvents, antibiotics, salts and other chemical compounds were purchased from Sigma.

*Streptomyces* media, microbiological methods and molecular biology protocols used herein are described in the *Streptomyces* practical genetics book by Kieser et al (2000). The Redirect system and protocols are described in deep detail in the redirect manual available at: http://strepdb.streptomyces.org.uk/redirect/protocol_V1_4.pdf. Some extra protocols used and improved during the development of this thesis are available as supplementary protocols S10.

# 6 Results Chapter I: Genome sequencing and analysis *Streptomyces lividans* 66

The results presented in this chapter have been published as: Cruz-Morales P, Vijgenboom E, Iruegas-Bocardo F, Girard G, Yañez-Guerra LA, Ramos-Aboites HE, Pernodet JL, Anné J, van Wezel GP, Barona-Gómez F. 2013. **The genome sequence of *Streptomyces lividans* 66 reveals a novel tRNA-dependent peptide biosynthetic system within a metal-related genomic island**. Genome Biol Evol.

The complete genome sequence and annotation of *S. lividans* 66 is available at http://www.ncbi.nlm.nih.gov/nuccore/509515407

# 6.1 Introduction

## 6.1.1 *S. lividans* 66, a perfect challenge for novel natural product genome mining approaches

As previously described in the introduction of this document, current genome mining approaches are highly efficient but limited in the identification of certain chemical families of Natural Products (NPs). Therefore an alternative bioinformatic approach able to predict novel classes of compounds and biosynthetic systems from genomic data is much needed. As a part of the research towards the development of such approach, the complete genome sequences of actinobacterial species, a taxonomic group renowned for their outstanding NP repertories, were mined for NPs, aiming for proof of concept (described in chapter II of the results section), and the discovery of novel biosynthetic pathways (described in Chapter III of the results section).

Among the actinobacterial genomes analyzed, the sequence of the most widely used streptomycete in industry and laboratories, *Streptomyces lividans* 66 (*SLP2*$^{+}$, *SLP3*$^{+}$) was obtained and included. *S. lividans 66* represented a perfect challenge for the novel NP mining approach developed during my research, this model organism has been studied for more than four decades using current and classic genetic, phenotypic, and biochemical approaches, however, no NP biosynthetic pathways have been identified other than those shared with the closely related model strain *S. coelicolor* M145 (*SCP1*$^{+}$, *SCP2*$^{+}$). Furthermore, standard microbiological protocols, several genetic tools for construction of gene knock outs, transformation, and plasmid systems have been developed for this strain (Kieser et al, 2000).

Since the sole functional and comparative analysis of the *S. lividans* 66 genome led to observations that will be highly important in the following sections of my thesis and in general for the *Streptomyces* research community, this first chapter is entirely dedicated to these analyses. In this chapter, the reader will find the comparative analysis of the genomes of *S. lividans* 66 and closely related strains, including *S. coelicolor* A3(2) (a plasmid less strain obtained from  the strain M145) and *S. lividans*

TK24 (a plasmid less strain obtained from the strain 66). Which led to the identification of a large genomic island with a mosaic structure, present in *S. lividans* 66 but not in the strain TK24. This genomic island harbors SLP3, the plasmid detected by David Hopwood 30 years ago which remained elusive until now, the entire island is rich in genes that encode for proteins functionally related to metal homeostasis. Transcriptional analysis of the response of *S. lividans* 66 to copper was used to corroborate a role of this large genomic island, including two SLP3-borne ‘cryptic’ peptide biosynthetic gene clusters, involved in metal homeostasis, one of which is presented in chapter III

### 6.1.2 Known genetic and phenotypic differences between *S. lividans* and *S. coelicolor*

Besides the distinctive traits that made *S. coelicolor* and *S. lividans* model organisms, other differences among these strains have been reported. *S. lividans* produces the same pigments as *S. coelicolor*, but only under certain conditions, and in contrast with *S. coelicolor*, it has been shown to lack agarase activity (Kieser et al, 2000) and to thiolate its DNA (Zhou et al, 2005). Moreover, *S. lividans* has been shown to be tolerant to high concentrations of mercury (Nakahara et al, 1985), whereas copper is required for its development (Keijser et al, 2000; Worral & Vijgenboom, 2010). Interestingly, these features are unique to the parental strain 66, (often also referred to as strain 1326), but are absent from or less prominent in the plasmid-less strain TK24, which was isolated after UV mutagenesis and protoplast regeneration (Hopwood et al, 1983).

During the early genetic characterization of *S. lividans*, the presence of two plasmids, termed SLP2 and SLP3, was inferred (Hopwood et al, 1983). However, only SLP2, a 50 Kbp linear molecule, has been physically isolated and characterized using pulsed field gel electrophoresis (Chen et al, 1993) and DNA sequencing (Huang et al, 2003). In contrast, SLP3 has remained elusive to date, despite clear early genetic evidence of its existence. Among this evidence, the *mer* genes responsible for resistance to mercury in *S. lividans* have been unambiguously linked to SLP3, confirming the mobile and conjugative nature of this element (Sedlmeier & Altenbuchner, 1992). Moreover, an amplifiable sequence termed AUD2 has been linked to the *mer* genes and thus to SLP3 plasmid (Eichenseer & Altenbuchner, 1994).

In addition to the report of the genome sequence of *S. coelicolor* M145, a strain obtained from *S. coelicolor* A3(2) that lacks its natural plasmids (Bentley et al, 2002), a draft genome sequence of *S. lividans* TK24 has been released and used for metabolic flux analysis (D'Huys et al, 2012). Genomic hybridization experiments using an *S. coelicolor* M145 microarray and DNA from different *S. lividans* strains, showing the absence of several *S. coelicolor* genes from *S. lividans*, have also been reported (Jayapal et al, 2007; Lewis et al, 2010). Most of these genes are clustered within genomic islands, although it was also concluded that genetic variation among these strains might actually span throughout the entire chromosome (Lewis et al, 2010).

## 6.2 Sequencing, assembly and annotation of the genome of *S. lividans* 66

The genomic sequence of *S. lividans* 66 was deciphered after a combination of second and third generation sequencing-platforms, including pyrosequencing and ligation-based approaches, and single molecule real-time sequencing, respectively. Hybrid *de novo* assembly of the sequences obtained after pyrosequencing shotgun (50 Mbp) and ligated paired-end sequencing runs (996 Mbp), yielded 1471 contigs that were sorted using the chromosomal sequence of the plasmid-less strains *S. coelicolor* M145 and *S. lividans* TK24 as references. The initial number of contigs was reduced to 999 contigs by iterative mapping and extension using the paired reads that were left out from the *de novo* assembly. To assist on the correct ordering, a mate-pair sequencing run, which led to 564 Mbp of sequence, was obtained. The sequence could then be organized in 154 scaffolds, leading to a genome sequence with 878 gaps. At this stage, construction of a fosmid library and gap closure after PCR was performed. The PCR-based efforts yielded an insignificant number of gaps being closed (Table S5), probably due to the repetitive nature of the sequence within the gaps, whereas ordering of the fosmid library is still work in progress.

Since these approaches are laborious and time-consuming, state-of-the-art sequencing technologies were explored, these technologies promise to massively generate long reads at a low cost (Koren et al, 2012). A library of single molecule real-time sequencing technology (PacBio), which yielded 67 Mb of sequence with a read average of 1.2 Kb, was obtained. Although the amount of useful sequence was reduced to 26 Mb after read correction, which significantly hampered the possibility of closing many gaps, the long reads obtained allowed us to confirm the correct order of the genome of 66. The assembly obtained after this process contains only 84 gaps with a depth of 170X, accounting for 99% of the predicted genome sequence of 66. Thus, a *S. lividans* 66 genomic sequence consisting of a chromosomal scaffold of 8,496,762 bp, predicted to encode SLP3, and a plasmid of 50,064 bp previously reported as SLP2 (Chen et al, 1993; Huang et al, 2003), was obtained.

**Table 6-1.** Selected genomic and phenotypic features of strains compared.

| **Strain** | **66** | **TK24** | **M145** |
|---|---|---|---|
| **Chromosome length** | 8,496,762 | 8,318,010 | 8,667,507 |
| **Contigs** | 85 | 333 | 1 |
| **(G+C) content** | 72.2 | 72.2 | 72.1 |
| **Non-coding RNAs** | 69 | 63 | 83 |
| **Proteins** | 8083 | 7551 | 7825 |
| **Metal sensitivity** [a] | $Cu^R$ $Hg^R$ $As^R$ $Zn^R$ | $Cu^S$ $Hg^S$ $As^S$ $Zn^S$ | $Cu^S$ $Hg^S$ $As^R$ $Zn^S$ |
| **Accession Number** | APVM00000000 | ACEY00000000 | AL645882.2 |

[a] Metal concentrations used to define resistance (R) or sensitivity (S) are provided in Figure 6-4, Hg resistance was previously reported by Nakahara et al, 1985 and Sedlmeier & Altenbuchner, 1992.

After automatic annotation and manual curation of the chromosomal scaffold of *S. lividans* 66, 8152 open reading frames were predicted, 8083 encoded proteins (named after the prefix SLI) and 69 for non-coding RNAs (Table 6-1). The relative differences of RNAs among *S. lividans* 66 (69), *S. lividans* TK24 (63) and *S. coelicolor* M145 (83) may be related to assembly and annotation difficulties of low (G + C) sequences. The annotated genome sequence of *S. lividans* 66 was deposited at DDBJ/EMBL/GenBank (BioProject id PRJNA168962, accession APVM00000000) and was used to perform whole-genome comparative analysis between closely related strains. These analyses lead to the discovery of 241 genes that are absent from *S. lividans* TK24, of which 136 genes are conserved between *S. lividans* 66 and *S. coelicolor* M145. Moreover, 367 genes in S. *lividans* lack orthologs in *S. coelicolor* (Table S6). Besides these differences, the syntenic regions between the three genomes have an average nucleotide identity of 99 %. As discussed further, the vast amount of these genes is held within genomic islands.

## 6.3 Differential genomic islands between *S. lividans* and *S. coelicolor*

Five genomic islands (>25Kbp; scoGI), and 18 genomic islets (<25 Kbp), that had been previously identified in *S. coelicolor* M145 could be confirmed, these had been annotated as absents as absent from *S. lividans* using microarray-based experiments (Jayapal et al, 2007; Lewis et al, 2010). Moreover, of the genes present in *S. lividans* 66 but not in *S. coelicolor* M145, 299 (81%) are contained within four novel genomic islands termed sliGI-1 to sliGI-4 (Figure 6-1). The obvious reason why these *S. lividans* islands could not be identified by previous efforts using microarray-based approaches is that specific oligonucleotides for the entire gene diversity of these strains could not be designed. Thus, our results complement previous results (Lewis et al, 2010), which concluded that the genetic differences among these strains span throughout their entire chromosomes. In addition, our sequence data show that the differential genomic islands of *S. coelicolor* and *S. lividans*, as summarized in Table 6-2, account for the majority of the genetic diversity amongst the strains compared.

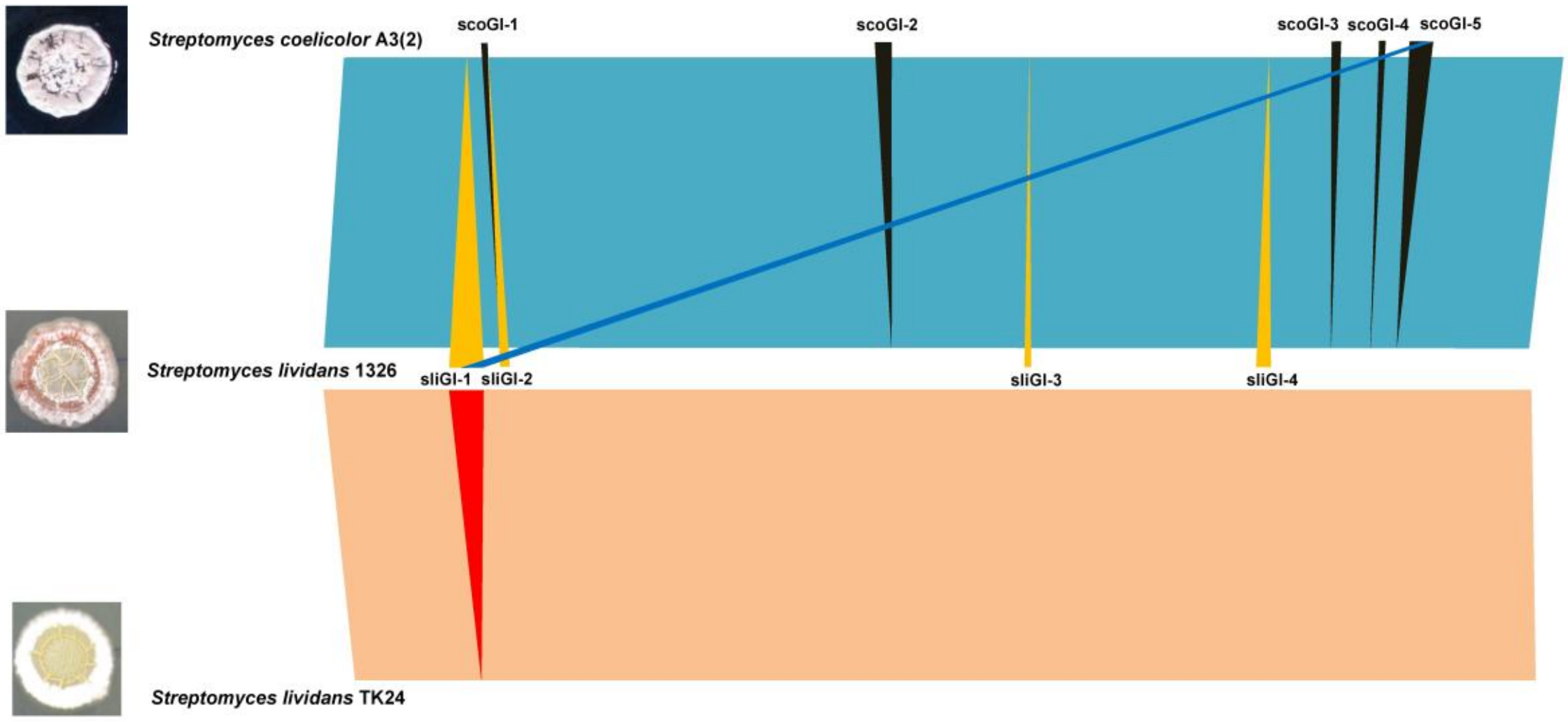

**Figure 6-1. Whole-genome comparison of *S. lividans* and *S. coelicolor*.** *S. lividans* 66 specific genome islands (SliGI) are shown in yellow triangles, and M145 (ScoGI) specific genome islands are shown in black triangles. A rearrangement between sliGI-1 and scoGI-5 is shown with a blue line, and the red triangle shows loss of sliGI-1 in TK24

SliGI-1, located at the left-arm of the chromosome, is the largest of the islands identified herein, with a total of 245 Kbp encoding 241 predicted genes (SLI0867-SLI1107). Of these genes, 138 are conserved in M145 within a single locus. Remarkably, however, this locus is actually at the right-arm of the chromosome of M145 (Figure 6-1), implying a recombination event and a chromosomal rearrangement in this strain as previously suggested (Weaver et al, 2004). Whether this recombination took place during plasmid curation of *S. coelicolor* A3(2), or earlier in the evolution of this strain, is an interesting question that remains to be answered. Moreover, sliGI-1 is the only genomic island absent from TK24. Therefore, is likely that sliGI-1 was deleted during laboratory adaptation of TK24, either during plasmid curing or afterwards, consistent with relaxation of purifying selection, as the functions encoded within this locus may not be required under laboratory conditions. Interestingly, the whole sliGI-1 region has an atypical (G + C) content of 68%, which is significantly lower than the average 71.2% for *Streptomyces* genomes as estimated herein using a total of thirty six selected genomes (Table S2), and is similar to the value found in plasmid SLP2 (Huang et al, 2003). This implies recent acquisition and a dynamic nature of sliGI-1.

**Table 6-2.** Genomic islands of *S. lividans* and *S. coelicolor*

| GI[a] | 66 genes | TK24 genes | A3(2) genes | Total Length (bp) | G+C (%) |
|---|---|---|---|---|---|
| sliGI-1 | SLI0867-1107 (SLP3SLI0867-0956) | - | SCO6808-6837 [b] SCO6841-6948[bc] | 243698 | 68.7 |
| scoGI-1 | | - | SCO0979-1000 | 26482 | 70.5 |
| sliGI-2 | SLI1222-1282 | SSPG06610-06554 | - | 59860 | 68.8 |
| scoGI-2 | SLI6339-6369[d] | SSPG01650-01677[d] | SCO3437-3539 | 107313 | 68.84 |
| sliGI-3 | SLI4708-4745 | SSPG03232-03202 | - | 27136 | 67.35 |
| sliGI-4 | SLI6282-6386 | SSPG01732-01632 | SCO3509-3533 [d] | 92503 | 67.74 |
| scoGI-3 | - | - | SCO6353-5405 | 57444 | 68.59 |
| scoGI-4 | - | - | SCO6625-6641 | 30332 | 68.61 |
| scoGI-5 | - | - | SCO6806-6953 | 153811 | 69.02 |

[a] *S. lividans* (sliGI) and *S. coelicolor* (scoGI).[b] within scoGI-5 [c] Several genes were differentially lost within this syntenic region [d] Genes present in sliGI-4 and ScoGI-2

## 6.4 SliGI-1 hosts plasmid SLP3

Early genetic and phenotypic evidence, together with a detailed sequence analysis and annotation of sliGI-1, allowed the identification of the plasmid SLP3 therein and its complete nucleotide sequence obtaining. Two lines of reasoning led to the identification of SLP3 as a region integrated into the chromosome of *S. lividans* 66 at the 5' end of sliGI-1.

First, mercury resistance, which is encoded by the *mer* operon (SLI0946-SLI0953), is absent from SLP3 minus strains, such as TK24. It has been shown that the mercury resistance phenotype can be rescued after conjugation between mercury sensitive strains, i.e. TK64, with *mer* plus strains, such as TK19 and 66 (Sedlemeier & Altenbuchner, 1992). Interestingly, a homologous *mer* operon has been found in the giant linear plasmids pRJ3L and pRJ28 from the mercury resistant *Streptomyces* strains CHR3T and CHR28T, respectively (Ravel et al, 1998), and in plasmid pPSED01 isolated from *Pseudonocardia dioxanivorans* (Sales et al, 2011). The above *Actinobacteria* were actually isolated from environments with heavy metal pollution, and when plasmids pRJ3L and pRJ28 were used to transform *S. lividans* TK24, the exconjugants became resistant to mercury. Indeed, integration of the *mer* operon into the chromosome of *S. lividans* has been previously confirmed (Ravel et al, 1998).

Second, Eichenseer and Altenbuchner (1994) described a 92 Kbp amplifiable element from *S. lividans* 66, called AUD2, which contained the *mer* operon responsible for mercury resistance. AUD2 is flanked by two insertion elements, called IS1373 (864 bp), which encode functional recombinases or InsA proteins (Volff & Altenbuchner, 1997). Two identical sequences predicted to encode transposases were found at the edges of a 94 Kbp region, which includes the *mer* operon. Thus, herein the region spanning from SLI0868 to SLI0957, which encode two identical InsA paralogs, is defined as the SLP3 amplifiable and mobile element of *S. lividans* 66 (Figure 6-2). Interestingly, in some *S. coelicolor* strains, the SLP3-borne *mer* operon was found not to be amplifiable, which has been previously explained by the fact that these strains only have one IS1373 element (Nakahara et al, 1985; Kieser et al, 2000).

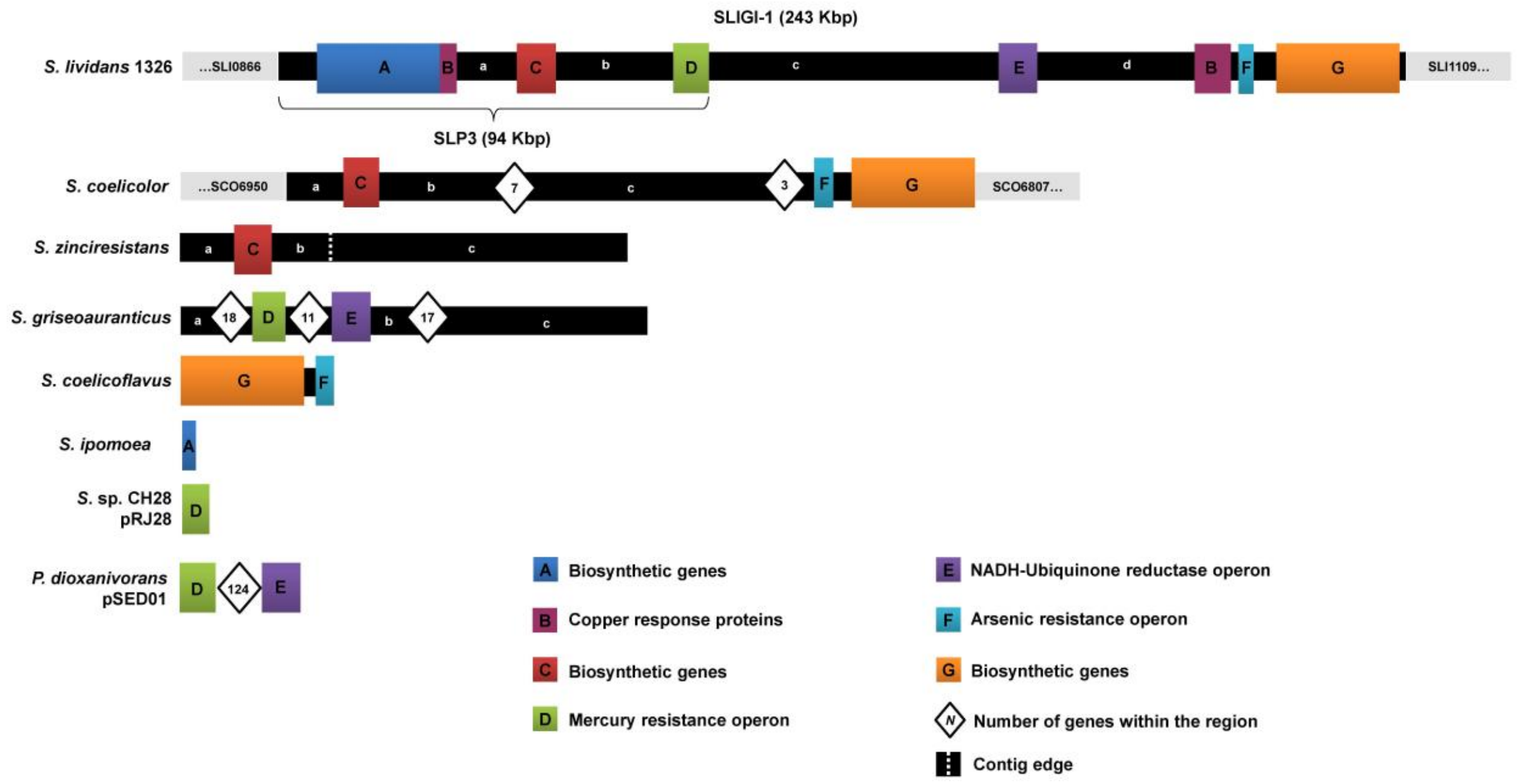

**Figure 6-2. Mosaic structure of Genomic Island 1.** SliGI-1 is represented as a thick black continuous line. Synteny blocks encoding metabolic functions implicated in metal response are shown with different colors and in uppercase letters. Regions that show certain degree of conservation between synteny blocks are marked with lowercase letters. The SLP3 mobile element is highlighted with a key. Conserved similar regions found in the genomes or plasmids of other actinomycetes are shown in the following lines. The biosynthetic systems depicted as A and G are described in chapter II of this thesis.

## 6.5 SliGI-1 is rich in genes that encode proteins functionally related to metal homeostasis

Annotation of SLP3 predicted ninety genes, including two ‘cryptic’ peptide biosynthetic gene clusters (SLI0883-96 and SLI0915-20). The first gene cluster is a previously unidentified biosynthetic system which is analyzed in chapter III. The second cluster codes for a lantibiotic biosynthetic system, also found conserved in *S. coelicolor* M145 and in *Streptomyces zinciresistens*, a taxonomically distant strain isolated from zinc and copper mine tailings (Lin et al, 2011a & 2011b). Remarkably, both biosynthetic systems are physically linked to a duplicate copper response system, consisting of a CopZ2 chaperon (SLI0895), a CsoR2 response regulator (SLI0893) and a CopA2 ATPase (SLI0896). Within these *cop* genes, a CsoR-like operator sequence, homologous to the sequence recognized by the main copper response regulator CsoR recently identified in *S. lividans* 66 (Dwarakanath et al, 2012), could also be identified (Figures 6-2 & 6-5).

Analysis of the remaining sliGI-1 genes also led to the identification of metabolic functions with a role on metal homeostasis. A duplicate NADH-ubiquinone reductase system (SLI1027-1034) known to reduce Cu(II) into Cu(I) (Rodríguez-Montelongo et al, 2006) could be identified. The gene context of this extra copper-reducing system could be found conserved in pPSED01 plasmid of *P. dioxanivorans*, as well as in the genome of *Streptomyces griseoaurianticus* (Figure 6-2). Other sliGI-1 metal-related genes include the previously reported *Streptomyces* arsenic resistance *ars* operon (SLI1077-80; Wang et al, 2006). The sliGI-1 *ars* operon is actually paralogous to a functional system conserved in the core region of the chromosome of many streptomycetes, including *S. lividans* and *S. coelicolor* (Hänel et al, 1989). Interestingly, the duplicate *ars* operon is physically linked to a polyketide biosynthetic gene cluster (SLI1088-SLI1103) that is conserved in M145 and in *S. coelicoflavus*, this biosynthetic system is revisited in chapter III. Finally, other copper-related genes could be found within SLI1047-1077, including two multi-copper oxidases (SLI1053 and SLI1071); extra copies of *copZ* and *copA* (SLI1063-4); and a paralog (SLI1067) of the previously characterized SLI4214 copper metallochaperone, or Sco1, involved in developmental switch and cytochrome c oxidase activity (Blundell et al., 2013) (Figure 6-2).

The high density of metal-related genes contained within sliGI-1, which seems to have a mosaic structure with elements from both closely and distantly related organisms, strongly hints towards a generalized functional role of this locus in metal homeostasis. Furthermore, the mosaic structure of sliGI-1, together with its anomalous (G + C) content, strongly argues in favor of recent acquisition. This suggests that *S. lividans* 66 originally inhabited a niche where metals played a role, and this may be related to loss of the entire locus in TK24. Indeed, previous reports (Ravel et al, 1998, Hänel et al, 1989), and our own data, demonstrate that TK24 is sensitive to metals, including zinc, arsenic, mercury and copper. TK24 metal sensitivity contrasts with the high metal tolerance exhibited by 66 for all four metals tested, and by M145 for zinc and arsenic (Table 6-1 & Figure 6-3).

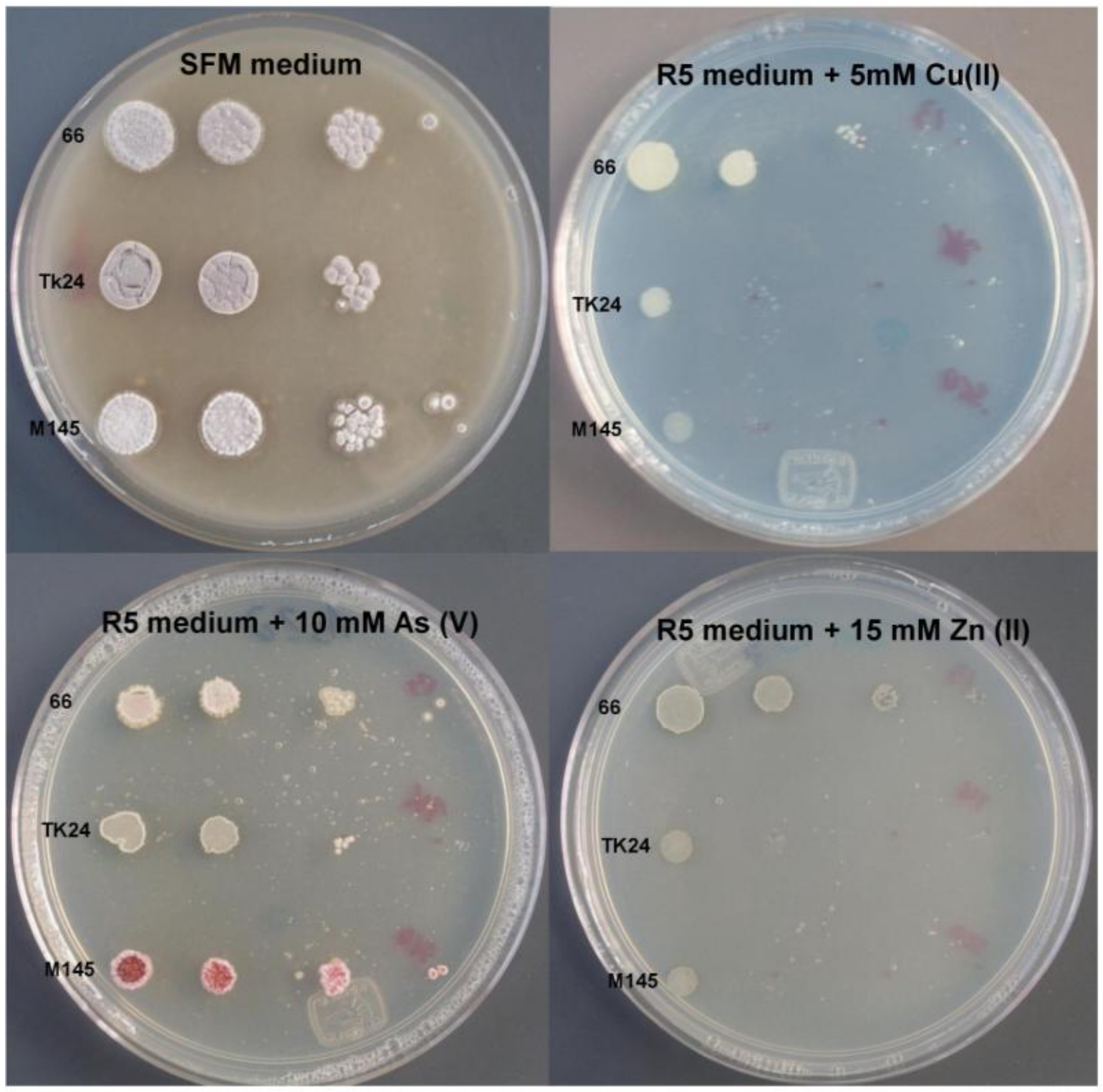

**Figure 6-3.** *Streptomyces lividans* 66 tolerance on Cu As and Zn and phenotypical comparison against TK24 and M145. As sensitivity in TK24 was addressed as the inability of this strain to produce aerial mycelium and general impaired growth.

## 6.6 Expression of SliGI-1 metal-related genes in response to copper

To corroborate the possible involvement of sliGI-1 in metal homeostasis the transcriptional response of *S. lividans* 66 under copper induction was assessed. For this analysis, RNAseq data recently reported by Dwarakanath et al, (2012) was used. The datasets analyzed come from cultures of 66 that were induced after addition of Cu(II) at a concentration of 400 µM for two hours during early log phase, and control cultures without added copper. After mapping the available RNA reads against our *S. lividans* 66 genome sequence, something that could not be done previously due to the lack of a proper reference sequence, differential expression of 137 genes throughout the entire chromosome was confirmed (Fold change ≥ 2, FDR corrected p-value < 0.05). Of these genes, 29 were induced and 108 were repressed (Figure 6-4).

**Table 6-3.** SliGI-1 genes induced in response to copper (II)

| Gene | Annotation | Fold change | FDR |
|---|---|---|---|
| **SLI0889** | Major facilitator superfamily MFS 1 | 6.46 | 3.57E-03 |
| **SLI0894** | Heavy metal-associated Domain-containing secreted protein | 2.65 | 0.01 |
| **SLI0895** | Copper chaperone | 6.51 | 2.22E-04 |
| **SLI1026** | Carbonic anhidrase | 3.62 | 0.05 |
| **SLI1041** | L,D-transpeptidase catalytic domain containing protein | 3.5 | 0.04 |
| **SLI1043** | Major facilitator superfamily transporter | 3.31 | 8.92E-05 |
| **SLI1044** | Cysteine synthase | 6.34 | 1.04E-07 |
| **SLI1052** | Multicopper oxidase | 3.65 | 5.19E-04 |
| **SLI1063** | Copper chaperone | 3.4 | 0.03 |
| **SLI1064** | Copper transporting ATPase | 3.03 | 1.58E-03 |
| **SLI1071** | Multicopper oxidase | 2.22 | 0.03 |

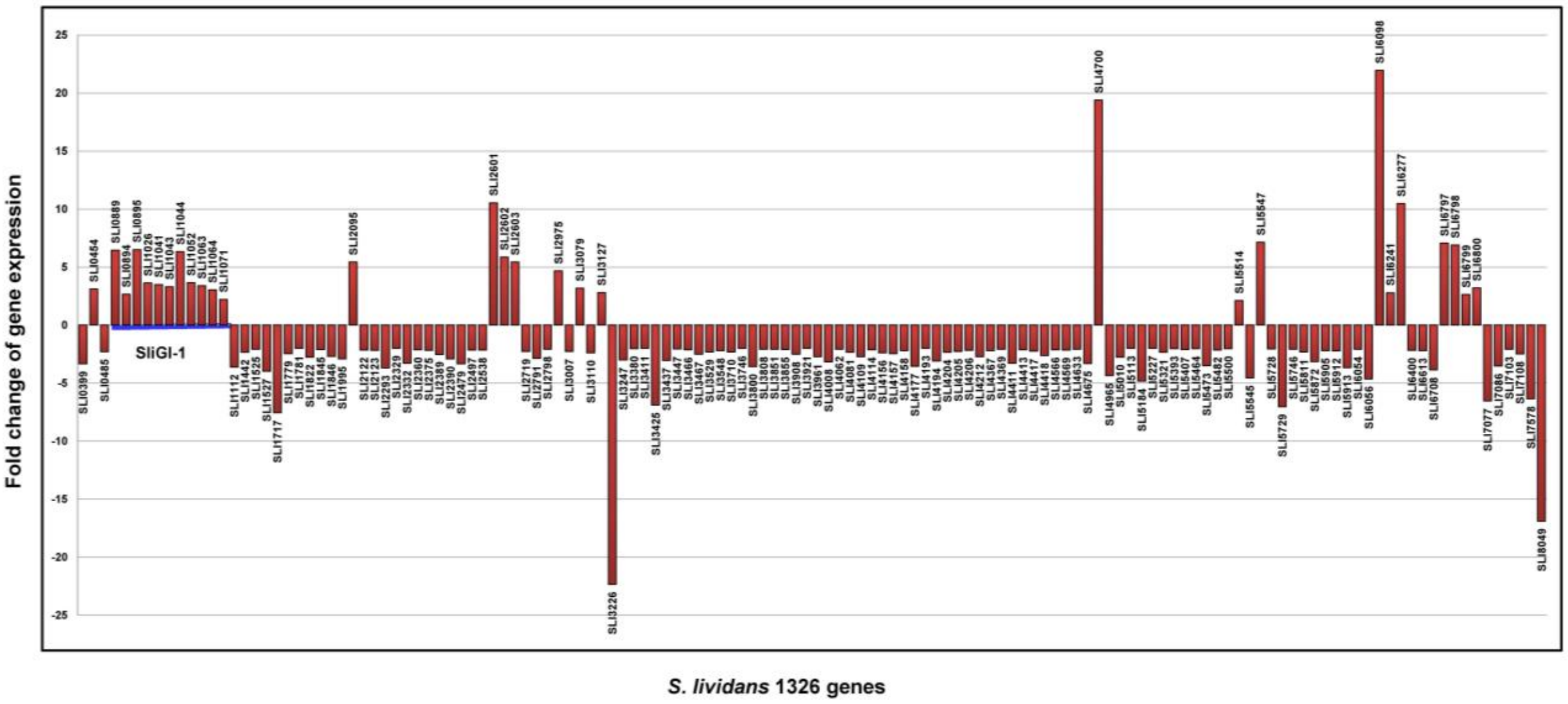

**Figure 6-4. Transcriptional response of *S. lividans* 66 to Copper.** The bar plot shows differentially expressed genes (Fold change ≥ 2; FDR corrected p-value < 0.05). Genes belonging to sliGI-1are highlighted in blue.

Of the observed copper-induced genes, eleven are actually encoded within sliGI-1 (Table 6-3), accounting for 38% of the total number of genes that are induced within only 2.9% of the chromosome. Strikingly, in contrast, none of the repressed genes are encoded within sliGI-1, supporting the proposed role of this locus in metal homeostasis. In agreement with the annotation, the induced genes include the above-mentioned duplicate copper response system, consisting of a CopZ2 chaperon (SLI0895), a CsoR2 response regulator (SLI0893), a CopA2 ATPase (SLI0896), and their predicted CsoR-like operator sequences (Figure 6-5).

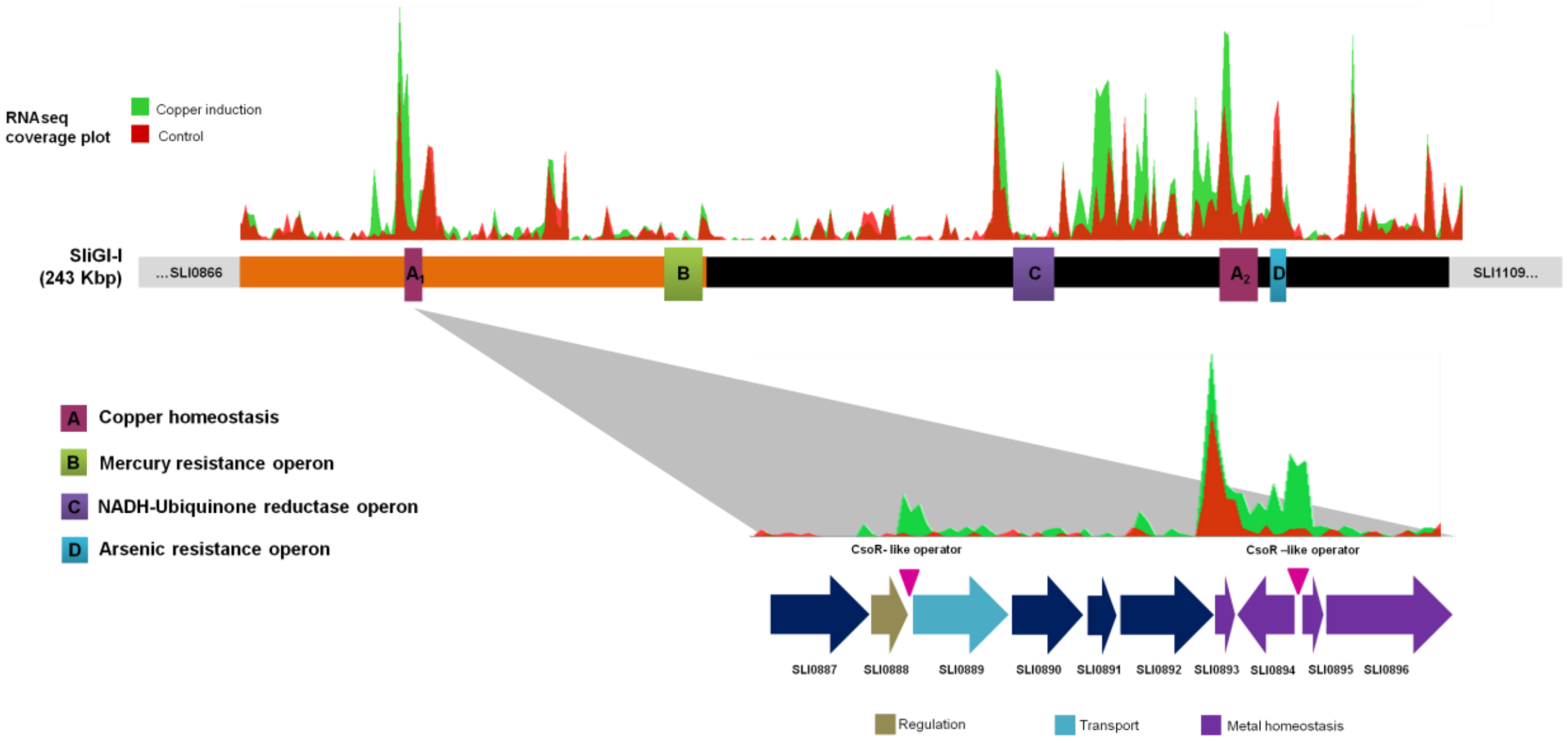

**Figure 6-5. Transcriptional response to copper of SliGI-1.** A coverage plot of the mapped RNAseq reads is shown on top of a map of SliGI-1. A zoomed region shows induction of transport and copper homeostasis proteins in SLP3 which is marked in orange, the localization of two CsoR-like operators is shown as pink triangles.

# 7 Results Chapter II: Development of EvoMining, an evolution-inspired genome mining tool

The data derived from the first stage of the strategy presented herein has been published as:

Barona-Gómez F, Cruz-Morales P, Noda-García L. 2012. **What can genome-scale metabolic network reconstructions do for prokaryotic systematics?** Antoine Van Leeuwenhoek; 101(1):35-43

The results derived from this chapter, including the complete mining strategy and proof of concept will be published, this manuscript is currently in preparation with the running tittle, "**EvoMining: evolution inspired genome mining for Natural Products**"

## 7.1 Introduction

The analysis of the first genome sequences of Natural Product (NP) producer strains, such as *S. coelicolor* (Bentley et al, 2002) *S. avermitilis* (Ikeda et al, 2003) and *S. griseus* (Ohnishi et al, 2008) published almost one decade ago, showed that practically all biosynthetic gene clusters, for known biosynthetic systems such as NRPSs and PKSs, can be detected solely on the basis of DNA sequence analysis. It also became apparent that several pathways identified after genome sequencing are cryptic, therefore their potential product(s) are unknown, implying that several new chemical structures could be found. Recent estimations indicate that the biosynthetic potential of the microorganisms has been under exploited (Clardy et al, 2006), raising the interest in microbes as the sources of therapeutically useful molecules. In addition, the drop in the costs of next-generation sequencing technology has resulted in an increase in the availability of microbial genomes, which together with the fast development of genome mining bioinformatic tools, has expanded the universe of known NPs. Although several novel molecules have been discovered using the genome-based NP mining approaches, as the currently available mining approaches are based on the identification of known NP biosynthetic systems, the identified compounds are chemically related. In other words, a high number of molecules is discovered but they are constrained in a limited space of chemical diversity. This side effect of modern NP mining approaches narrows down the boundaries of the chemical space explored in the search for interesting molecules. Therefore, novel approaches that boost the discovery of not only new molecules, but also novel classes and unprecedented biosynthetic pathways, are much needed.

In this chapter, the development of an original an efficient genome-mining system capable of identifying such NP biosynthetic pathways is described. This approach exploits evolutionary ideas and methods such as enzyme family expansion and phylogenomic analysis. A working example of this approach for the discovery of a novel compound lacking typical NP biosynthetic enzymes is presented as proof-of-concept.

## 7.2 Evolution inspired natural products Genome Mining Rationale

For the development of a bioinformatic approach able to identify potential novel biosynthetic systems, including those whose chemical or biosynthetic nature has not been observed previously an evolutionary perspective was used. The evolutionary ideas upon which this approach was developed can be summarized as follows:

1. Enzyme families ***broaden their substrate specificity*** while they share structural and mechanistic attributes (e.g. catalytic residues, transition states, active site geometry). This implies a common evolutionary origin (Gerlt and Babbitt, 2001) and thus the phylogenetic reconstruction of an enzyme family (namely the topology of the tree) should reflect this evolutionary process.

2. The ***expansion of the enzymatic repertoire*** after evolution of homologous (divergent) and non-homologous (convergent) enzymes, including the presence of complete alternative metabolic pathways, released the expanded enzymes from their original physiological constraints, allowing functional divergence. These expansions may result from gene duplication events or Horizontal Gene Transfer (HGT) (Iwasaki an Takagi, 2009; Ohno, 1970; Ramsay et al, 2009; Treangen and Rocha, 2011).

3. NP biosynthetic pathways ***evolved relatively recently***, probably from genes involved in ***central metabolism*** (Vinning, 1992; Morowitz 1999). This notion is supported by the observation that several (if not all) of the NP biosynthetic enzymes have homologues in central metabolic pathways (Heinzelmann et al, 2001; Blodgett et al, 2007; Asamizu et al, 2012).

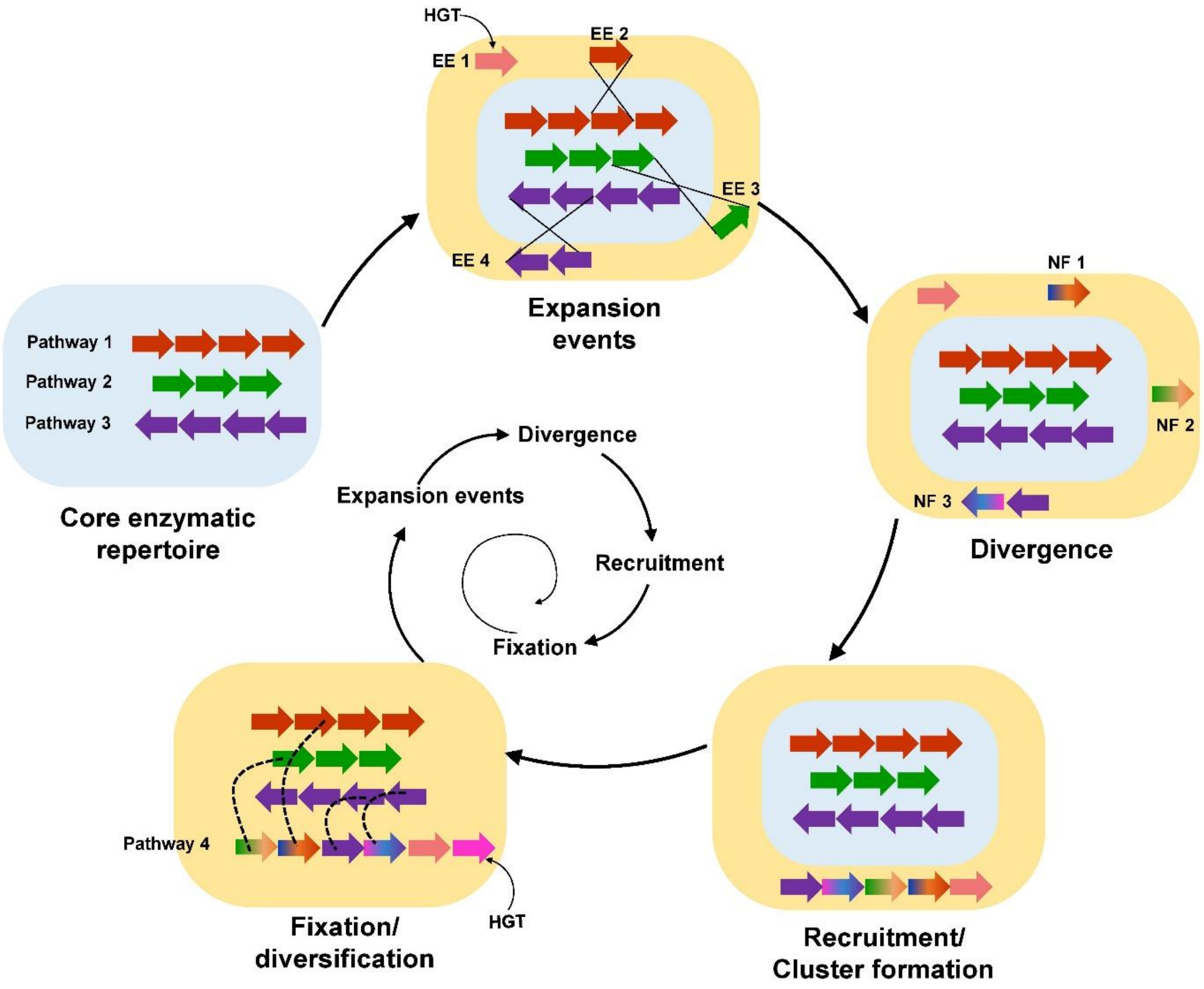

**Figure 7-1**. **Evolutionary model for the evolution of NP biosynthetic pathways**. In a first stage, the metabolic repertoire expands trough horizontal gene transfer or gene duplication, then the new genes evolve new substrate specificity while retaining their reaction mechanism. Some of the new enzymatic functions are recruited to perform their function within new biosynthetic pathways, the genes involved in these novel pathways form clusters to improve their coordinate expression, finally the gene clusters can acquire new features either in the same specie or in others after being horizontally transferred. This process can be iteratively repeat to increase the metabolic repertoire.

Based on the previous ideas, it was speculated that the expansion of central metabolism in lineages of NP-producing organisms followed divergence of substrate specificity. This in turn gave rise to enzyme sub-families that were recruited and specialized in NP biosynthesis (Figure 7-1). A phylogenomic analysis of expanded enzyme families in central metabolism should reflect this process and could be exploited for the identification of novel NP biosynthetic systems (Figure 7-2). Therefore, to identify novel NP biosynthetic pathways within genome sequences following the above described evolutionary precepts, a phylogenomic analysis including the following three stages was postulated:

> **Stage 1.** Detection of enzyme family expansions within central metabolism using as source of query sequence state-of-the-art Genome-Scale Metabolic-network Reconstructions (GSMR).
>
> **Stage 2.** Phylogenetic analysis and systematic identification of the recruited enzyme subfamilies (clades) using as benchmark members of expanded families known to have been recruited into known NP biosynthetic pathways
>
> **Stage 3.** Gene context analysis of recruited enzyme subfamilies with special emphasis on members not identified as part of known NP biosynthetic systems, but that seem to be part of gene clusters.

This strategy was captured in a bioinformatic pipeline called EvoMining (for evolution-inspired NP genome mining), which was developed to systematically predict novel pathways for the biosynthesis of NPs (Figure 7-2). EvoMining was applied using databases that were assembled and used as proof of concept. These DBs consisted of:

**DB1. Precursor supply central Pathways (PSCP)**, including eleven central metabolic pathways, which are involved in NPs precursor supply (Table S1). PSCP are highly conserved and their function is relatively easy to discern among the different homologues involved in each enzymatic step, i. e. fatty acid biosynthesis, was excluded because of the difficulty in ascribing function with confidence to the many homologues that takes part in this biosynthetic pathway. The metabolic reconstructions of *Streptomyces coelicolor* (Borodina et al, 2005) *Corynebacterium glutamicum* (Kjeldsen et al, 2009) and *Mycobacterium tuberculosis* (Jamshidi et al, 2007) were used as the source of queries sequences of the enzymes involved in each of the eleven PSCPs. Using three GSMRs from distantly related organisms, yet from the same taxonomic lineage, ensures a broad metabolic coverage including analogous and highly divergent enzyme sequences. A total of 499 non-redundant amino acid sequences are included in this database

**DB2. Genome sequences database,** including 141 actinobacterial genomes collected from public databases (Table S2). This DB was rich in genome sequences from the genus *Streptomyces*, the most proficient producers of NPs known to date.

**DB3. Natural products Biosynthesis database,** including the enzymatic repertoire present in 129 characterized NP biosynthetic pathways extracted from the literature (Table S3). The sequences in this database were manually curated to exclude sequences corresponding to PKS and NRPS enzymatic assembly complexes, as well as transport, resistance and regulatory proteins, this process was done manually based in the literature available for each biosynthetic pathway in the database. The database includes 1969 amino acid sequences involved in NP biosynthesis.

EvoMining, was used to systematically identify the enzymatic expansions (EE) within PSCP from *Actinobacteria* (the method for EE identification is described in section 7.3). The sequences of the members of **actinobaterial expanded enzyme families** were stored in **DB4**. The enzyme families within DB4, which included NP-related members or recruitments were identified by matching DB4 and DB3, the **recruited enzyme families** were stored in **DB5**. The phylogenetic reconstructions of the enzyme families in DB5 were used for the identification of enzymes potentially involved in NP biosynthesis and the genome context of potential new NP enzymes was analyzed to finally predict novel NP biosynthetic systems (Figure 7-2). The results of the application of the EvoMining approach using the above databases are described in detail in the following section.

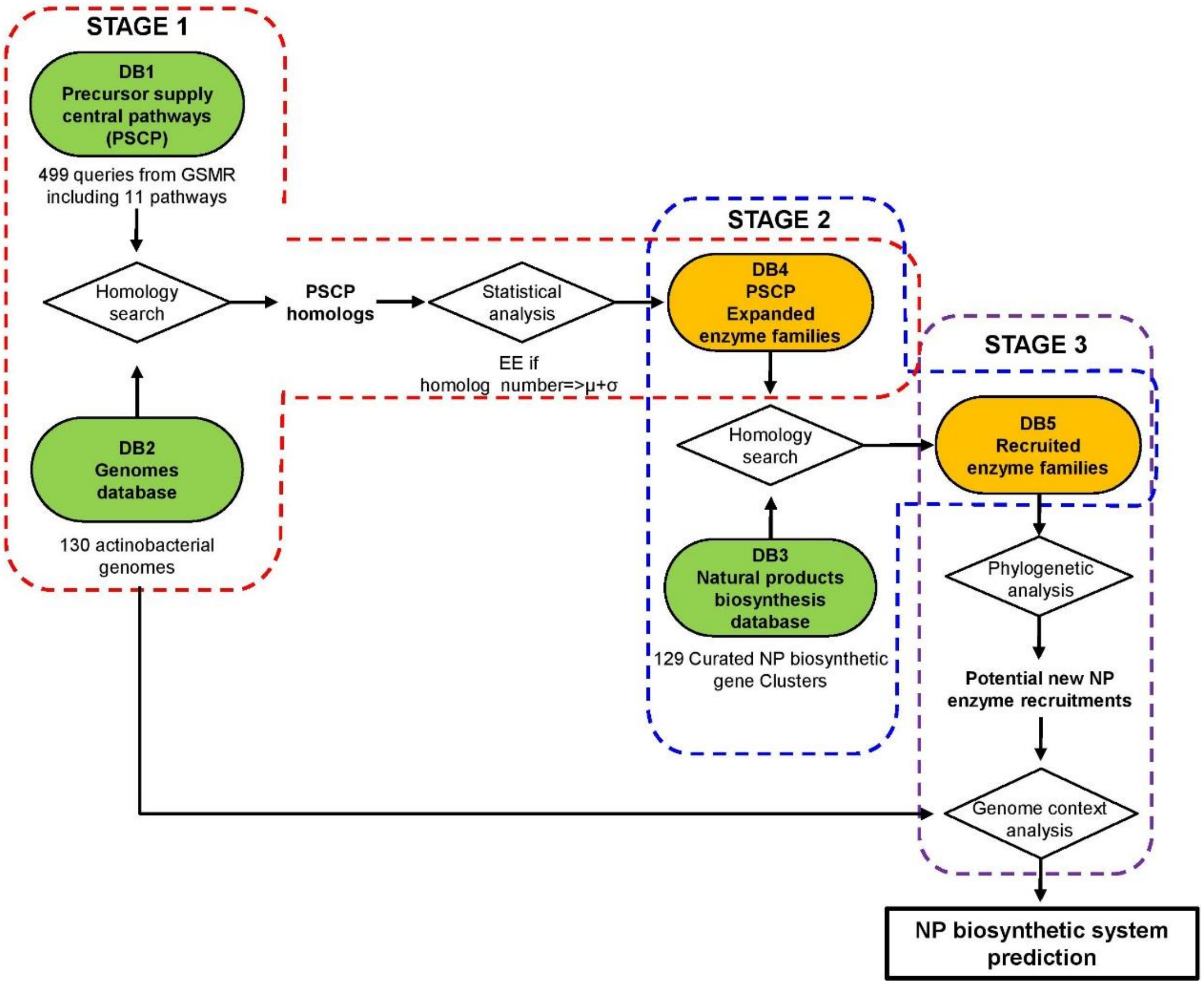

**Figure 7-2.** The EvoMining bioinformatic pipeline for identification of NP biosynthetic pathways in microbial genomes. Input databases are coloured in green, and internal databases obtained during the analysis are coloured in yellow. The stages of the analysis described in the following section are delimited in dashed lines in different colours.

## 7.3 Stage One: Enzymatic Expansion (EE) Survey

The first stage of the EvoMining pipeline (Figure 7-2) consisted of the identification of enzyme expansions in PSCP of the organisms whose genomes were available in DB2. The sequences in DB1 were used as the sources of queries for homology searches using BlastP (Altschul et al, 1990).

The homologues of the enzymes queries involved in a single metabolic step in each of the 11 PSCP were systematically identified within the actinobacterial genome database. A set of homologues associated with a metabolic step is called an enzyme family from this point onwards. The enzyme families were recorded in DB4 (PSCP expanded families) when the number of homologues in a given genome was equal or higher than the average number of homologues per genome plus the standard deviation in the same enzyme family. This statistical treatment enabled the identification of specific and particularly significant expansions in a phylogenetic context.

The result of this analysis is represented as a heatplot of EEs (figure 7-3). The rows are sorted using a species tree. Columns represent the steps for each pathway in the order they appear in the GSMRs. Each cell represents the number of homologues found for each metabolic step within central metabolism. The metabolic steps with a number of homologues significantly higher than the background are highlighted in red, whereas lower values (not expanded) are shown as empty cells. The enzyme families with significant expansions were selected for the second stage of the pipeline.

An important observation derived from this analysis is that the suborders *Micromonosporineae, Streptomycineae Pseudonocardineae*, and the *Nocardiaceae* family, show an overrepresentation of EEs. This observation is in agreement with the fact that these actinobacterial families are actually the most proficient NP producers. Based in this observation it is tempting to speculate that NP biosynthetic diversity and the occurrence of EEs may be related traits (Barona-Gómez, Cruz-Morales et al, 2012). This result is congruent with the evolutionary foundations of this approach: in proficient NP-producing species more enzymes are needed to synthesize these compounds, and if the origin of this enzyme repertoire is central metabolism, then the expansions will be more frequent in such organisms.

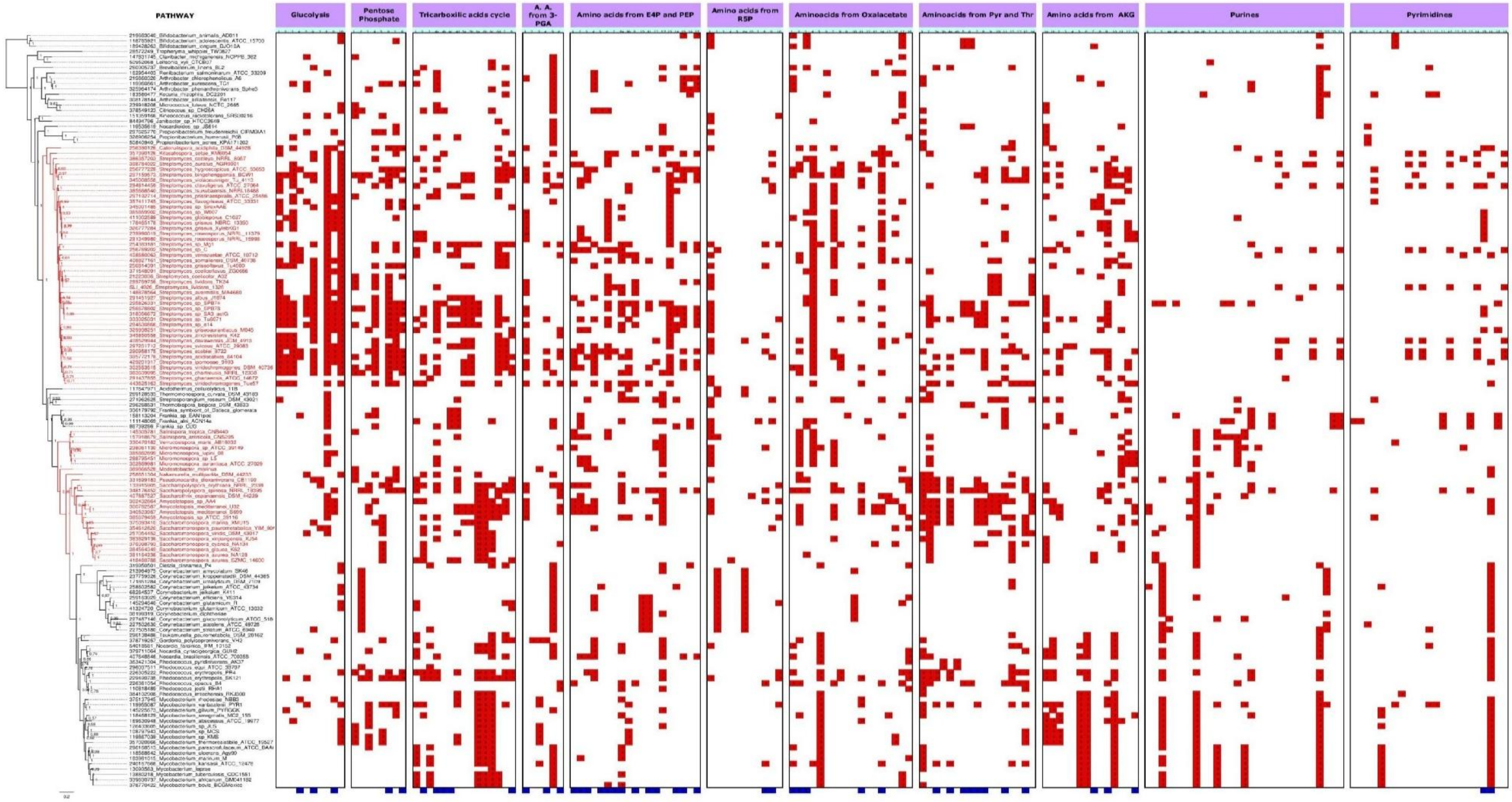

**Figure 7-3. Actinobacterial EEs with taxonomic resolution.** The order of the columns is the same as the order of the enzymatic conversions within each metabolic pathway indicated at the top. The heat plot shows those expansions statistically relevant when compared with the background number of homologues for each enzyme in the pathway. The rows are phylogenetically sorted using as guide a species phylogenetic tree. Clades of NP-producing actinobacterial families are highlighted in red. The blue cells in the bottom of the plot indicate the enzyme families (columns) with members recruited in NP biosynthesis. An electronic version of this heat plot is provided as supplementary table S4.

## 7.4 Stage Two: Recruitment events amongst NP biosynthesis

This stage was dedicated to the identification of enzyme families that include homologues of enzymes recruited within the context of NP biosynthesis. For this purpose, the sequences of previously characterized enzymes involved in the biosynthesis of NPs (Collected in the DB3; table S3) were used as queries for homology searches in each expanded enzyme family identified in stage one. When recruitment events were identified within an enzyme family, the protein sequences of the family members were retrieved from the genome database and stored in DB5 (Recruited enzyme families) for analysis in stage three.

This analysis led to the identification of sixty recruitment events in thirty-three expanded central enzyme families (table 7-1). Eighteen of the recruited families were involved in at least two and up to eleven different NP pathways, including different classes of compounds. A remarkable example of multiple recruitments is the case of the asparagine synthetase family, which is involved in the biosynthesis of polyketides, glycopeptides, alkaloids, glycolipids, and thiopeptides such as A-74528, Bleomycin, Zorbamycin, Limphostyn, Fredericamyin, Lysolipin, Moenomycin, Thiostrepton, Tetracycline-SF2575 and Piericidin A1 (A complete list of references is available in table S3). This result implies that the enzymatic function of a recruited enzyme family is independent of the chemical class of the final product in a biosynthetic pathway were the recruitment occurred

From the enzyme expansion profile and its relationship with the patterns of enzyme recruitment two contrasting expansion scenarios were identified:

(i) Within *Streptomyces*, the genus with the higher number of expansions, and the best represented group in the genome database, only three enzyme families have been significantly expanded in most strains (i. e. above 93%), namely, the pyruvate kinase, phosphoglycerate mutase, and aspartate semialdehyde dehydrogenase families. The same scenario is observed in other enzyme families across different genus in the *Actinobacteria* phylum.

(ii) Recruited EE are not conserved amongst the whole *Actinobacteria* phylum or even at the genus level: The expanded enzyme families with recruitments are present in less than 65% of the *Streptomyces* genomes analysed (figure 7-3; table 7-1).

The whole-genus conservation of enzyme expansions contrasts with the strain-specific distribution of NPs and the recruitment of enzyme families for their biosynthesis. Such anti-correlation is consistent with the role of NPs as adaptive traits, which seem to evolve in particular niches (Firn, 2010), and suggests genus-specific central metabolic adaptations for the conserved EEs. In the case of *Streptomyces*, they may be linked to adaptations needed for production of NPs. To further explore this observation, phylogenomic and experimental analysis was performed to analyse the genus-specific central metabolic EEs. The results of those analyses are the main subject of the chapter IV of this document.

**Table 7-1. EEs recruited within the context of NP biosynthesis.**

| Enzyme | EE Frequency[a] | E.C. | Known Pathways[b] |
|---|---|---|---|
| Homoserine O-succinyl transferase | 3.2 | 2.3.1.46 | D-Cycloserine |
| Anthranilate phosphoribosyl transferase | 3.2 | 2.4.2.18 | CDA |
| 3-dehydroquinate dehydratase | 3.2 | 4.2.1.10 | Rifamycin |
| Histidinol-phosphatase | 4.5 | 3.1.3.15 | Puromycin |
| Indole-3-glycerol-phosphate synthase | 4.5 | 4.1.1.48 | Calcium-dependent antibiotic |
| 3-isopropylmalate dehydrogenase | 4.5 | 1.1.1.85 | FR900098, Triostin-A |
| Aspartate kinase | 9.1 | 2.7.2.4 | Grixazone, |
| Thymidylate synthase | 9.7 | 2.1.1.148 | Polyoxin |
| Chorismate synthase | 9.7 | 4.2.3.5 | Esmeraldin, Prenilated phenazines, |
| Acetylglutamate kinase | 12.9 | 2.7.2.8 | Polyoxin |
| 3-dehydroquinate synthase | 14 | 4.2.3.4 | Acarbose, Validamycin, Gentamycin, Kanamycin, Rifamycin, Neomycin, Yatakemycin |
| Prephenate dehydrogenase | 14 | 1.3.1.12 | A-40926, A-47934, Balhimycin, Clorobiocin, Complestatin, Enduracidin, Novobiocin, CDA |
| Asparagine synthetase | 14 | 6.3.5.4 | A-74528, Bleomycin, Zorbamycin, Lymphostin, Fredericamyin, Lysolipin, Moenomycin, Thiostrepton, Tetracycline-SF2575, Piericidin A1 |
| Malate dehydrogenase | 16.1 | 1.1.1.37 | Dehydrophos |
| 2-isopropylmalate synthase | 16.1 | 2.3.3.13 | FR900098 , Fosfomycin (in *P. syringeae*) |
| Histidinol-phosphate transaminase | 18 | 2.6.1.9 | Phoslactomycin, Thiostrepton |
| 3-deoxy-7-phosphoheptulonate synthase | 18 | 2.5.1.54 | Balhimycin, Esmeraldine, Tetracycline-SF2575, Prenilated phenazines, Tomamycin, Rifamycin, CDA |
| Anthranilate synthase | 18 | 4.1.3.27 | Candicidin, Tetracycline-SF2575, Neoaureothin, CDA |
| Aconitate hidratase | 23 | 4.2.1.3 | FR900098, Phosphinothricin |
| Cysteine synthase | 23 | 2.5.1.47 | Napsamycin, Pacidamycin, Rapamycin, D-Cycloserine |
| Phosphoglycerate dehydrogenase | 23 | 1.1.1.95 | Teicoplanin, A-47934, Phosphinothricin, Daptomycin |
| dUTP diphosphatase | 32 | 3.6.1.23 | Puromycin |
| Dihydroxy-acid dehydratase | 32 | 4.2.1.9 | Salinosporamide A |
| Acetolactate synthase | 32 | 2.2.1.6 | Valanimycin, |
| Enolase | 36 | 4.2.1.11 | Phosphinothricin |
| Succinyl diaminopimelate transaminase | 41 | 2.6.1.17 | Clavulanic acid, Napsamycin, Pacidamycin, Spectinomycin, Validamycin, Acarbose, Thiocoraline |
| Acetyl ornithine aminotransferase | 41 | 2.6.1.11 | Napsamycin, Validamycin, Acarbose, Liposidomycin, Kanamycin, Clavulanic acid, Caprazamycin, Pacidamycin, Rifamycin, Spectinomycin. |
| Transketolase | 41.9 | 2.2.1.1 | Rifamycin |
| 3-phosphoshikimate 1-carboxyvinyltransferase | 45 | 2.5.1.19 | Asukamycin, Phoslactomycin, Prenilated phenazines, |
| Glyceraldehyde-3-phosphate dehydrogenase | 55 | 1.2.1.12 | Pentalenolactone |
| Fructose-bisphosphate aldolase | 64 | 4.1.2.13 | Clorobiocin |
| Threonyl-tRNA synthetase | | 6.1.1.3 | Oxazolomycin |
| Seryl-tRNA synthetase | | 6.1.1.11 | Valanimycin |
| Leucyl-Phenylalanyl tRNA Protein transferase | | 2.3.3.6 | None |
| Peptidoglycan branched peptide synthesis protein | | 2.3.2.0 | Pacidamycin, Napsamycin |
| Lysine-tRNA ligase | | 2.3.2.3 | Valanimycin |

**[a] EE Frequency in this table was calculated only for the genus *Streptomyces*. [b] The complete list of references is provided in table S1**

## 7.5 Stage three: Phylogenetic reconstruction of recruited enzyme families and genome context analysis.

***Phylogenetic reconstruction.*** After the identification of the recruitment events (DB5) performed in the stage 2 of the EvoMining pipeline (Table 7-1), the phylogenetic relations of the members of each enzyme family were reconstructed using bayesian inference methods that stimate the posterior probabilities. The identification of novel recruitment candidates, which were likely to be involved in new NP biosynthetic pathways, was done after identification of clades including known recruitments which evolutionary rate is consistent with a recent evolutionary history (i.e. long branch length) were called recruitments clades, the members of the enzyme family present in recruitment clades which have not been linked to the production of known NPs, were expected to be involved in NP biosynthetic pathways and therefore were selected for genome context analysis.

***Genome context analysis.*** The gene context analysis aimed for identification of gene neighbourhoods with an operon-like structure, i.e potentially transcriptionally- coupled genes and genes transcribed from the same DNA strand. Amongst these genes the occurrence of other biosynthetic enzymes, including but not limited to NRPS and PKS megasynthase, as well as transport and regulatory genes, were seen as good indicators of the existence of a NP biosynthetic gene cluster. Several biosynthetic pathways were predicted after this stage of the analysis and are presented in the following section.

The phylogenetic reconstructions of the each of the 36 enzyme families with recruitments (DB5) were obtained (supplementary trees S11 and figure 7-4). In most phylogenetic reconstructions, single and neat recruitment clades were identified, usually including all the known NP-related members. In several cases these clades had long branches, indicating functional divergence. In contrast, the homologues related to central metabolism followed a clade distribution similar to that in a species tree, in other words the divergence among central homologues is related to the speciation process. In few cases the phylogenetic reconstructions were too complex or their branches had low statistical support to make accurate predictions.

Several recruitment clades included homologues of unknown functions, given their location in the phylogenies and their protein evolutionary rate, is likely that these homologues are not performing functions in central metabolic processes and rather are involved in more recently evolved adaptive metabolism, or in other words, in NP biosynthesis. From these recruited homologues novel biosynthetic pathways were predicted after the gene context analysis. A list of selected predictions, which is derived from the analysis of the evolutionary history of the 36 enzyme families with recruitments is presented in table 7-2. These predictions are the final result of the complete analysis including the three stages of EvoMining.

The predictions in table 7-2 are by no means the result of comprehensive mining, but rather a survey of the results obtained using EvoMining, where the selected candidates for gene context analysis were found in branches located very close to known recruitments, and where the genome context was highly informative or even obvious (i.e. a NRPS or PKS coding gene was close to the candidate coding gene). Several candidate recruitments with less obvious genome contexts but strong phylogenetic support are still under analysis. Given the difficulty to come up with a biosynthetic prediction from these candidates and their strong phylogenetic signal of NP recruitment, it is tempting to speculate that these candidates are involved in biosynthetic pathways with previously unseen biosynthetic features which may only be solved after experimental characterization. Nevertheless, this set of predictions provides a glimpse of the effectiveness of the EvoMining approach and its potential for the identification of unprecedented biosynthetic system, confirming that the use of NRPSs, PKSs or other typical NP biosynthetic enzymes queries for NP genome mining can be avoided and that the analysis of the evolutionary history of other enzyme families can provide an unbiased method for genome mining which can also detect pathways that include NRPS, PKS and other well-known biosynthetic systems

Table 7-2. EvoMining predictions of known and novel NP biosynthetic pathways.

| **Central metabolic homologue** | **Organism** | **NP class** | **GI** |
|---|---|---|---|
| Homoserine O-succinyl transferase | *Amycolatopsis* Sp AA4 | D-cycloserine-like | 302435098 |
| | *S. espanaensis* | D-Cycloserine-like | 407884165 |
| 3-dehydroquinate dehydratase | *S.* sp e14 | PKS | 294631941 |
| | *S. erythraea* | PKS | 133912030 |
| | *M.* sp L5 | Unknown | 288796624 |
| | *S. ipomoeae* 9103 | PKS | 429195850 |
| | *S. roseosporus* NRRL15998 | PKS | 291346163 |
| | *S. violaceousniger* Tu4113 | PKS | 345012821 |
| Histidinol-phosphatase | *S. clavuligerus* ATCC 27064 | NRPS | 294815597 |
| 3-isopropylmalate dehydrogenase | *S. acidiscabies* 84104 | NRPS | 395772829 |
| | *S. violaceusniger* Tu4113 | NRPS | 345007516 |
| | *S.cluavuligerus* ATCC27064 | NRPS | 294816460 |
| | *S. coelicoflavus* ZG0656 | NRPS | 371552324 |
| | *M. kansasii* ATCC12478 | Unknown | 240169798 |
| | *S. roseosporus* NRRL11379 | Phosphonate-NRPS | 239988954 |
| | *K. radiotolerans* SRS30216 | Unknown | 151360961 |
| | *S. scabiei* | NRPS | 290954956 |
| Aspartate kinase | *P. dioxanivorans* CB1190 | NRPS | 31696814 |
| | *S. cattleya* NRRP8057 | Unknown | 386352248 |
| | *F. alni* ACN14a | Unknown | 111149371 |
| | *S. spinosa* | Unknown | 348168987 |
| | *F.* sp CCI3 | NRPS | 86740773 |
| | *S. griseoflavus* Tu4000 | Phenazine | 256812705 |
| | *S. azurea* | Unknown | 381165333 |
| Thymidylate synthase | *S. viridochromogenes* DSM 40736 | Unknown | 302553285 |
| Chorismate synthase | *S.acidiscabies* 84104 | Thaxtomin | 395768714 |
| | *K. setae* KM6054 | NRPS | 357392712 |
| | *S.* sp SA3 actG | Phenazine | 318059818 |
| Acetyl glutamate kinase | *S. hygroscopicus* ATCC 53653 | NRPS | 256782072 |
| | *S. acidiscabies* | NRPS | 395774558 |
| | *S.* sp Sirex AA | NRPS | 344998249 |
| | *S. flavogriseus* | Unknown | 357415334 |
| 3-dehydroquinate synthase | *M.marinum* M | PKS | 183982342 |
| | *S. davawensis* | NRPS | 408534222 |
| | *S. griseus* Xyleb | Unknown | 326774441 |
| | *S. coelicoflavus* ZG0656 | Acarvostatin | 371552208 |
| | *S. clavuligerus* ATCC27064 | PKS-NRPS | 294817508 |
| | *S.* sp Mg1 | NRPS | 254385349 |
| | *S. chartreusis* NRRL 12338 | Glycosyde | 383639061 |
| Prephenate dehydrogenase | *F.* sp CcI3 | NRPS | 86741149 |
| | *S. roseum* | NRPS-PKS | 271968231 |
| | *S. tsukubaensis* | NRPS | 385671607 |
| | *S.* sp e14 | NRPS-PKS | 294632600 |
| | *S. hygroscopicus* 53653 | Unknown | 256774129 |
| | *S. auratus* AGR0001 | Unknown | 398787525 |
| | *S. violaceusniger* | NRPS-PKS | 345016020 |
| Asparagine synthase | *N. multipartita* DSM442332 | Unknown | 258654594 |
| | *N. brasiliensis* ATCC700358 | PKS | 407645039 |
| | *S. auratus* AGR0001 | PKS-NRPS | 398787684 |
| | *S. zinciresistens* K42 | NRPS | 345848170 |
| | *S. roseum* DSM 43021 | PKS | 271965620 |
| | *S. acidiscabies* 84-104 | PKS | 395773795 |
| Malate dehydrogenase | *G. polyisoprenivorans* VH2 | Unknown | 378719943 |
| | *S. ipomoeae* 91-03 | Thiopeptide | 429202858 |
| 2-isopropylmalate synthase | *S. roseosporus* NRRL15998 | Phoshonate NRPS | 291348895 |
| | *S. roseosporus* NRRL11379 | Phoshonate NRPS | 239988956 |
| Anthranilate synthase | *S. roseum* DSM 43021 | NRPS | 271965230 |
| | *S. albus* J1074 | PKS | 291449793 |
| Aconitate hidratase | *S. roseosporus* NRRL 15998 | NRPS-Phosphonate | 91348897 |
| Phosphoglycerate dehydrogenase | *N. brasiliensis* ATCC700358 | Phosphonate-NRPS | 407645452 |
| | *F. alni* ACN14a | Phosphonate | 111153248 |
| Enolase | *S. sviceus* | Phosphonate | 494621398 |
| Transketolase | *S. azurea* | PKS-like | 418461294 |
| 3-phosphoshikimate 1-carboxyvinyltransferase | *S. coelicoflavus* | Unknown | 371546377 |
| | *S. coelicolor* | Unknown | 21225112 |
| | *Acidothermus cellulolyticus* | Unknown | 117648862 |
| | *S.auratus* AGR000 | Unknown | 398789709 |
| | *S.*sp. SPB74 | Polyketide | 197698457 |
| | *S.*sp SA3_actG | Phenazine | 318059820 |
| | *S.* sp Mg1 | Phenazine | 254383413 |
| | *S. viridochromogenes* | Unknown | 443627401 |
| | *F.* sp CcI3 | Unknown | 86742701 |
| Glyceraldehyde-3-P dehydrogenase | *S. bingchenggensis* BCW1 | NRPS | 297154351 |

## 7.6 Proof-of-Concept: A novel NP biosynthetic pathway predicted after a member of the enolase enzyme family

### 7.6.1 Phylogenetic analysis of the actinobacterial enolase enzyme family

Amongst the predictions made using EvoMining (Table 7-2), in this sub-section we focus on a novel biosynthetic pathway that was predicted after recruitment of an enzyme belonging to the enolase enzyme family. Enolase is a glycolytic enzyme that catalyzes the dehydratation of 2-Phosphoglycerate (2-PGA) to produce phosphoenolpyruvate (PEP) in an $Mg^{++}$-dependent reaction. Two divergent members of the family, present in *Streptomyces viridochromogenes* and *Streptomyces hygroscopicus*, have been previously identified within the biosynthetic gene cluster of phosphinothricin (PTT), a phosphonate NP. Every actinomycete genome analyzed was found to have at least one gene annotated as an enolase. Moreover, 17% of the actinobacterial species analyzed were found to have at least two enolase-like genes. This figure rose up to 47% when only the *Streptomyces* genomes were considered. From the phylogenetic reconstruction of the actinobacterial enolases, three clades could be distinguished (Figure 7-4). Clade 1 includes homologues in all actinobacterial species in the database. Clade 2 includes EEs that could not be link to enzyme recruitments within the context of NP biosynthesis. Finally, clade 3 was found to include two enolases involved in phosphonate biosynthesis, as well as other family members unlinked to a NP.

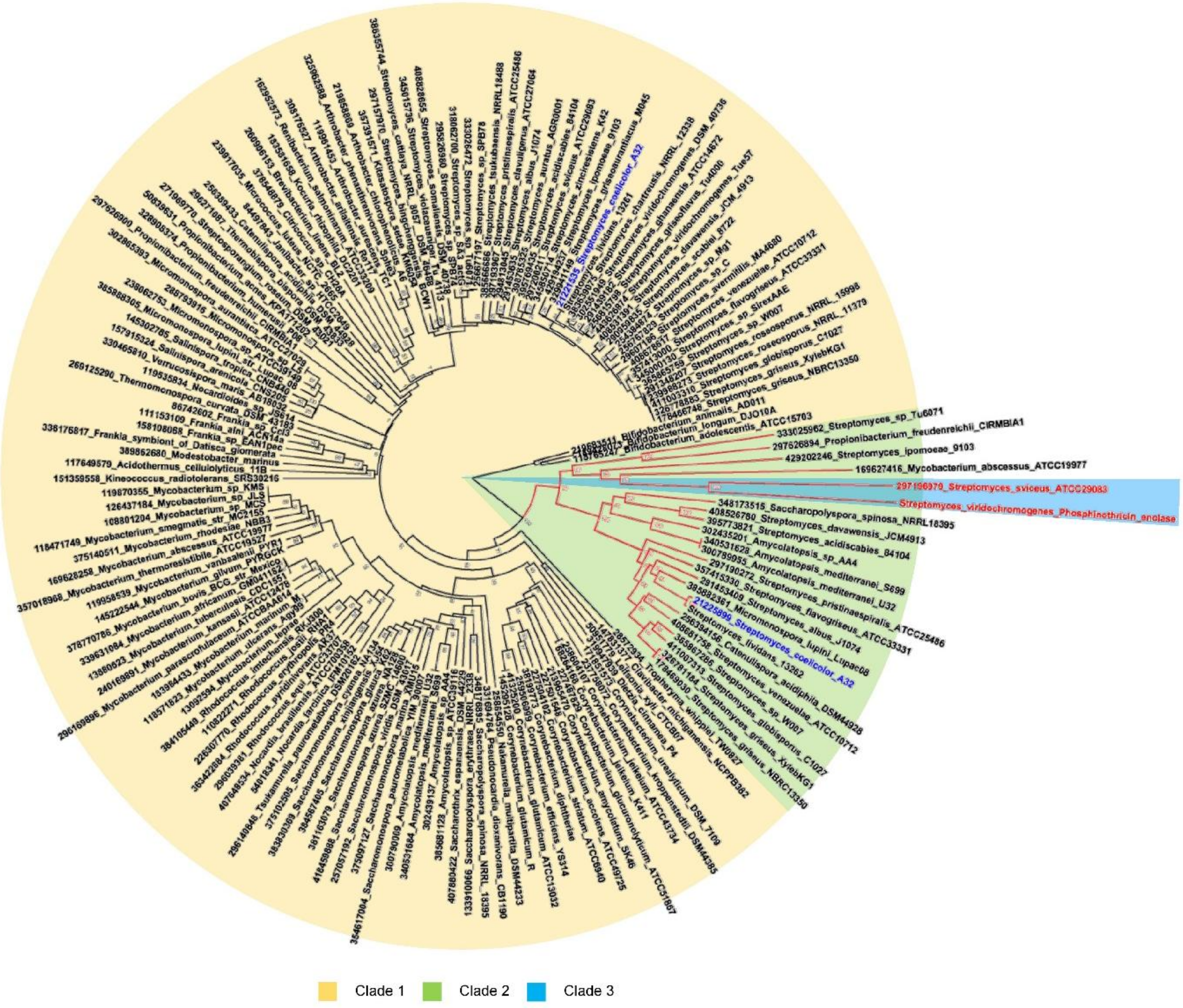

**Figure 7-4. Phylogenetic reconstruction of the enolase family in actinomycetes.** The 3 main clades corresponding to central, expanded and recruited enolases are highlighted. The recruited enolases of *S. sviceus* and *S. viridochromogenes* are highlighted in red characters. The *Streptomyces hygroscopicus* enolase involved in PTT biosynthesis is not included in the phylogenetic reconstruction due to the low quality of its sequence, however preliminary phylogenetic reconstructions locate this homologue within clade 3.

Overall, clade 1 shows a conserved genome context and was found to be conserved in all organisms analyzed. Furthermore, the topology of the branches within the clade correlates well with the species tree shown in figure 7-1. Therefore, it could be inferred that the clade includes the enolases involved in central metabolism. Clade 2 was found to include expanded enolases from *Streptomyces, Amycolatopsis, Propionebacterium* and *Catenulispora* genomes. These enolase genes were found encoded in genomic regions of these organisms that lacked conservation (e. g. the *Streptomyces* chromosome arms). The homologues in clade 2 have 57% average sequence identity with their corresponding homologues in Clade 1. For all of them the known active site residues important for dehydratation of 2PGA and the binding of $Mg^{2+}$ ions (Zhang et al, 1997) could be identified in their sequences (figure 7-6A). The topology of the tree indicates that this allele was acquired prior to the divergence of *Streptomyces*, *Amycolatopsis*, *Propionebacterium* and *Catenulispora* genera.

To determine if the role of the enolases belonging to clades 1 and 2 are linked to central metabolism, the gene coding for the enolase orthologs from the model organism *S. coelicolor* were knocked out. These enolases are encoded by locus SCO3096 and SCO7638, respectively. Both mutants could be obtained following standard mutagenesis protocols as described in Materials & Methods, and were found to be able to grow in rich defined media (R5) and minimal media with glucose as sole carbon source (figure 7-5). This experiment suggests that both enolases, from clades 1 and 2, can overtake the physiological role related to a glycolytic enolase.

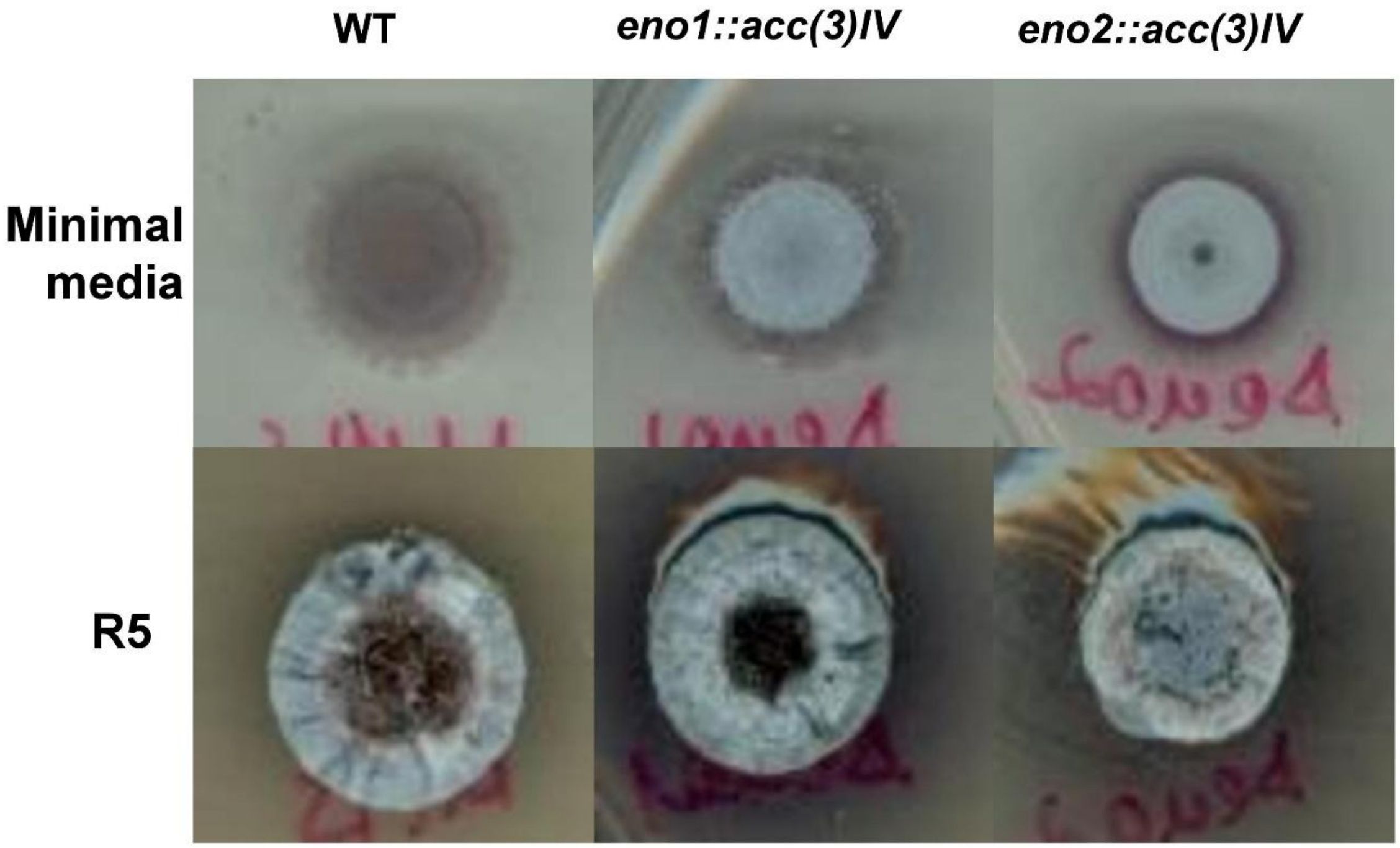

**Figure 7-5. Phenotypes of the *S. coelicolor* mutants lacking *eno*1 and *eno*2.** The wild-type parental strain and the mutants were grown on minimal media with glucose as the sole carbon source and on rich define media (R5) for 72 hrs.

The highly divergent Clade 3 includes the expanded enolases of *Streptomyces sviceus* and *S. viridochromogenes*. The latter is actually known to be part of the PTT biosynthetic gene cluster. The PTT enolase or carboxyphosphoenolpyruvate synthase (CPS) shares 33% sequence identity with its glycolytic counterpart from the same organism, *S. viridochromogenes*. A detailed sequence analysis showed only few changes in the active site residues. To identify the 3D position of these changes a structural model of the CPS from *S. viridochromogenes* was obtained and compared with the thoroughly characterized crystal structure of the yeast enolase (PDB entry 2ONE; Zhang et al, 1997). This structural analysis points towards a change in the active site residue E211S (numbering of yeast enolase) in CPS. This change could potentially allow the formyl moiety, bound to the phosphor atom in the substrate of CPS, to be accommodated. The C-P bond is the only difference between the substrate / product of the glycolytic and NP enolases (Blodgett et al, 2005; figure 7-4B and 4C).

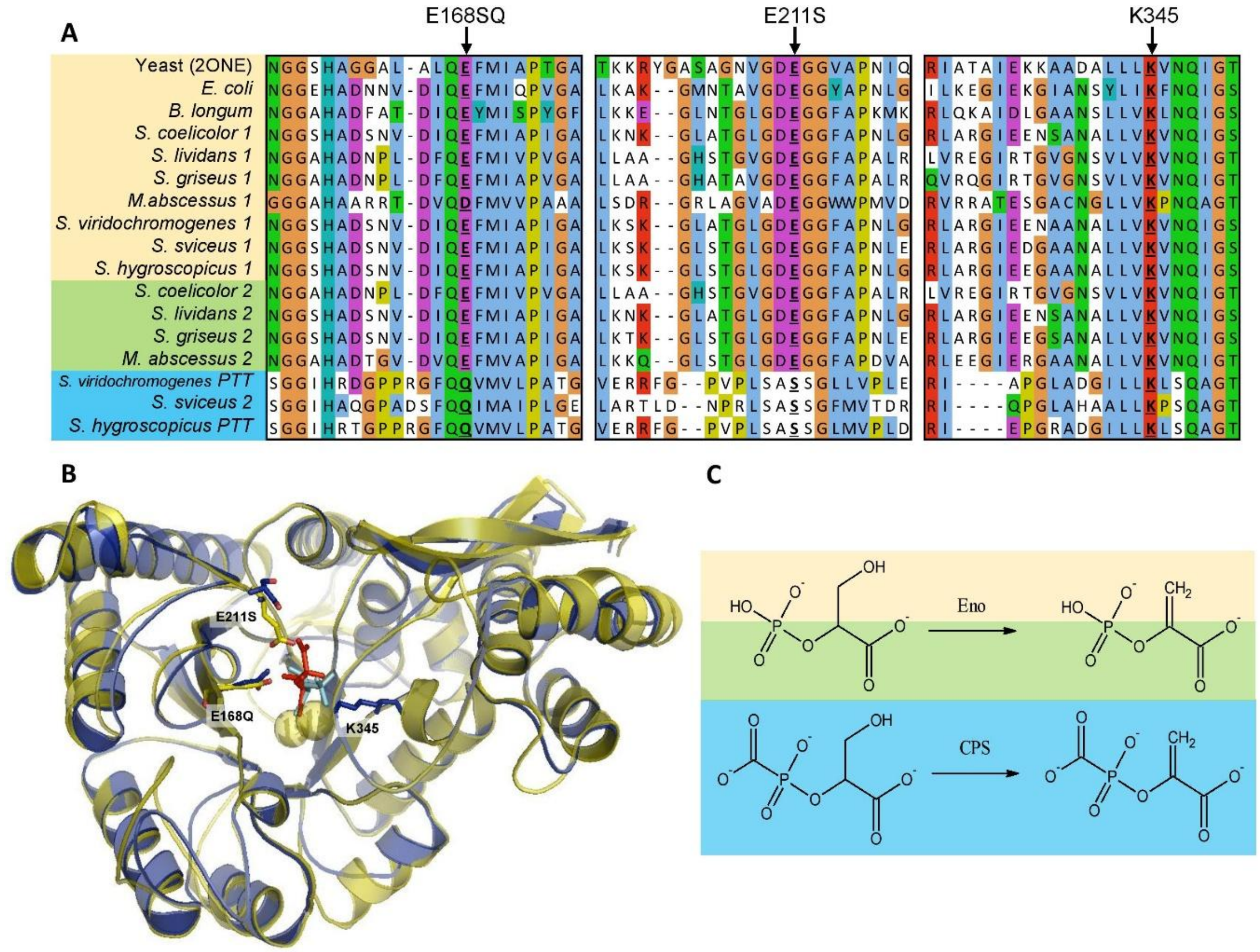

**Figure 7-6. Structure-function analysis of enolases and carboxyphosphoenolpyruvate synthases (CPS). A.** Sequence alignment of enolases from various organisms and CPS, the amino acid numbers are relative to the yeast enolase. The catalytic residues are indicated at the top and the colors are the same than those in the phylogenetic reconstruction in figure 7-4 **B.** Comparison of the yeast enolase crystal structure bound with its product phosphoenolpyruvate (PEP) and a structural model of the CPS from *S. viridochromogenes* and its substrate carboxy-phosphoenolpyruvate (CPEP), K345 the conserved catalytic base, and the mutations in the catalytic acid E211S and the catalytic water molecule holder E168Q are indicated and shown in sticks. **C.** Reactions catalysed by the glycolytic enolases and the CPSs.

## 7.6.2 Prediction of a novel phosphonate natural product in *S. sviceus*

It has previously been reported that several *Streptomyce*s species are able to produce phosphonates (Metcalf and Van der Donk, 2009). However, the production of PTT (or any phosphonate NP) in *S. sviceus* has not been reported to date. Due to the location of this enolase homologue within clade 3, i.e. the recruitments clade, this system was selected for gene context analysis. The genome sequence of *S. sviceus* has not been completed (GenBank accession: CM000951.1 and BioProject PRJNA59513), and as many genomic projects currently available in the public databases, it has been deposited as a draft genome. The quality of this draft genome sequence implies that the sequence is fragmented in several scaffolds, including many gaps. In fact, six gaps were located in the region of interest, including one at the 5' end of the enolase homologue, leading to a partial sequence. Interestingly, no PKSs, NRPSs or C-P bond forming enzymes (a mutase-decarboxylase pair; Metcalf and Van der Donk, 2009) could be found in the gene neighborhood in the original genome draft.

Using the strong phylogenetic signal of the recruited enolases as a benchmark, a NP biosynthetic gene cluster could be predicted. To confirm this prediction, the six gaps in the region were closed by iterative PCR amplification and sequencing. After confirmation of the contig order and gap filling, the region (table 7-3) was manually annotated. The annotation and functional predictions confirmed the presence of a NP biosynthetic gene cluster encoding a pathway that shares common steps with PTT biosynthesis, this cluster and its genes were named *psv* after **P**hosponate and ***sv****iceus*. The boundaries of the gene cluster were predicted on the basis of the gene organization and the products of the genes, according to these criteria, the *psv* gene cluster includes 24 genes. The conserved genes between the *psv* and the PTT biosynthetic gene clusters relate to the formation of phosphinopyruvate (a phosphinate) from phosphoenolpyruvate (Blodgett et al, 2003; figure 7-7), these reactions are potentially catalyzed by the enzymatic products of *psv*9,15-21 and 23, including the mutase-decarboxylase pair present in all *Streptomyces* phosphonate biosynthetic systems (Figure 7-7).

A second building block in the *psv* encoded biosynthetic pathway was predicted to be aminolevulinate (ALA): *psv13* encodes for a homologue of the central metabolic enzyme Glutamate-1-semialdehyde aminotransferase, which isomerizes glutamate semialdehyde to produce ALA, a precursor in heme biosynthesis (Petrícek et al, 2006). This precursor is predicted to be condensed with coenzyme A by the action of a 2-amino-3-ketobutyrate coenzyme-A ligase, encoded in *psv14*, which is potentially transcriptionally coupled with *psv13*. Homologues of this enzyme are involved in the biosynthesis of 2-amino-3-hydroxycyclopent-2-enone (also known as $C_5N$) a moiety found in several polyketide NPs (Zhang et al, 2010). In C5N formation, aminolevulinate-CoA is further cyclized by an aminolevulinate synthase, which condenses Succinyl-CoA and glycine. However, in the *psv* gene cluster no similar enzymatic functions could be annotated. Therefore it was predicted that the product of *psv*13-14 is ALA-CoA. No other enzymes responsible of building block incorporation were identified within the *psv* gene cluster.

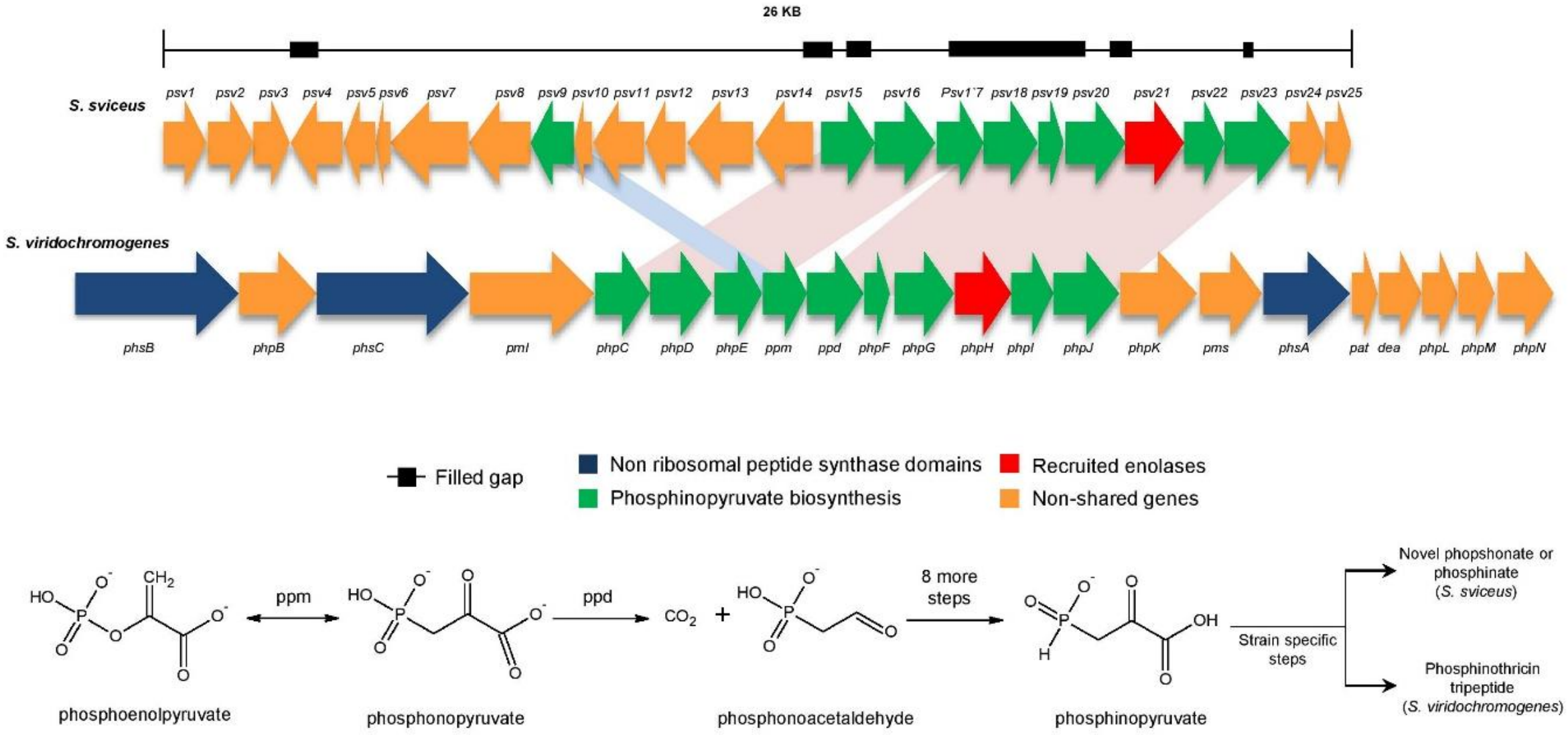

**Figure 7-7.** The *psv* gene cluster that encodes a novel biosynthetic pathway for a cryptic phosphonate NP identified using EvoMining on the genome of *S. sviceus*. At the top, the gene cluster organization is presented and compared with the PTT gene cluster of *S. viridochromogenes* at the bottom the common biosynthetic steps between the PTT and PSV pathways are shown

The assembly of the aminolevulinate-CoA and the phosphinate intermediary could be performed by the products of *psv*5-7 that encode enzymes predicted to form a small assembly system. In this system, a phosphopantetheinyl transferase (*psv*5) attaches a phosphopanteteinyl group to *psv*6. the psv6-phosphopanteteinyl complex is then used as a carrier protein for the assembly of the two predicted intermediaries in a similar fashion than that of PKSs and NRPSs. In this putative assembly complex, ALA-CoA is the starter molecule and is transferred to the carrier protein by a phosphonate-acyl tansferase encoded in *psv*7, forming an ALA-carrier protein complex and releasing CoA. The condensation of ALA and the phosphinate intermediariy could then occurs after the transfer of the phosphinate to the ALA-carrier protein complex as an extender unit, this reaction will be catalyzed again by Psv7, the only acyl transferase-like enzyme in the cluster. The release of the assembled ALA-phosphinate molecule must be performed by reductases or thiosterases, however, no thiosterases could be identified within the *psv* gene cluster. Instead, a phosphite dehydrogenase (PTHD) homologue was found encoded in *psv*4, PTDH can oxidaze phosphinates and produce phosphonates in an NAD dependent reaction (Relyea and Van der Donk, 2005). Since the release of the product is caused by its oxidation, it was predicted that the final product of the pathway is a small phosphonate NP (250-300 Da).

Two important observations derived from the previous biosynthetic prediction stress the advantages of the EvoMining approach: (i) the presence the phosphonopyruvate decarboxylase, a key enzyme in phosphonate biosynthesis (*psv18)* as well as other phosphonate biosynthetic enzymes *(psv19* and *psv20*) was only detected after closing the gaps in the original sequence, these enzymes would be the typical hits in an homology-based NP genome mining approach  (ii) The *psv* gene cluster lacks of NRPS or PKS coding genes. Therefore, using conventional genome mining approaches, such as the search for typical NP biosynthetic enzymes, would make the prediction of the *psv* gene cluster very difficult.

**Table 7-3.** Annotation of the *psv* gene cluster in *Streptomyces sviceus* ATCC29083

| Gene | Locus tag | Original Annotation | Predicted function | Length[1] | Closest homologue[2] | ID[3] |
|---|---|---|---|---|---|---|
| *psv1* | SSEG_06268 | Transcriptional regulator | LysR family transcriptional regulator | 306 | LysR family transcriptional regulator *Nostoc punctiforme* PCC 73102 WP_012411263.1 | 36% |
| *psv2* | SSEG_09941 | ABC transporter | ABC transporter ATPase subunit | 326 | ABC transporter *Catenulispora acidiphila* DSM 44928 YP_003114674.1 | 63% |
| *psv3* | SSEG_09940 | ABC transporter | ABC-type multidrug transporter permease | 261 | ABC transporter *Catenulispora acidiphila* DSM 44928 YP_003114675.1 | 63% |
| *psv4* | SSEG_06265 | Frame-shifted CDS | Phosphonate dehydrogenase | 372 | D-isomer specific 2-hydroxyacid dehydrogenase *Cyanothece* sp. ATCC 51142 YP_001803974.1 | 45% |
| *psv5* | SSEG_09939 | Phosphopantetheinyl transferase | Phosphopantetheinyl transferase | 223 | 4'-phosphopantetheinyl transferase *Nocardia asteroides* WP_019048095.1 | 39% |
| *psv6* | SSEG_09938 | Acyl carrier protein | Phosphopantetheine-binding protein | 106 | Hypothetical protein *Actinokineospora enzanensisWP*_018685819.1 | 46% |
| *psv7* | SSEG_06262 | AMP-dependent synthetase and ligase | Phosphonate-acyltransferase | 556 | Hypothetical protein *Salinispora pacifica* WP_018724810.1 | 50% |
| *psv8* | SSEG_06261 | Manganese transporter MntH | Manganese transporter MntH | 439 | mn2+/fe2+ transporter, nramp family *Micromonospora* sp. L5 YP_004081943.1 | 52% |
| *psv9* | SSEG_09937 | Phosphonopyruvate hydrolase | Phosphoenolpyruvate phosphomutase | 309 | Phosphoenolpyruvate phosphomutase *Saccharopolyspora spinosa* WP_010311250.1 | 61% |
| *psv10* | SSEG_09936 | Putative Ferredoxin | Rieske (2Fe-2S) iron-sulfur domain protein | 126 | Rieske (2Fe-2S) iron-sulfur domain-containing protein *Pseudonocardia dioxanivorans* CB1190 YP_004991190.1 | 46% |
| *psv11* | SSEG_09935 | Amidohydrolase 2 | Metallo-dependent amidohydrolase | 361 | Hypothetical protein *Paenibacillus daejeonensis* WP_020617226.1 | 50% |
| *psv12* | SSEG_09934 | 2-hydroxycyclohexanecarboxyl-CoA dehydrogenase | Short chain dehydrogenase/reductase family | 284 | Alcohol dehydrogenase *Nocardiopsis halotolerans* WP_017571421.1 | 44% |
| *psv13* | SSEG_09933 | Glutamate-1-semialdehyde-2,1-aminomutase | Glutamate-1-semialdehyde aminotransferase | 468 | Glutamate-1-semialdehyde aminotransferase *Pseudomonas mendocina* NK-01 YP_004381962.1 | 36% |
| *psv14* | SSEG_09932 | 2-amino-3-ketobutyrate coenzyme A ligase | Aminolevulinate-coenzyme A ligase | 412 | 8-amino-7-oxononanoate synthase *Pontibacter* sp. BAB1700 WP_007656444.1 | 58% |
| *psv15* | Within a gap | - | Putative alcohol dehydrogenase | 382 | Alcohol dehydrogenase *Streptomyces rimosus* WP_003983090.1 | 41% |
| *psv16* | SEG_10418 | Hydroxyethylphosphonate dioxygenase | Hydroxyethylphosphonate dioxygenase | 439 | 2-hydroxyethylphosphonate dioxygenase phpD *Streptomyces viridochromogenes* Q5IW40.2 | 46% |
| *psv17* | SSEG_10417 | D-3-phosphoglycerate dehydrogenase | 3-phosphoglycerate dehydrogenase | 338 | D-3-phosphoglycerate dehydrogenase *Frankia alni* ACN14a YP_716510.1 | 62% |
| *psv18* | Within a gap | - | Phosphonopyruvate decarboxylase | 383 | phosphonopyruvate decarboxylase *Nocardia brasiliensis* ATCC 700358 YP_006809213.1 | 63% |
| *psv19* | Within a gap | - | Nicotinamide mononucleotide adenylyltransferase | 184 | Nicotinamide mononucleotide adenylyltransferase phpF *Streptomyces viridochromogenes* WP_003988634.1 | 74% |
| *psv20* | SEG_10416 | 2,3-bisphosphoglycerate-independent phosphoglycerate mutase (Partial CDS) | 2,3-bisphosphoglycerate-independent phosphoglycerate mutase | 427 | PhpG *Streptomyces viridochromogenes* AAU00082.2 | 61% |
| *psv21* | SSEG_10415 | C-terminal TIM barrel domain-containing protein (Partial CDS) | Enolase | 421 | Carboxyphosphoenolpyruvate synthase *Streptomyces viridochromogenes* CAJ14048.1 | 50% |
| *psv22* | SSEG_10414 | carboxyvinyl-carboxyphosphonate phosphorylmutase | Carboxyphosphonoenolpyruvate mutase | 286 | Carboxyphosphonoenolpyruvate mutase *Streptomyces hygroscopicus* CAA48139.1 | 80% |
| *psv23* | SSEG_10413 | Conserved hypothetical protein (Partial CDS) | Aldehyde dehydrogenase | 462 | Hypothetical protein *Amycolatopsis nigrescens* WP_020670096.1 | 48% |
| *psv24* | SSEG_08119 | Beta-lactamase domain-containing protein | Beta-lactamase domain-containing protein | 255 | Aldehyde dehydrogenase PhpJ *Streptomyces viridochromogenes* WP_003988630.1 | 69% |

[1]Number of amino acids in the protein; [2]Identified using BlastP; [3] Percentage of amino acid sequence identity based in the BlastP alignment

# 8 Results Chapter III: Discovery of novel Natural Product biosynthetic pathways in *S. lividans* 66 using EvoMining

## 8.1 Introduction

In this chapter, the potential of EvoMining, the genome mining approach described in the previous chapter, is evaluated after using it for the identification of novel biosynthetic pathways in the complete genome sequence of *S. lividans* 66. The whole genome sequence of this model streptomycete, and its analysis, was presented in chapter I. As a point of comparison, *S. lividans* 66 was also mined for natural products using AntiSMASH (Medema et al, 2011), a genome mining tool that integrates current knowledge on Natural Products, and represents the most currently used approach. The analysis of *S. lividans* 66 genome sequence using EvoMining led to the identification of biosynthetic pathways that were not detected using AntiSMASH, and have not been reported elsewhere. Detailed genetic, biosynthetic and functional analyses of the pathways predicted by EvoMining are presented in this chapter. The remarkable observation that *S. lividans* 66, a strain that has been under study for at least 40 years, is yet delivering exciting new biosynthetic discoveries, validates the use of EvoMining as a powerful approach. Moreover, these results support the idea that an important fraction of the natural product and biosynthetic repertoire is still awaiting to be discovered, a process that can be boosted by the use of appropriate tools with predictive power.

## 8.2 Mining of *S. lividans 66* for Natural Products

To test the potential of EvoMining the genome of *S. lividans* 66 (Cruz-Morales et al, 2013; Accession number: APVM01000000) was mined using the same criteria and databases described in chapter I. To compare the performance of EvoMining with current state-of-the art Natural Product genome mining tools, *S. lividans* 66 was also investigated using AntiSMASH (Medema et al, 2011), via its web interface (http://antismash.secondarymetabolites.org/). Using AntiSMASH, 27 NP biosynthetic gene clusters were identified (Table S7). As expected, antiSMASH was efficient for detection of biosynthetic systems of known classes of metabolites, including 7 PKSs, 4 NRPSs, 3 lantipeptides, 5 terpenes, amongst others. However, since *S. lividans* 66 is closely related to *S. coelicolor*, most of the biosynthetic systems detected with AntiSMASH were reported during the first efforts of genome mining using this model organism (Bentley et al, 2002; Nett et al, 2009). These predictions, include several pathways and metabolites that have been structurally elucidated. A comparison of the predicted NP biosynthetic gene clusters using AntiSMASH and the predictions based simply in high sequence identity (> 95 %) between orthologous genes in *S. lividans* 66 and *S. coelicolor* is shown in Table 8-1. Strikingly, the biosynthetic gene clusters for two known compounds, the thiopeptide SapB and the polyketide germicidin were not detected by AntiSMASH within *S. lividans* 66. This emphasizes the empirical and knowledge-based nature of approaches based in sequence similarity searches, such as antiSMASH.

In comparison, given that EvoMining does not rely in known NP enzyme sequences, the use of this approach for the identification of potential biosynthetic systems in the genome of *S. lividans* 66, yielded nine hits from six biosynthetic gene clusters. At this point is evident that both approaches are complementary, while AntiSMASH is highly efficient in the prediction of well-known biosynthetic systems, EvoMining Allows the prediction of novel pathways. (Table 8-2). Indeed, none of the EvoMining hits were reported by AntiSMASH, although five are linked to NRPSs that were detected by this approach. Also, six of these hits were paralogs from central metabolic pathways that have been previously implicated in precursor supply of NPs. These recruitments, together with the three completely novel hits, are discussed below.

**Table 8-1.** Natural Products biosynthetic repertoire of *S. lividans* 66 as detected by AntiSMASH.

| Biosynthetic system | AntiSMASH prediction | Natural Product | Gene numbers | References |
|---|---|---|---|---|
| 1 | T4PKS-t1pPKS | Eicosapentaenoic acid | SLI0039-SLI0076 | Nett et al, 2009 |
| 2 | Terpene | Isorenieratene[a] | SLI0118-SLI0144 | Takano et al, 2005 |
| 3 | Lantipeptide | Lantibiotic | SLI0208-SLI0231 | Nett et al, 2009 |
| 4 | **none** | Deoxysugar | SLI0339-SLI0359 | Bentley et al, 2002; Nett et al 2009 |
| 5 | NRPS | Coelichelin[a] | SLI0449-SLI0459 | Lautru et al, 2005 |
| 6 | Bacteriocin | Bacteriocin | SLI0730-SLI0737 | Nett et al, 2009 |
| 7 | Other | | SLI0864-SLI0898 | |
| 8 | Lantipeptide | Lantibiotic | SLI0902-SLI0926 | Nett et al, 2009 |
| 9 | T1PKS | Polyketide | SLI1064-SLI1104 | Bentley et al, 2002; Nett et al 2009 |
| 10 | T3PKS | Tetrahydroxynaphtalene[a], flaviolin[a] | SLI1463-SLI1505 | Izumikawa et al, 2003; Austin et al, 2004; Nett et al, 2009 |
| 11 | **none** | Aromatic polyketide | SLI1546-SLI1554 | Bentley et al, 2002; Nett et al 2009 |
| 12 | Ectoine | 5-Hydroxyectoine[a] | SLI2171-SLI2182 | Bursy et al, 2008 |
| 13 | Melanin | Melanin | SLI3043-SLI3055 | Nett et al, 2009 |
| 14 | Siderophore | Desferrioxamine[a] | SLI3127-SLI3132 | Barona-Gómez et al, 2004 |
| 15 | NRPS | CDA[a] | SLI3572-SLI3618 | Hojati et al, 2002 |
| 16 | Lantipeptide | | SLI4448-SLI4474 | - |
| 17 | T2PKS | Actinorhodin[a] | SLI5345-SLI5390 | Malpartida et al, 1984 |
| 18 | Terpene | Albaflavenone[a] | SLI5510-SLI5512 | Lin et al, 2006 |
| 19 | T2PKS | Gray Spore pigment[a] | SLI5608-SLI5624 | Yu et al, 1998 |
| 20 | Siderophore | Siderophore | SLI6063-SLI6067 | Bentley et al, 2002 |
| 21 | T1PKS | Prodiginine[a] | SLI6151-SLI6186 | Cerdeno et al, 2001 |
| 22 | Bacteriocin | | SLI6435-SLI6444 | - |
| 23 | Terpene | Geosmin[a] | SLI6466 | Gust et al, 2003 |
| 24 | Siderophore | | SLI6220-SLI6232 | - |
| 25 | t1PKS | Coelimycin P1 | SLI6653-SLI6674 | Gómez-Escribano et al, 2012 |
| 26 | NRPS | Dipeptide | SLI6675-SLI6797 | Bentley et al, 2002 |
| 27 | **none** | SapB lantipeptide[a] | SLI7025-SLI7029 | Kodani et al, 2004 |
| 28 | Terpene | Hopene[a] | SLI7101-SLI7117 | Poralla et al, 2000 |
| 29 | **none** | Germicidin[a] | SLI7437 | Song et al, 2006 |
| 30 | NRPS | Coelibactin | SLI7894-SLI7924 | Bentley et al, 2002 |
| 31 | Terpene | 2-Methylisoborneol[a] | SLI7931-SLI7932 | Wang et al, 2008 |

[a] The product of orthologous biosynthetic gene clusters from *S. coelicolor*, when the NP has been previously elucidated, is reported in this column. Otherwise, the product is a prediction based in the early genome mining reports of Bentley et al (2002) and Nett et al (2009).

The first novel hit is a member of the L/F-tRNA-protein transferase enzyme family (LFT; E.C 2.3.2.6) coded in SLI0884. This enzyme family is involved in the N-rule proteolytic pathway, and has been previously postulated as a potential target for enzyme recruitment from central metabolism to NP biosynthesis (Zhang et al, 2011). The postulation of this enzyme as an interesting catalyst for NPs biosynthesis is based in the recent findings of tRNA dependent peptide synthases, from other enzyme families, within the context of NP biosynthesis. The LFT enzyme family was thus manually included into the EvoMining database of central metabolic expanded enzymes as complement to the enzymes included in the PSCP introduced in previous chapter.

**Table 8-2.** Enzymes involved in NP biosynthesis detected by EvoMining

| Enzyme family | E.C. | Locus | Known recruitments in NP biosynthesis | biosynthetic pathway in *S. lividans* 66 |
|---|---|---|---|---|
| **L/F protein transferase** | 2.3.2.6 | SLI0884 | None | Novel |
| **3-Phosphoshikimate 1-carboxyvinyltransferase** | 2.5.1.19 | SLI1096 | Phosplactomycin, asukamycin | Novel |
| **FemXAB** | | SLI1569 | Valanimycin | Novel |
| **3-deoxy 7 phosphoheptulosonate synthase** | 2.5.1.54 | SLI3566 | | CDA (Hojati et al, 2002) |
| **Anthranilate phosphoribosil transferase** | 2.4.2.18 | SLI3567 | | |
| **Anthranilate synthase** | 4.1.3.27 | SLI3569 | | |
| **Prephenate dehydrogenase** | 1.3.1.12 | SLI3579 | | |
| **Phosphoglycerate dehydrogenase** | 1.1.1.95 | SLI3837 | | Vancomycin resistance; (Hong et al, 2004) |
| **Anthranilate synthase** | 4.1.3.27 | SLI7923 | | Coelibactin (Bentley et al, 2002 ) |

Although antiSMASH predicted an NRPS-based biosynthetic gene cluster in the region were the recruited LFT is located, according to this prediction, the presumed gene cluster includes 30 genes of which 21 are classified as *other genes*, including the LFT itself. Of these genes only 9 are annotated as biosynthetic genes. Thus, it seems that this prediction results from the identification of a single adenylation domain within the locus where SLI0883 is present. Since no other adenylation domains or known biosynthetic systems are present in this region AntiSMASH's prediction is rather ambiguous. The exclusion of the LFT by AntiSMASH and its direct identification as a hit by EvoMining reflects on the fundamental differences between these two approaches.

The second EvoMining novel hit is a member of the 3-Phosphoshikimate 1-carboxyvinyltransferase enzyme family (AroA or EPSP; EC 2.5.1.19) encoded in SLI1096. The AroA family members linked to central metabolism participate in the biosynthesis of aromatic amino acids. Also, two AroA family members have been previously shown to participate in the biosynthesis of Asukamycin (Rui et al, 2010) and phoslactomycin (Chen et al, 2012), and therefore EvoMining identified AroA as an expanded enzyme family that has been previously recruited. The identification of SLI1096 by EvoMining led to the discovery of a second novel biosynthetic pathway in *S. lividans* 66, which is analyzed in detail in the following sections. Again, AntiSMASH did not detect this biosynthetic system.

The third EvoMining novel hit is a member of the FemXAB enzyme family encoded in SLI1569. The members of this enzyme family participate in cell wall biosynthesis (Hegde and Shrader, 2001), and a member of this enzyme family has been previously described to be involved in the biosynthesis of valanimycin, an azoxy group-containing NP. As described in the following section this EvoMining hit lead to the prediction of yet another new biosynthetic pathway, although experimental validation is not reported. As in the previous cases, AntiSMASH did not detect this biosynthetic system.

The six remaining EvoMining hits are highly conserved orthologs of pathways previously elucidated in *S. coelicolor*, and generally implicated in precursor supply for non-ribosomal peptide biosynthesis. Four hits are linked to the non-ribosomal peptide calcium-dependent antibiotic biosynthetic system (Hojati et al, 2002): (i) 3-deoxy 7 phosphoheptulosonate synthase (2.5.1.54; SLI3566); (ii) anthranilate phosphoribosyl transferase (2.4.2.18; SLI3567); (iii) anthranilate synthase (4.1.3.27; SLI3569); and (iv) prephenate dehydrogenase (1.3.1.12; SLI3579). An extra anthranilate synthase enzyme family member (4.1.3.27; SLI7623) was found to be linked to the 'virtual' non-ribosomal peptide coelibactin (Bentley et al, 2002), and the last hit was a phosphoglycerate dehydrogenase family member (1.1.1.95; SLI3837) that is linked to the biosynthetic pathway that provides resistance to glycopeptide antibiotics by modifying the cell wall (Hong et al, 2004). The phylogenetic reconstructions of these enzyme families can be found as supplementary material S11.

It should be noted that all EvoMining predictions were obtained independently of the use of NRPS / PKS sequence motifs, which are safe indicators of NP biosynthetic systems. This observation demonstrates the validity of the approach. When convergence with PKS/NRPS-containing biosynthetic systems occurs, this relates to detection of enzyme recruitments within a biosynthetic gene cluster that may include them or not. In other words, EvoMining is an unbiased method blind to specific chemical families of compounds or biosynthetic systems, but which can indeed detect well-known biosynthetic systems.

The following sections are dedicated to the analysis of the three EvoMining novel hits involved in new biosynthetic pathways. For each one of these cases phylogenomics, biosynthetic predictions, and in two cases, experimental characterization is reported. In one of such cases a novel metabolite was purified and identified, setting the basis for the full elucidation of an unprecedented biosynthetic pathway in *S. lividans* 66.

## 8.3 Phylogenomic analysis of the actinobacterial Leucyl-Phenylalanyl-tRNA-Protein Transferase (LFT) enzyme family

The LFT enzymes catalyze the transfer of leucine or phenylalanine from a charged aminoacyl-tRNA, to an N-terminal basic residue of a protein, usually an arginine or lysine via the N-end rule protein degradation pathway (Leibowitz and Soffer, 1969; Watanabe et al, 2007). Following the prediction by Zhang et al (2011) the LFT family was included in the expansion with recruitments using enzymes from non-actinobacterial organisms such as *E. coli* as queries. Only 13 LFT enzyme family members could be found within the actinobacterial genome database. Among them, only 2 *Streptomyces* strains, *S. ipomoea* (Bioproject PRJNA183480) and *S. lividans* 66, were found. The branches of these enzymes form a single clade with the longest branch in the phylogeny, indicating a different evolutionary rate than the other family members (Figure 8-1).

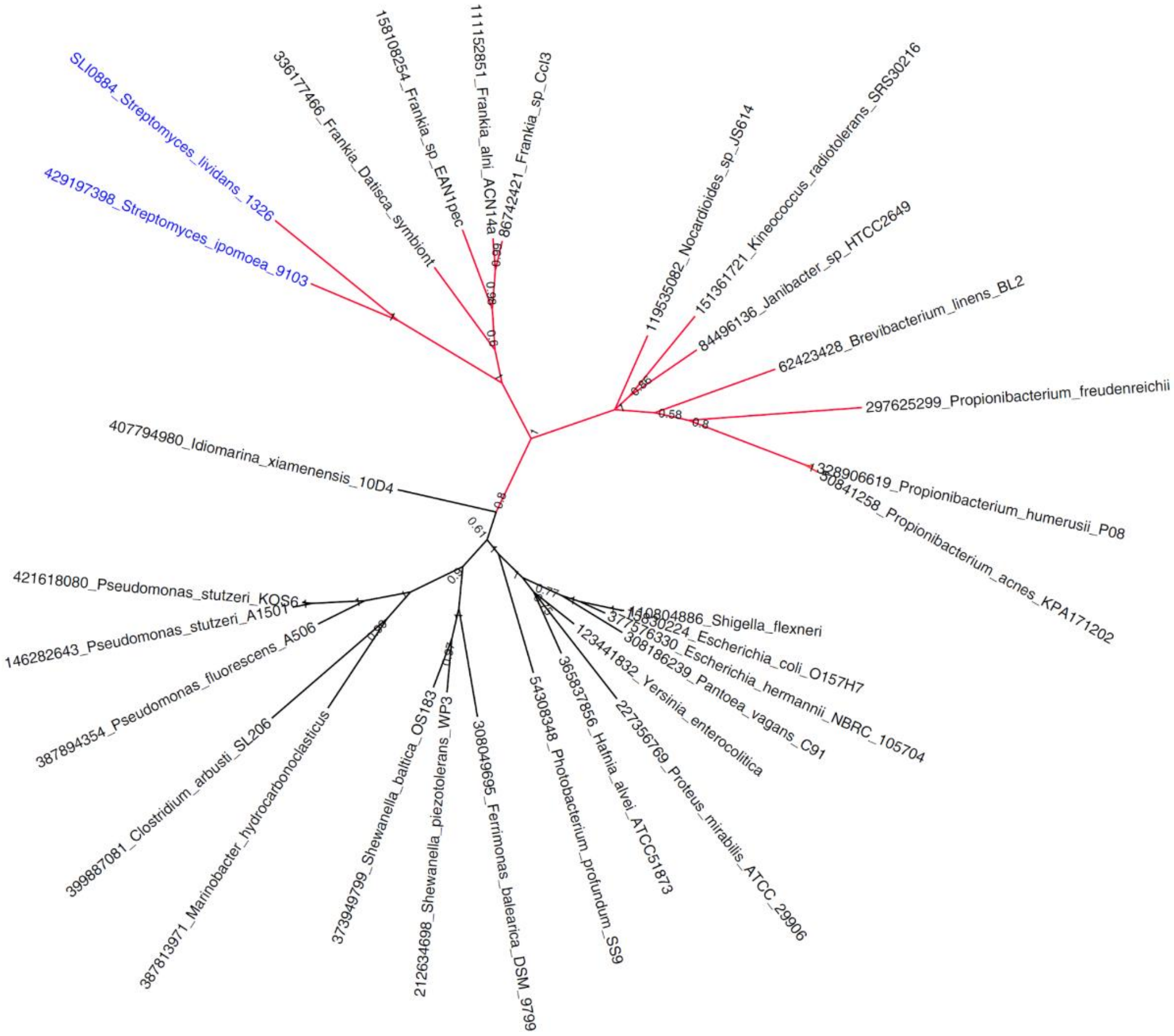

**Figure 8-1. Phylogenetic reconstruction of the LFT enzyme family. The phylogenetic tree was constructed using MrBayes, posterior probabilities are shown at the nodes.** The actinobacterial members of the LFT enzyme family are shown in red branches; the *Streptomyces* LFTs are shown in blue. The accession number of the enzyme family members are indicated before the name of each organism.

The *S. lividans* LFT (SLI0884) is located within SLP3, a mobile genetic element inserted in the chromosome (Cruz-Morales et al, 2013), and shares 35% amino acid sequence identity with the enzyme Aat (locus tag ECK0876) from *Escherichia coli*. SLI0884 is located downstream of an unusual NRPS-like protein containing a single adenylation domain (A). The vicinity with an NRPS coding gene is conserved in *S. ipomoea* where its LFT occurs (GI ZP_19189297). The closest homologues outside *Streptomyces* was found in *Frankia* species, although their gene context is not related to any NRPS or NP biosynthetic enzyme, and thus a proteolytic role could well be assumed. A similar scenario was encountered outside the *Actinobacteria*.

The unusual *S. lividans* 66 NRPS, SLI0883, encodes a single adenylation (A) domain predicted to recognize arginine; a Phosphopantetheinyl Carrier Protein (PCP); and a Reductase (R) domain. However, no condensation (C) domain, Thioesterase (TE) domain, or additional adenylation domain, could be identified in or closely associated with this unusual biosynthetic gene cluster (SLI0883-SLI0892). The six predicted additional biosynthetic genes, all transcribed in the same direction and potentially transcriptionally coupled, encode putative tailoring enzymes (Figure 8-2 and Table 8-3).

Given that SLI0883 and SLI0884 are potentially transcriptionally coupled, and that a single A domain linked to PCP and R domains seem unlikely to be able to produce a peptide bond, it was proposed that the LFT homologue present in SLP3 accounts for the lack of both A and C domains. Indeed, recent mechanistic data for the *E. coli* enzyme homologue suggest an analogous condensation mechanism as found in ribosomes and C domains (Fung et al, 2011).

**Table 8-3.** Annotation of the NRPS-LFT biosynthetic gene cluster of *S. lividans* 66

| Gene | Predicted Function |
|---|---|
| SLI0883 | Non-ribosomal peptide synthetase |
| SLI0884 | Leucyl/Phenylalanyl-tRNA-protein transferase |
| SLI0885 | N-Acetyltransferase |
| SLI0886 | O-methyltransferase, family 2 |
| SLI0887 | Butyryl-CoA dehydrogenase |
| SLI0888 | TenA family transcriptional regulator |
| SLI0889 | Major facilitator superfamily MFS1 |
| SLI0890 | Putative cytochrome P450 monooxygenase |
| SLI0891 | Pyridoxamine 5'-phosphate oxidase-related, FMN-binding |
| SLI0892 | Unknown |
| SLI0893 | CsoR-like DUF156 |
| SLI0894 | Heavy metal-associated domain-containing secreted protein |

## 8.3.1 Prediction of a unprecedented NRPS-LFT hybrid peptide biosynthetic system

Based in the nature of the biosynthetic enzymes included in the gene cluster, the following biosynthetic logic was predicted:

It is proposed that the enzyme product of SLI0884 would attach a Leu or Phe residue, provided by the cognate aminoacyl-tRNA, to an Arg residue, while the latter is bound to the PCP of SLI0883. The emerging peptide will then be released by the action of the reductase (R domain) upon the thioester group, as previously found in myxochelin biosynthesis (Li et al, 2008). A reductive cleavage of the predicted metabolite would presumably lead to an aldehyde that might then be further reduced to an alcohol group rather than yielding the compound containing a free carboxylic acid (Figure 8-2).

The proposed biosynthetic logic, as depicted in Figure 8-2, is compatible with the generation of a peptide molecule with free pairs of electrons and thus with the ability to chelate metals. This hypothesis is actually also supported from the phenotyping, functional genomic and transcriptional analyses of the *S. lividans* SLP3 plasmid described in chapter II, which linked this mobile genetic element with metal metabolism. Moreover, this proposal complements the growing universe of aminoacyl-tRNA transferases involved in amide bond formation within peptide biosynthesis (Garg et al, 2008; Gondry et al, 2009; Zhang et al, 2011). Given the peptidic nature of the metabolite potentially produced by this biosynthetic system, from now onwards, this compound will be referred to as **Livipeptin**. Thus, in order to demonstrate the existence of livipeptin, including its possible biological role and chemical nature, experimental characterization was conducted.

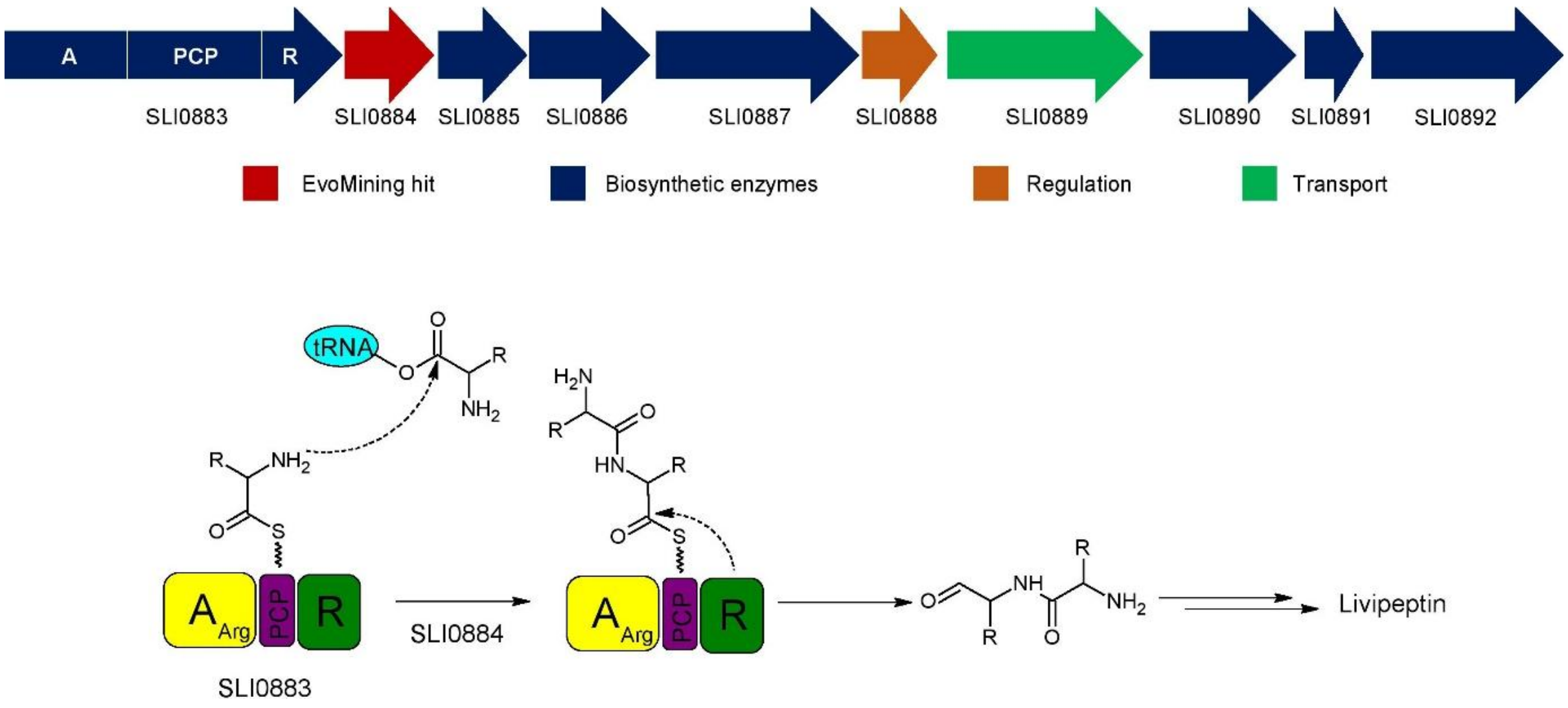

**Figure 8-2. Livipeptin biosynthesis.** Proposed mechanism of peptide bond formation by the NRPS-tRNA biosynthetic hybrid system.

## 8.3.2 Experimental characterization of livipeptin pathway

A knockout mutant of the SLI0883-0884-0885 gene cluster was constructed using in-frame gene replacement. An apramycin resistance cassette was used as the selection marker, and the cassette was cloned into the delivery plasmid pWHM3. After integration of the cassette by double crossover the plasmid was curated after consecutive rounds of propagation, as previously reported (van Wezel et al, 2005). Integrity of the mutant SLI883-5::*acc (3)IV* was confirmed by PCR and by checking for the Apramycin$^R$ and Thiostrepton$^S$ phenotypes.

Previous work in *S. lividans 66*, presented in chapter one, demonstrates a functional relationship between SLP3 plasmid and metal homeostasis after trancriptomic analysis. It was also shown that the livipeptin biosynthetic gene cluster is silent after addition of copper induction and in standard growth conditions. Therefore, to identify the conditions were livipeptin biosynthesis would be expressed a screening in different metal-stress conditions was performed.

The confirmed mutants were grown on metal-depleted solid media, and on solid media supplemented with different metals, transition metals and metalloids, namely, Na, Mg, K, Ca, Mn,

Fe, Co, Ni, Cu Zn, and As. The metals were supplemented in concentrations ranging from 10 µM to 300 mM. The plates were inoculated with fresh spores of wild type *Streptomyces lividans* 66 and the SLI883-884::*acc (3)IV* mutant, using droplets containing 10E1 to 10E6 spores. Unexpectedly, after 72 hours of incubation, *S. lividans* 66 was unable to grow in the presence of up to 200 mM of KCl and $MgCl_2$, while the mutant SLI883-884::*acc (3)IV* was able to grow under the same conditions. This suppression phenotype was confirmed when Minimal, R5, ISP2, SFM and LB solid media was used (Figure 8-3).

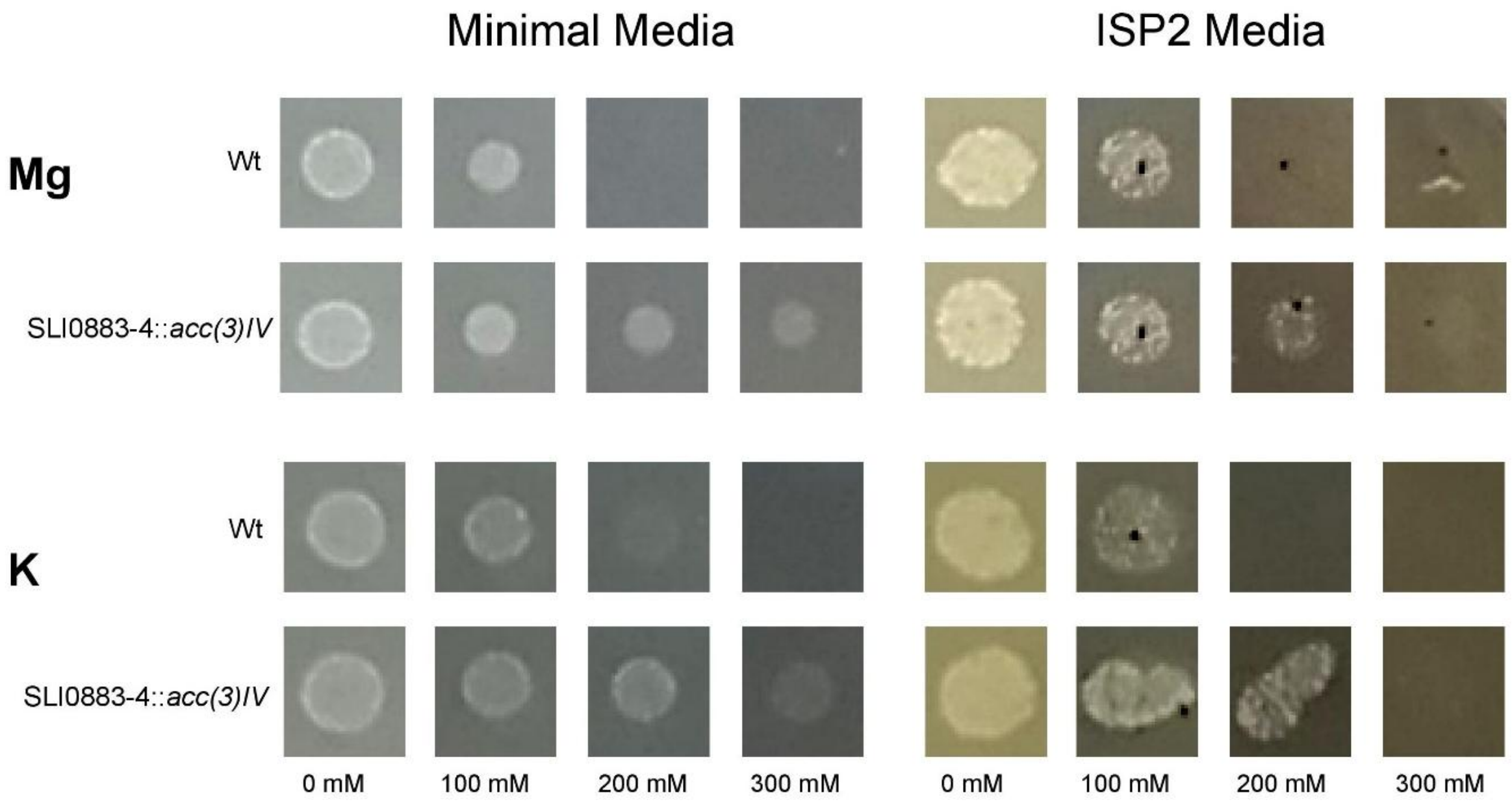

**Figure 8-3.** *S. lividans* 66 response to Mg and K. Wild-type and SLI883-5 minus mutant were plated on minimal and ISP2 solid media supplemented with different concentrations of $MgCl_2$ and KCl. 200 mM of both metals have a highly toxic effect on the WT strains, while the mutant strain is able to grow in up to 300 mM.

Based on the previous results, the expression of SLI0883 and SLI0884 in *S. lividans* 66 was tested by RT-PCR after addition of $Mg^{++}$ and $K^+$ into liquid cultures. As shown in Figure 8-4, expression of these genes was induced in the presence of 300 mM KCL after four hours. This result suggest that the peptide biosynthetic system is turned on in the presence of this metal ions. Whether induction of expression by $K^+$ is directly connected to the biological role of livipeptin, or is a consequence of a pleiotropic and more general mechanism, remains unknown. Nevertheless, the culture of *S. lividans* 66 under these conditions was exploited for the identification of this compound by means of using comparative analysis of the metabolite profiles between Wt and mutant strains. Overall, these experiments, which will be reported elsewhere once a patent protecting the industrial property of this discovery is submitted (winter of 2013), lead to the identification and partial structural elucidation of livipeptin.

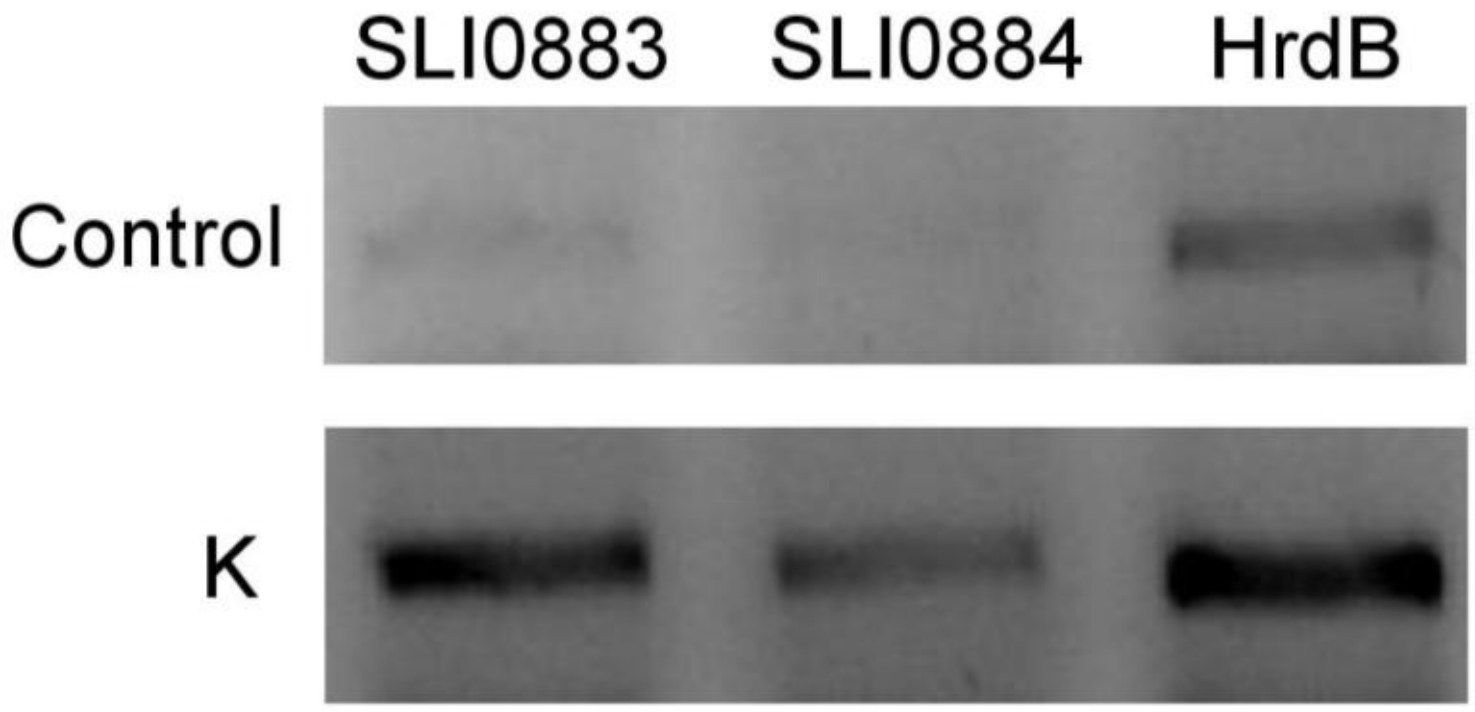

**Figure 8-4.** *S. lividans* 66 transcriptional response to K. *S. lividans* 66 was inoculated in liquid minimal media, and after 66 hours of incubation KCl was added to the cultures. Total RNA was extracted after four hours of the addition of the metals and used for RT-PCR analysis. HrdB is a conserved Streptomyces housekeeping sigma factor used as control for the experiment

## 8.4 Phylogenomic analysis of the actinobacterial AroA enzyme family

AroA catalyzes the transfer of a vinyl group from phosphoenolpyruvate (PEP) to 3-phosphoshikimate to form 5-enolpyruvylshikimate-3-phosphate, releasing a phosphate molecule (McMurry and Begley, 2005; Figure 8-5). AroA participates in the shikimate pathway that is a common pathway for the biosynthesis of aromatic amino acids and other metabolites.

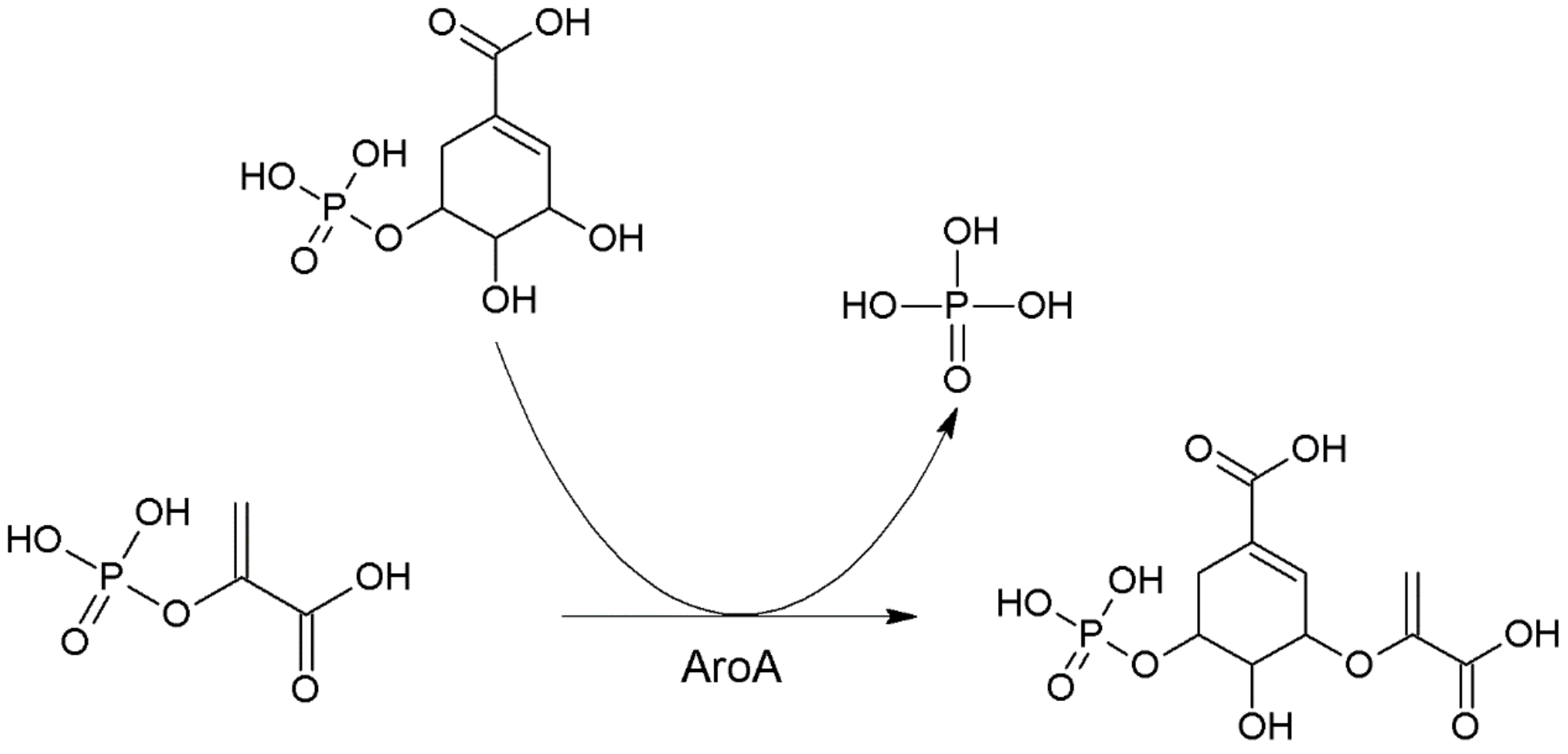

**Figure 8-5**. The reaction catalyzed by the 3-Phosphoshikimate 1-carboxyvinyltransferase in central metabolism (AroA, E.C. 2.5.1.19).

AroA was found to be expanded in 45% of the *Streptomyces* genomes analyzed, which contrasts with 24% of all *Actinobacteria* genomes included in the database. This expansion is limited to only two homologues without exception, SLI5501 and SLI1096, in the genome of *S. lividans* 66. According to the phylogenenomic analysis of the AroA enzyme family (Fig. 6) the homologue encoded by SLI5501 could be linked to central metabolism. This homologues forms a major clade together with central metabolic enzymes of all other strains included in the database that only encode for one AroA enzyme. Further support of this role is indicated by the topology of the clade, which is highly similar to that of the species-tree, suggesting functional conservation. The little divergence among the homologues in this clade can be attributed to the evolutionary rate of speciation (see Fig S8).

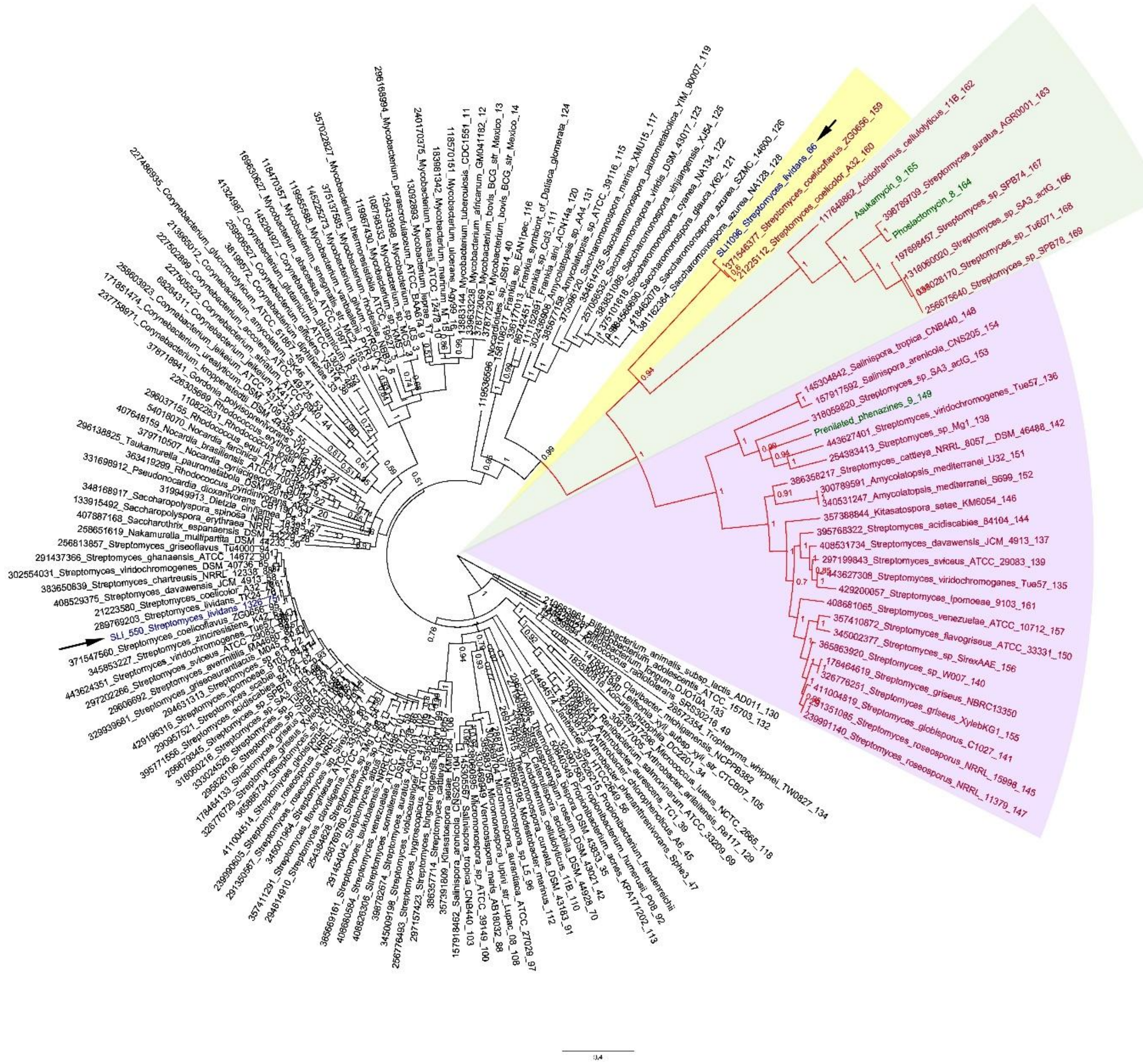

Figure 8-6**.** Phylogenetic reconstruction of the AroA enzyme family in Actinobacteria. The clades including expanded homologues are highlighted; sub-clade 1 is shown in yellow, sub-clade 2 in green and sub-clade 3 in pink. The known recruited homologues are indicated with green characters, and arrows indicate the location of the homologues found in S. lividans 66. The accession number of each sequence is indicated before the species names.

The scenario described above contrasts with the recruited AroA enzymes, including SLI1096. These AroA enzymes group together in a single clade that includes 35 family members encompassing the genera *Salinispora* (2), *Amycolatopsis* (2), *Kitasatospora* (1) and *Acidothermus* (1). The remaining 29 homologues, accounting for 85% of the enzyme expansions, come from the genus *Streptomyces*. The clade including these recruited AroA enzymes could be divided into three sub-clades. The first sub-clade includes the expanded AroA enzymes from *S. lividans*, *S. coelicolor* and *S. coelicoflavus* (Guo et al, 2012; GenBank Accession: AHGS00000000). These three species are closely related and belong to the *S. violaceoruber* sub-clade. The AroA enzymes from these organisms, specifically SLI1096 from *S. lividans*, were selected for further analysis, as described further below.

The second sub-clade includes two family members related to biosynthetic pathways of two polyketide compounds, asukamycin (Rui et al, 2010) and phoslactomycin (Chen et al, 2012). Interestingly, both biosynthetic pathways involve cyclohexenyl-CoA starter units in their assembly lines, opening the possibility that the AroA homologues are involved in precursor supply biosynthetic steps. The third sub-clade includes biosynthetic AroA homologues involved in the biosynthesis of the aromatic phenazine (Seeger et al, 2011).

Several family members from the first and second sub-clades could be linked to cryptic NP biosynthetic gene clusters using the criteria described in chapter II. A few examples of these include GI: 197698457 in *Streptomyces* sp. SPB74, which is predicted to participate in a polyketide biosynthetic pathway; GI: 318059820 in *Streptomyces* sp SA3_actG and GI: 254383413 in *Streptomyces* sp Mg1 predicted to participate in phenazine biosynthetic pathways; and GI: 117648862 in *Acidothermus cellulolyticus* and GI: 443627401 in *Streptomyces viridochromogenes*, which participate in the biosynthesis of metabolites of unknown classes (also see table 7-2).

SLI1096 is located within a large genome island in *S. lividans* 66 that has been called SliGI-1 (Cruz-Morales et al, 2013; Chapter I) and several Kbp at the 3' of a PKS system (SLI1088 and SLI1089). This PKS system, also present in *S. coelicolor* and *S. coelicoflavus,* has been previously annotated as an orphan PKS that seems to act independently of a biosynthetic context provided by a gene cluster (Nett et al, 2009; Bentley et al, 2002). To further establish the potential boundaries of a putative gene cluster, the gene context of the expanded AroA homologues of *S. coelicoflavus* and *S. coelicolor* were analyzed and compared with that of *S. lividans* 66. Interestingly, as discussed in Chapter 1, the gene neighborhood in these three species is highly conserved. The syntenic region of these three strains spans from SLI1077 till SLI1103. This observation strongly suggests that the PKS, the AroA enzyme, and other biosynthetic genes in the neighborhood of this locus are functionally linked and they form a biosynthetic gene cluster (Table 8-4; Figure 8-7).

### 8.4.1 Biosynthetic pathway prediction for SLI1077-1103

The manual annotation of the gene cluster in *S. lividans* 66 (Table 8-4) revealed the presence of a 2,3-bisphosphoglycerate-independent phosphoenolpyruvate mutase (SLI1097 PPM), downstream and possibly transcriptionally coupled to the *aroA* homologue. These observations suggest a functional link between the enzymatic products of these genes. Interestingly, a phosphonopyruvate decarboxylase coding gene (PPD; SLI1091) is located within the gene cluster. The combination of mutase-decarboxylase enzymes is a conserved biosynthetic feature of NPs containing C-P bonds (Metcalf and van der Donk, 2009), opening the possibility for a phosphonate or a similar molecule as the product of this biosynthetic gene cluster.

Regarding non-enzymatic functions encoded within the gene cluster, a set of ABC transporters coding genes, originally annotated as phosphonate transporters, could be found (SLI1100 and SLI1101). Four Arsenic tolerance-related genes are also located towards the 5' end of the gene cluster; these genes are paralogous of the main arsenic tolerance system encoded in the *ars* operon, located at the core of the *S. lividans* chromosome (SLI1077-80; Wang et al, 2006). The gene cluster also codes for regulatory proteins, mainly arsenic responsive repressor proteins (SLI1078, SLI1092, SLI1102 and SLI1103). This, in addition to the presence of an arsenic resistance operon (SLI1077-1080), suggests a link between arsenic and a putative biological function of this gene cluster. To conciliate the presence of phosphonate biosynthetic-like genes and transporters, polyketide biosynthetic enzymes and arsenic resistance genes in a single biosynthetic system, the possibility of a biosynthetic pathway for a polyketide compound, which would incorporate arsenic atoms into its structure, using a biosynthetic strategy similar to the one used for phosphonate biosynthesis, was postulated together with Luis Yanez-Guerra (Master's thesis in preparation).

First, arsenate and phosphate are highly similar; their chemical and thermodynamic properties are almost identical. These properties cause phosphate-utilizing enzymes to have almost similar affinities and kinetic parameters for phosphate and arsenate, with the biologically relevant difference that arsenate derived products are more labile. Indeed, as exemplified in the two next points, arsenic compounds are commonly used as analogs of native substrates in kinetic and mechanistic studies of phosphate enzymes (Elias et al, 2012; Tawfik and Viola, 2011).

**Table 8-4.** Functional Annotation of the biosynthetic gene cluster in *S. lividans* 66.

| Locus | Proposed function |
|---|---|
| SLI1077 | ArsB heavy metal resistance transport membrane protein |
| SLI1078 | ArsR-family transcriptional regulator |
| SLI1079 | ArsC Heavy metal reductase |
| SLI1080 | ArsT Thioredoxin reductase |
| SLI1081 | B12 binding domain of Methylmalonyl-CoA mutase |
| SLI1082 | Methylmalonyl-CoA mutase |
| SLI1083 | Uknown function |
| SLI1084 | Arsenical resistance operon repressor |
| SLI1085 | Putative oxidoreductase |
| SLI1086 | FAD dependent oxidoreductase |
| SLI1087 | Uknown function |
| SLI1088 | Modular polyketide synthase |
| SLI1089 | 3-oxoacyl-ACP synthase III |
| SLI1090 | DedA protein |
| SLI1091 | Arsonopyruvate decarboxylase |
| SLI1092 | ArsR-family transcriptional regulator |
| SLI1093 | Major facilitator superfamily permease |
| SLI1094 | CTP synthase-like amino transferase |
| SLI1095 | Anaerobic dehydrogenases, typically selenocysteine-containing |
| SLI1096 | 5-Enolpyruvylshikimate-3-phosphate synthase |
| SLI1097 | 2,3-bisphosphoglycerate-independent phosphoenolpyruvate mutase |
| SLI1098 | Uknown function |
| SLI1099 | Periplasmic phosphate-binding protein |
| SLI1100 | ABC transporter permease |
| SLI1101 | ABC transporter ATP-binding protein |
| SLI1102 | ArsR Transcriptional regulator |
| SLI1103 | ArsR Transcriptional regulator |

Second, previous studies have demonstrated that AroA is able to catalyze a reaction in the opposite direction of the biosynthesis of aromatic amino acids, meaning the formation of PEP and 3-Phosphoshikimate from enolpyruvil shikimate 3-phosphate and phosphate (Zhang and Berti, 2006). However, since phosphate is an intrinsically non-reactive substrate, the demonstration of the backwards reaction catalyzed by AroA required the use of phosphate analogues including arsenate. Zhang and Berti (2006) demonstrated that arsenate and enolpyruvil shikimate 3-phosphate could react to produce arsenoenolpyruvate (AEP), a labile analog of PEP, which is spontaneously broken down into pyruvate and arsenate. This key insight was highly important for the integration of the enzymatic functions within the biosynthetic gene cluster is the contribution of Luis Yanez-Guerra (Master thesis in preparation).

Third, it has been demonstrated that the phosphoenolpyruvate mutase PPM, an enzyme responsible for the isomerization of PEP to produce phosphonopyruvate, a key step in phosphonate biosynthesis (Metcalf and Van der Donk, 2009) is capable of recognizing AEP as a substrate. Although at low catalytic efficiency, the formation of 3-arsonopyruvate by this enzyme, a product analog of the phosphonopyruvate intermediary in phosphonate NPs biosynthesis, has been indeed demonstrated (Chawla et al, 1995).

The previous evidence is used herein to postulate a novel biosynthetic pathway including the enzymatic potential encoded within SLI1077-SLI1103 (Figure 8-7). In the postulated biosynthetic pathway, SLI1096, the AroA homologue detected by EvoMining is responsible for the formation of AEP from arsenate and enolpyruvil shikimate 3-phosphate; in the following biosynthetic step, AEP, an ester of arsenate, is isomerized by the action of SLI1097, a PPM homologue, producing arsonopyruvate, a C-As containing molecule, which is decarboxylated by SLI1091, a phosphonopyruvate decarboxylase homologue. The activation energy of this reaction could be similar to that of C-P bond formation by mutase-decarboxylase coupled enzymes, favoring the formation of arsonoacetaldehyde.

Two possibilities rise from this point onwards; the first one is the formation of 2-Aminoethylarsonic acid (AEA), by the amination of the arsenic-containing metabolite by SLI1094, an aminotransferase. AEA is an analog of the most abundant phosphonate in nature, 2-aminoethylphosphonic acid, which is present in the membranes of eukaryotic organisms (McGrath et al, 2013). Interestingly, it has also been demonstrated that AEA can be metabolized by *Pseudomonas aeroginosa* by the promiscuous action of an aminotransferase that recognizes with high affinity the arsenic-containing analog of its native substrate (Lacoste et al, 1992).

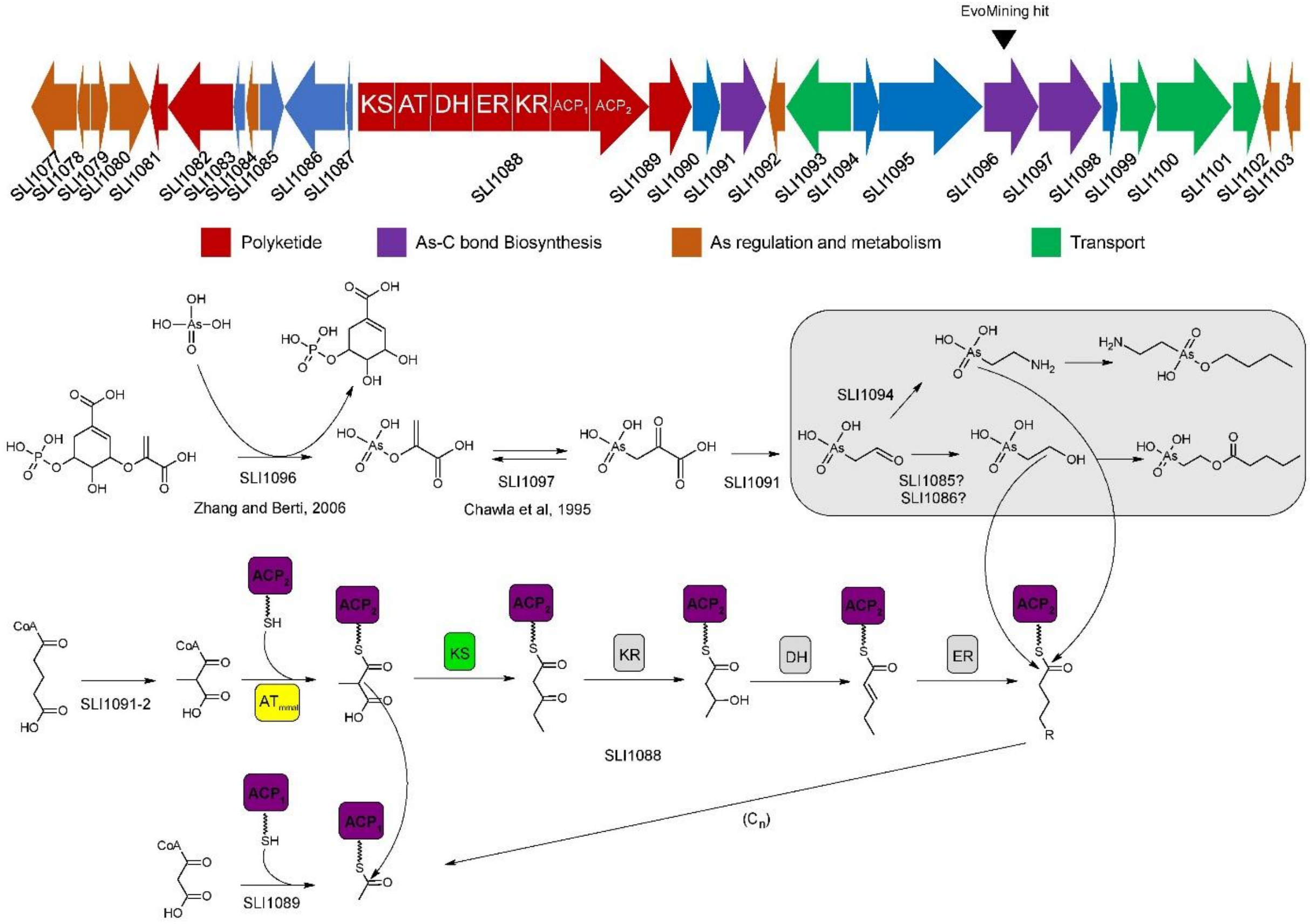

**Figure 8-7. Arsonolividin gene cluster organization and biosynthetic pathway predictions**. The figure depicts the proposed mechanism of As incorporation via the coupled reactions of SLI1096-97 (AroA-PPM), as well as the assembly of a polyketide that will be elongated iteratively and reduced until a fatty acid-like hydrophobic chain is assembled. This moiety is proposed to be condensed with the As containing precursor molecule in a reaction that should release the polyketide chain from the PKS complex. A possible condensation mechanism for the synthesis of a polyketide-As-C intermediary is shown within a grey box.

A second possibility for arsonoacetaldehyde is its condensation with a polyketide molecule produced by SLI1068-69, a type I iterative polyketide synthase. To infer a possible product of the PKS system within the gene cluster, a domain analysis was performed using the PKS-NRPS web tool (Bachmann and Ravel, 2009 http://nrps.igs.umaryland.edu/nrps/). This analysis revealed the presence of 7 domains (figure 8-7) encoded within SLI1088, including a single domain for acyl transfer with specificity for Methyl-malonyl-CoA (AT), a condensation (KS) domain, as well as modification domains including single dehydratase (DH), enoyl reductase (ER) and ketoreductase (KR) domains. Moreover, at least two acyl carrier protein (ACP) domains could be identified. According to the assembly line derived from the combination of these domains, a single condensation reaction would lead to a small (7 Carbon skeleton) saturated carbon chain, which may resemble the structure of a fatty acid hydrophobic tail.

The condensation reaction between arsonoacetaldehyde and the PKS product may occur upon the action of an oxidoreductase (SLI1085, SLI1086 or SLI1095), which can release the polyketide molecule from the ACP while forming a putative molecule that may resemble the structural characteristics and properties of a fatty acid, however, a possible mechanism for the condensation of an arsonate intermediary and the polyketide chain is yet to be elucidated. The proposed biosynthetic assembly line and the predicted product synthesized by this PKS enzymatic complex are depicted in Figure 8-7.

An alternative possibility is that the biosynthetic gene cluster is dedicated to the formation of a phosphonate compound, associated or not with the PKS complex. This possibility could be supported by the fact that AroA is a target of a synthetic phosphonate, known as glyphosate, which is widely used as an herbicide. It has been reported that the resistance determinant of this compound can be a glyphosate insensitive version of AroA (Comai et al, 1983). However, this possibility does not account for the high number of functions related to arsenic encoded within this gene cluster. Recent experiments performed by Luis Yanez-Guerra support the hypothesis of an arsenic-containing compound as the product of this pathway, therefore this possibility has been ruled out. From this point onwards the putative product of the pathway will be called **Arsonolividin** (From **Arson**ate and ***livid****ans*).

After the identification of the arsonolividin biosynthetic gene cluster in *S. lividans* 66, *S. coelicolor*, and *S. colicoflavus* using the EvoMining approach, similar biosynthetic systems could be identified by conventional methods, i.e. mining for similar enzymes using blast and the GenBank NR database. After the analysis, 5 new biosynthetic gene clusters genomes were identified. From a total of eight arsonolividin-like clusters (Figure 8-8) those identified in *S. lividans* 66, *S. coelicolor*, *S. coelicoflavus*, *Streptomyces* sp Bole5 (GenBank accession: AREI00000000), *Streptomyces* sp CNS335 (GenBank accession: ARHS00000000), *Streptomyces sp* CNY243 (ARHU00000000), and *Saccharomonospora saliphila* (GenBank accession: AICY00000000), are highly syntenic. A divergent biosynthetic gene cluster, present in *S. fulvissiumus* (GenBank accession: NC_021177) and including homologues of AroA, PPM, and arsenic resistance genes, but instead of a PKS includes a PKS-NRPS hybrid system, could also be found.

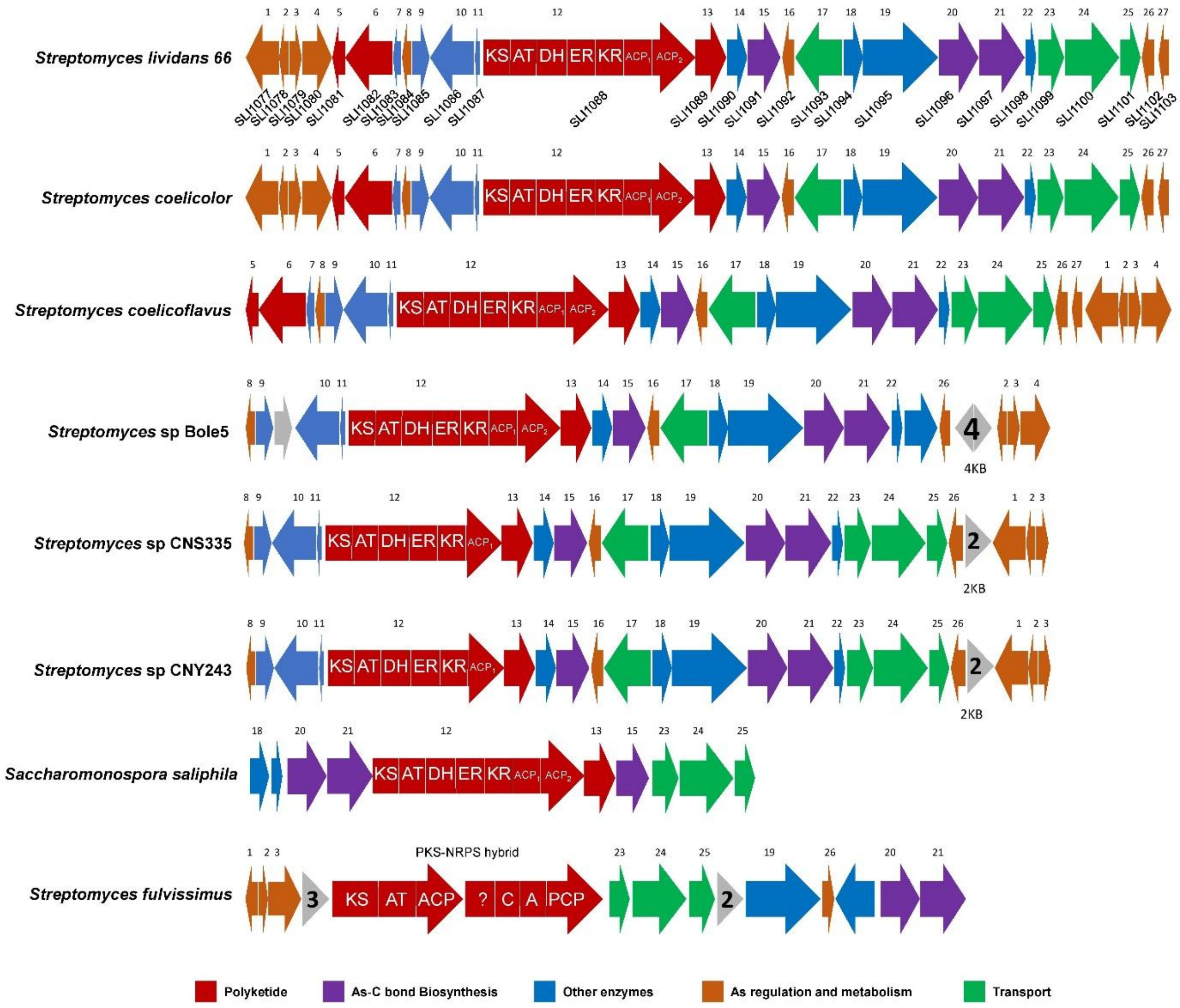

**Figure 8-8. Arsonolividin-like biosynthetic gene clusters found in *Actinobacteria* species.** Conserved features are shown in different colors. The numbers in the top of each gene indicate the equivalence to orthologs in the *S. lividans* 66 arsonolividin biosynthetic gene cluster. Triangles with numbers indicate the presence of a certain number of genes with no orthologs in the *S. lividans* 66 arsonolividin and the direction in which they are transcribed.

## 8.5 Phylogenetic analysis of the actinobacterial FemXAB enzyme family

The FemXAB enzyme family members are tRNA-dependent amino acid transferases responsible for the formation of interchain peptides during peptidoglycan biosynthesis (Hedge and Shrader, 2000). The phylogenetic reconstruction of the FemXAB family members revealed a sub-clade with enzyme recruitments (Figure 8-11) that includes VlmA from the valanimycin biosynthetic pathway (Garg et al, 2008), as well as thirteen other homologues from the genera *Streptomyces* (7), *Kitasatospora* (2), *Frankia* (2), *Nakamurella* (1), and *Mycobacterium* (1). SLI1569 from *S. lividans* 66, originally annotated as a conserved hypothetical protein, was found within this sub-clade (Figure 8-11).

The FemXAB enzyme family belongs to a superfamily that includes several enzymes of unknown function called *Domain of unknown function* 482 (DUF482). Using Gene Ontology terms, the DUF482 superfamily has been linked to nucleotide binding, acyl transfer, cell wall organization and peptidoglycan biosynthesis (Gooneseker et al, 2010). Interestingly, this superfamily also includes the L/F transferase enzyme family (Watanabe et al, 2007).

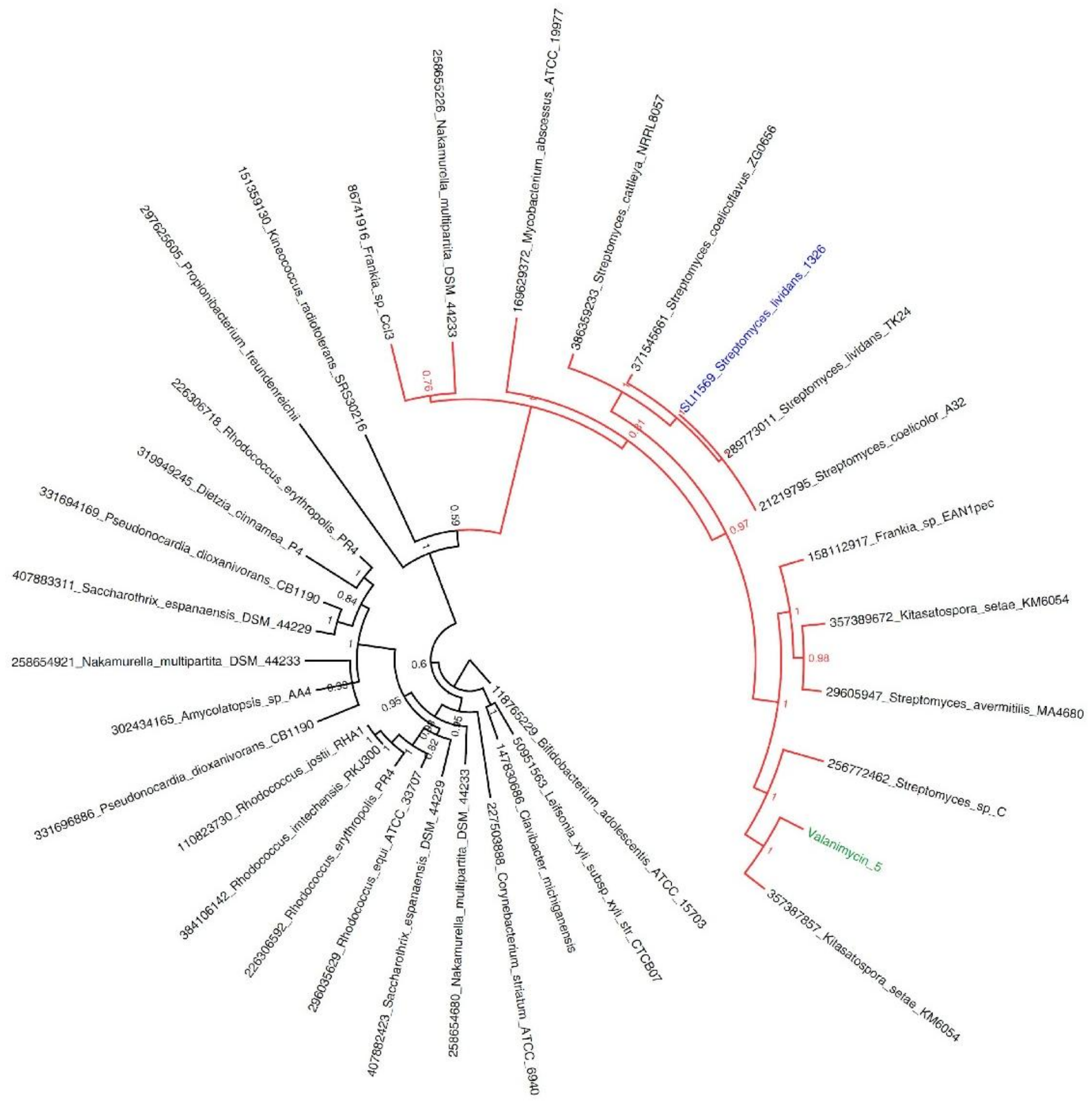

**Figure 8-9. Phylogenetic reconstruction of the actinobacterial FemXAB enzyme family**. The phylogenetic tree was constructed using MrBayes, the posterior probabilities are shown at each node. The expanded clade is shown in red; the valanimycin FemXAB family member VlmA is shown in green and the homologue found in *S. lividans* 66 is highlighted in blue.

## 8.5.1 Genome context analysis of SLI1569 functional annotation

Following the EvoMining approach, the genome context of SLI1569, which shares 88% amino acid sequence identity with VlmA, was analyzed in order to determine its involvement in NP biosynthesis. The region where SLI1569 is encoded included ten genes (Figure 8-12). Among these, the coding sequence automatically annotated with the locus tag SLI1566 was excluded since the sequence was determined to be a miss-prediction after manual curation. The remaining nine genes are potentially transcribed in the same direction, seven downstream of SLI1569 (SLI560-SLI1568), and one upstream (SLI1560), just next to a regulatory gene in the opposite direction (SLI1571). This operon-like structure, including overlapping start and stop codons and a regulatory element in the opposite direction, strongly suggests a functional association of the enzymes encoded by SLI1560-1571 (Table 8-5).

**Table 8-5.** Functional annotation of the *femXAB* biosynthetic gene cluster in *S. lividans* 66

| Locus tag | Predicted Function |
|---|---|
| **SLI1560** | Methyltransferase |
| **SLI1561** | Diacylglycerol O-acyl transferase |
| **SLI1562** | FAD dependent oxidoreductase |
| **SLI1563** | Oxidoreducatse |
| **SLI1564** | Cyclase dehydratase |
| **SLI1565** | Ornithine aminotransferase |
| **SLI1567** | Ketoreductase |
| **SLI1568** | amine N-hydroxylase |
| **SLI1569** | *O*-seryl-isobutylhydroxylamine synthase, DUF482 |
| **SLI1570** | Integral membrane protein |
| **SLI1571** | Transcriptional regulator |
| **SLI1560** | Methyltransferase |
| **SLI1561** | Diacylglycerol O-acyl transferase |
| **SLI1562** | FAD dependent oxidoreductase |

SLI1568, which was automatically annotated as an oxidoreductase, shares 45% amino acid sequence identity with VlmH, an isobutylamine N-Hydroxylase also involved in the biosynthesis of Valanimycin. VlmH hydroxylates the amine group of isobutylamide (i.e. a decarboxylated valine residue) to produce hydroxybutylamide. This compound is condensed with a serine residue via an ester bond at the hydroxylamine group generated by VlmH (Garg et al, 2008). Despite the presence of VlmA and VlmH homologues, other enzymatic functions present in the valanimycin biosynthetic pathway could not be found, indicating common biosynthetic steps but a different final product. Other biosynthetic functions could be annotated for the remaining genes in the neighborhood, including a methyltransferae (SLI1560), a Diacylglycerol *O*-acyl transferase (SLI1561), as well as other three reductases (SLI1562-3 and SLI1567).

Moreover, two kbp downstream of SLI1569, a homologue of the ornithine aminotransferase enzyme (OAT; E.C. 2.6.1.13) is encoded by SLI1565. OAT reversibly catalyzes the amination of glutamate 5-semialdehyde to produce ornithine. For this reaction, glutamate acts as the amine donor. The glutamate-5-semialdehyde is produced spontaneously from pyrroline-5-carboxylate (P5C). P5C can in turn be produced by the oxidation of proline by the pyrroline-5 carboxylate reductase. All these reactions are reversible (McMurry and Begley, 2005). SLI1570 encodes the only transport protein within the region. Given the gene organization and the combination of functions, it was predicted that the region spanning from SLI1560 till SLI1571 is a biosynthetic gene cluster directing the synthesis of a NP with an *O*-seryl-hydroxylamine ester intermediary.

### 8.5.2 Biosynthetic pathway prediction for SLI1560-SLI1571

On the basis of the presence of VlmH-VlmA homologues, i.e. SLI1568-69, within this predicted biosynthetic gene cluster, the following biosynthetic logic is postulated. First, a seryl group could be transferred from the aminoacyl-tRNA cognate to a hydroxyl-amine containing molecule by SLI1569. Second, SLI1568 could serve as the enzyme responsible for the hydroxylation of a primary amide-containing molecule. In valanimycin biosynthesis this substrate is obtained after the decarboxylation of valine. However, no decarboxylase coding genes could be detected amongst SLI560-SLI1568. This opens the possibility for a different molecule to be hydroxylated by SLI1568, which could act as the acceptor of the seryl group. Third, based in the gene content of the biosynthetic cluster, the more appealing possibility for the second substrate of SLI1569 is hydroxyl-ornithine. OAT (SLI1565) is predicted to incorporate glutamate-5-semialdehyde from central metabolism to produce ornithine, which then could be hydroxylated by SLI1568, producing hydroxyl-ornithine. This molecule may act as the acceptor of the seryl group transferred by SLI1569, forming the ester O-seryl-ornithyl-hydroxylamine in a tRNA-dependent manner.

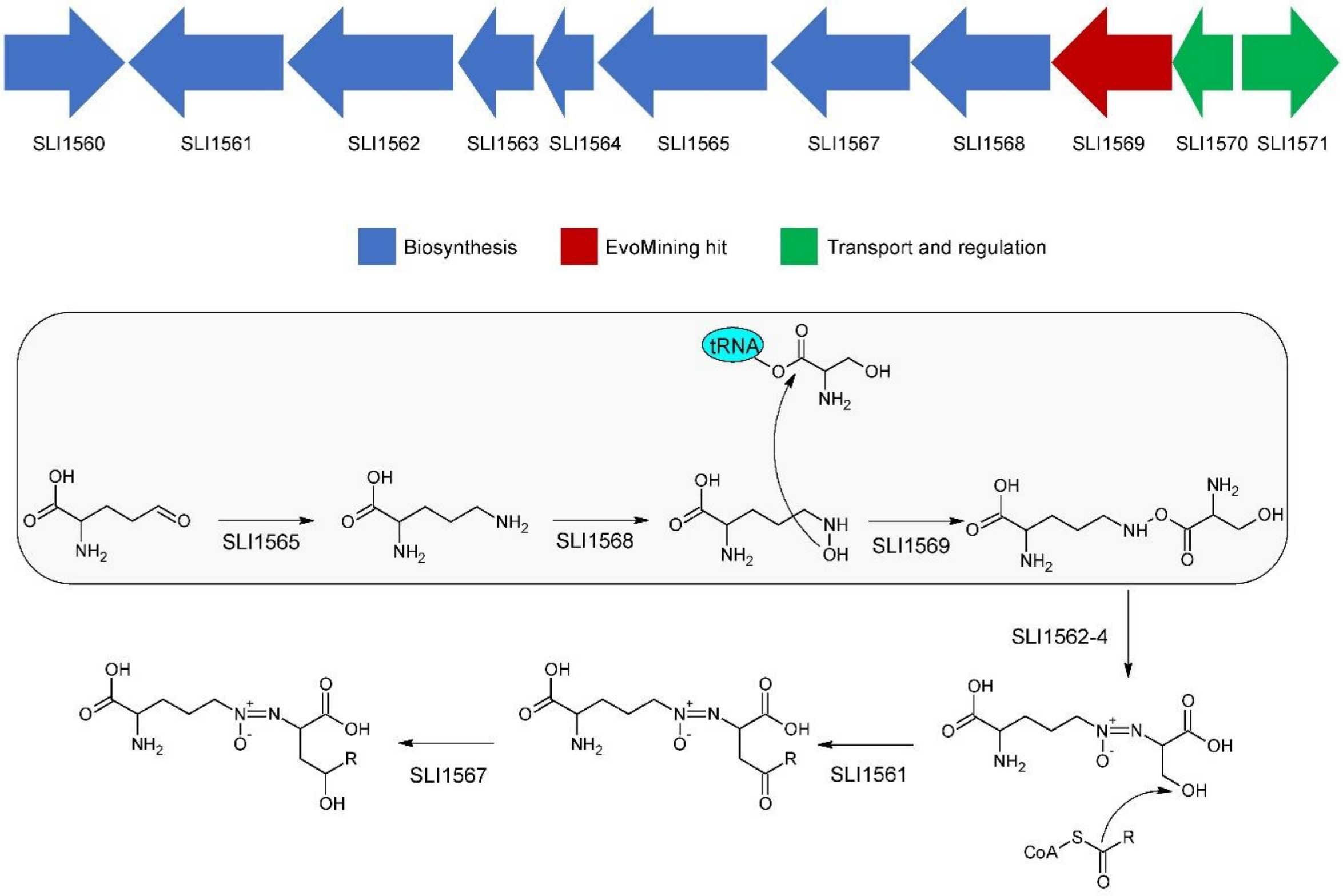

**Figure 8-10. Predicted biosynthetic pathway for an O-seryl-hydroxylamine ester intermediary.** The upper part shows the gene cluster organization for the predicted pathway. The proposed biosynthetic pathway intermediaries and final product are shown below. The proposed mechanism of tRNA dependent O-seryl-hydroxylamine ester formation is highlighted in a grey box.

Further modifications, such as the formation of an azoxy group, are also envisaged. How this group is formed is still unclear. The functions in the valanymycin and SLI560-SLI1568 biosynthetic gene clusters that cannot be connected to the formation of the di-amino acid ester intermediary, as they are mainly oxidoreductases (SLI1562-4), may catalyze oxidations related to formation of the azoxy group. Finally, it is likely that at some point during the biosynthetic pathway the transfer of an acyl group by SLI1561 will occur. This could take place at an available hydroxyl group, followed by the reduction of this group by SLI1567, a keto reductase. Since this biosynthetic gene cluster includes a previously unknown combination of enzymes, the prediction of their concerted functions indicates that its metabolic product is a novel molecule. Further experimental analysis is required for the confirmation of this biosynthetic prediction.

# 9 Results Chapter IV: *Streptomyces* conserved enzymatic expansions linking central and “secondary” metabolism

## 9.1 Introduction

EvoMining, an evolution-inspired genome mining strategy for the prediction of novel NP biosynthetic pathways presented in chapter II, includes a stage in which enzyme expansions are identified in the genomes of a selected group of organisms. Subsequently, this allows addressing the NP biosynthetic potential of enzyme families originated in central metabolic pathways. The enzyme expansions discussed in this Chapter were identified using the EvoMining approach. For this purpose metabolic reconstructions and publically available genome sequences were used. Indeed, the enzyme expansion survey revealed conserved enzymatic expansions, including the last enzymes of the glycolytic pathway (Barona-Gómez et al, 2012).

Phylogenomic analyses indicated that the last common ancestor of *Streptomyces* must have acquired these glycolytic expansions before speciation. Given that *Streptomyces* are proficient NP producers it is tempting to speculate that evolution of an expanded glycolytic node and specialization of NP biosynthesis are linked. This hypothesis is the main subject of this chapter, where *in silico* and *in vivo* analyses, including comparative genomics and phylogenetics, transcriptional analysis, gene over-expression, gene knockout experiments, together with phenotype characterization, were conducted. The results obtained herein point towards a major physiological role for this node involved in carbon and nitrogen utilization with development and NP production.

## 9.2 Enzyme expansions in the *Actinobacteria* phylum

Given that a single metabolic function can be performed by enzymes with different evolutionary histories (Omelchenko et al, 2010), as different protein sequences and folds can encode identical enzyme activities, the concept of “enzymatic expansion” (EE) was introduced in chapter II. EEs actually account for the potential enzymatic capability of an organism to catalyze a metabolic conversion with one or more enzymes (Barona-Gómez et al, 2012). As part of the EvoMining analysis presented in chapter II, EEs were identified within the eleven central metabolic pathways that were selected as proof-of-concept. These eleven PSCP are highly conserved and their function is relatively easy to annotate, even when different homologues may be involved in each enzymatic step (Supplementary material S1). Fatty acid biosynthesis, for instance, was excluded because of the difficulty in ascribing function with confidence to the many homologues that take part in this pathway.

The enzyme homologues involved in a single metabolic step within the eleven PSCP were systematically retrieved from the actinobacterial database (Supplementary material S2). An enzymatic expansion was called when the number of enzymes capable of performing a single metabolic step present in a given genome was equal or higher than the average plus the standard deviation of the homologues linked to the same metabolic step in the whole actinobacterial family. This enabled the identification of specific and particularly significant expansions that could be resolved within the taxonomic relationships of the genomes analyzed. After statistical processing of the data, the enzymatic expansions were identified and represented graphically as a plot (Figure 7-3; Barona-Gómez et al, 2012).

The analysis of the EE profiles of actinobacterial central metabolism revealed that events of EEs are overrepresented in the clades corresponding to the *Micromosnosporaceae* and *Pseudonocardiaceae* families, as well as in the *Streptomycineae* suborder. Remarkably, these three A*ctinobacteria* families account for approximately 48% of the production of bioactive known NPs (Marinelli, 2009). This observation coincides with the idea that expansion of central metabolism is an important source of raw material for the evolution of new NP biosynthetic pathways (see Chapter II; Vining, 1992). Furthermore the enzymatic expansions show similar intra-family patterns, while the pattern differs among families.

## 9.3 Conserved EEs in central metabolism of *Streptomyces*

The genus *Streptomyces* includes the most proficient and widely studied NP producing strains. Several examples of enzyme families that have been recruited in NP biosynthetic pathways have been identified (see Chapter II). However, the most conserved expansions, present in 90-100 % of the members of the genus, could not be linked to NPs. Among the highly conserved and non-recruited EEs are the glycolytic enzymes phosphoglycerate mutase (PGM) and pyruvate kinase (PYK), as well as the amino acid biosynthetic enzyme aspartate-semialdehyde dehydrogenase (ASD). These functions occur in nodes within the metabolic network (Figure 9-2). Could it be that these EEs may be related to adaptation of central metabolism for the production of NPs synthesised by these species? In other words, could it be that these EEs are important for the metabolic diversification and radiation of the genus *Streptomyces*? These questions relate to previously investigated mechanisms involved in mediating the interaction between central and peripheral metabolic pathways (Hodgson, 2000; van Wezel and McDowall, 2011; Barona-Gómez & Cruz-Morales et al, 2012).

## 9.4 Phylogenomic analysis of conserved EEs in *Streptomyces* metabolism

The comparative analysis of the genome context of the genes encoding PGM, PYK and ASD enzymes in *Streptomyces* showed a high degree of synteny. These EEs are actually located within the “core” region of the chromosome (Figure 9-1 B), where central metabolic functions have been systematically identified (Bentley et al, 2002; Choulet et al, 2006). A concatenated phylogenetic analysis of the expanded enzyme families showed a similar clade distribution between the genes present in most *Actinobacteria* (core genes), and their expanded partners only conserved in *Streptomyces* (Figure 9-1 A). These observations suggest an ancestral origin for this set of expanded enzymes, probably going back to the last common ancestor of the genus *Streptomyces*.

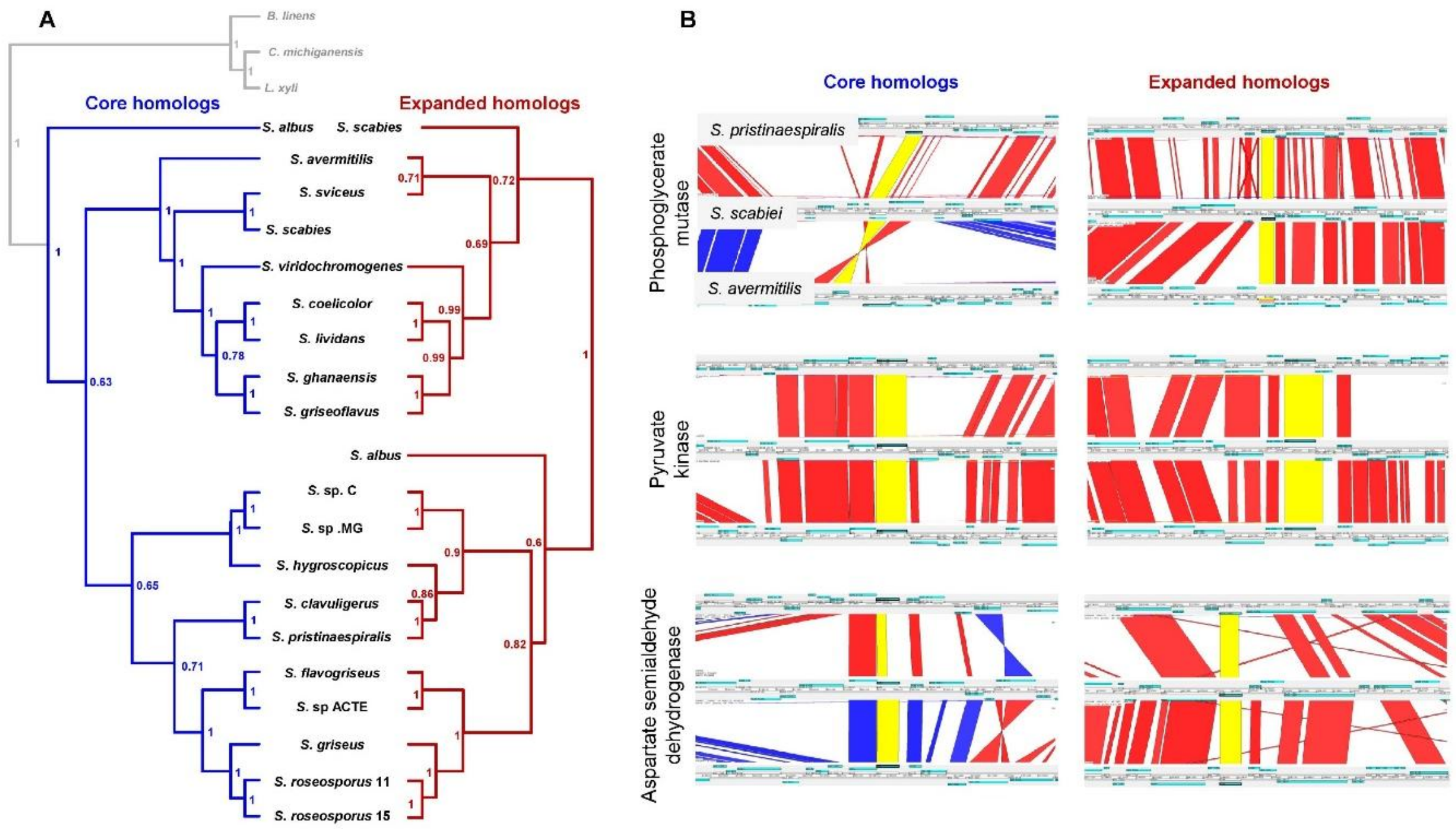

**Figure 9-1. Phylogenomic analysis of conserved EEs in *Streptomyces*. (A) Phylogenetic tree of concatenated EEs sequences.** The homologues present in all *Actinobacteria* are marked as core, while the *Streptomyces* specific expansions are indicated as expanded homologues. **(B) Genome context analysis of conserved EEs (PGM, PYK and ASD).** Artemis comparative tools view of the genomes of *S. pristinospiralis*, *S. scabiei* and *S. avermitilis*. The regions were their orthologous loci are located are marked in yellow rectangles. This view is representative of the genome context of these genes in *Streptomyces*.

ASD produces aspartate semialdehyde, a precursor for cell wall biosynthesis and the production of lysine, methionine and threonine. Moreover, the latter amino acid is a precursor for the biosynthesis of isoleucine. The lower part of the glycolytic pathway, which includes reactions catalyzed by PGM and PYK (Figure 9-2), is also an important metabolic node responsible for the production of ATP via the oxidation of carbohydrates, as well as the production of key precursors, such as phosphoenolpyruvate (PEP) and pyruvate (PYR). For instance, PYR is converted into acetyl-CoA, which is used for the production of fatty acids, or further oxidized to produce reduction power and ATP in the tricarboxylic acids cycle (TCA). PYR can also be used for alanine biosynthesis, in addition to taking part in the PEP-PYR-Oxalacetate anaplerotic node. Interestingly, in *Streptomyces*, this node has been linked to the accumulation of TCA intermediaries for polyketide biosynthesis (Rodriguez et al, 2012).

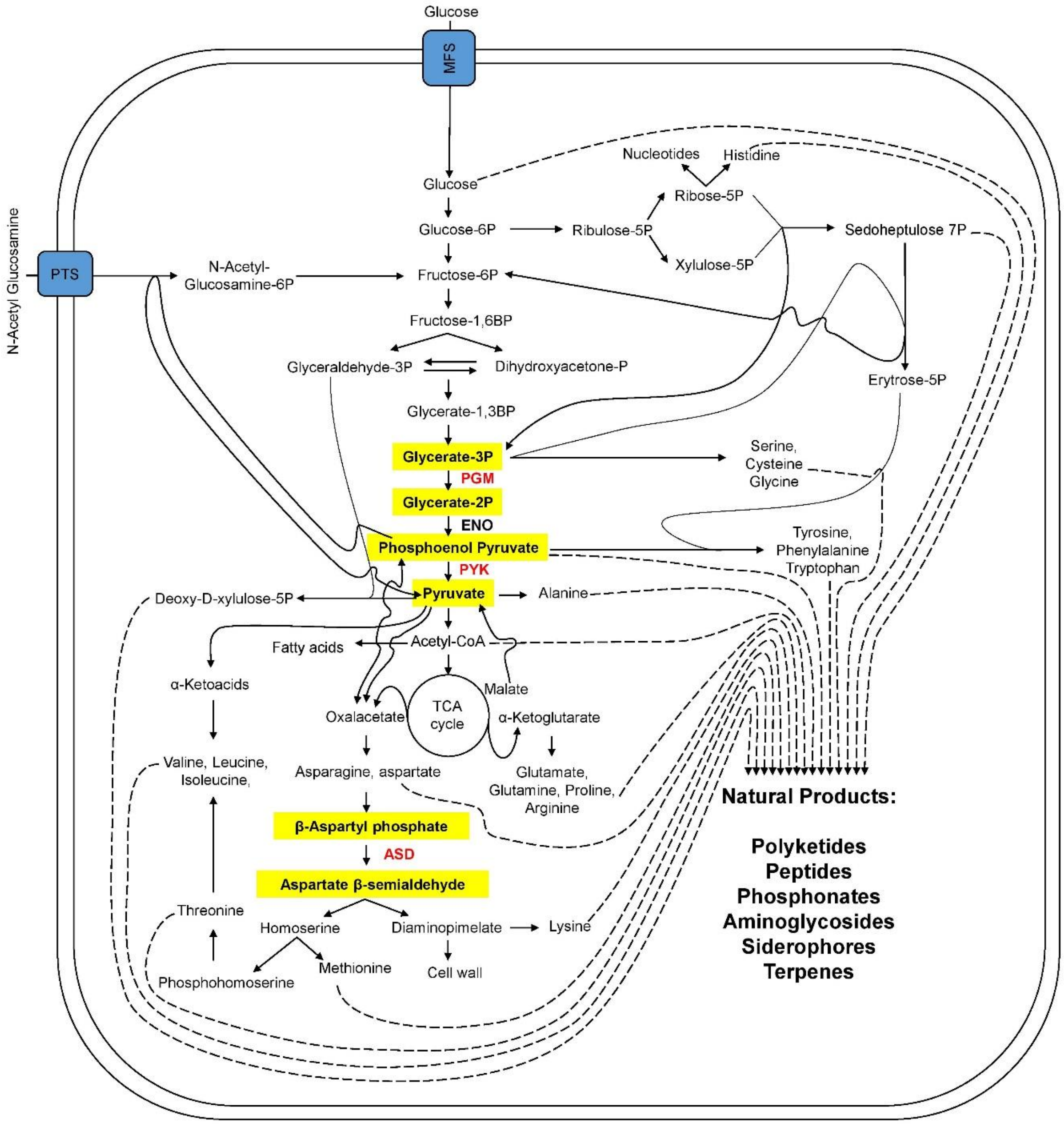

**Figure 9-2. *Streptomyces* PSCP and their connections with NP biosynthesis.** The conserved EEs are indicated in red, their substrates and products are highlighted in yellow. The precursor-NP links are indicated as dashed arrows.

However, it should be noted that the PGM and PYK reactions in the glycolytic pathway are linked by the activity of the enolase (ENO), which catalyzes the conversion of 2-phosphoglycerate - the product of PGM - into PEP – substrate of PYK. Therefore, the degree and pattern of expansion of this enzyme was revisited. ENO was found to be expanded in 47% of the *Streptomyces* analyzed, and as discussed in Chapter II, it has also been recruited for phosphonate biosynthesis. The differences in the degree of conservation between ENO and PGM/PYK could be due to a limited taxonomic coverage, although it may also reflect on an adaptive process. Nonetheless given its central position in the metabolic network, linking PGM with PYK, expansion of ENO was also selected for further analysis.

### 9.4.1 Phylogenetic analysis of glycolytic EEs in *Streptomyces*

**ENO.** The analysis of this enzyme family is presented in Chapter II (Figure 7-4). In summary, three clades could be distinguished in this phylogenetic reconstruction. The first clade corresponds to the core homologues. The second clade includes ENO homologues with high sequence identity at the amino acid level (57% in average) and it corresponds to the expanded homologues. The last clade, discussed in Chapter II, includes a small number of sequences that have been recruited by NP biosynthesis.

**PGM.** The phylogenetic reconstructions using these sequences (Figure 9-3) showed that PGM homologues are clearly divided into two groups of orthologs, which only share 22% of amino acid sequence identity. However, both groups were found to be 99% conserved in the *Streptomyces* strains that were included in the analysis. This expansion was also found in some members of the *Nocardiaceae* family, although not as conserved as in *Streptomyces*. Given its very low identity, and the high divergence of the expanded PGMs, it is possible that functional divergence has occurred within the expanded homologues. However, if a new function has evolved, this has been fixed in the entire genus *Streptomyces* and in several other NP producing strains.

**PYK.** The phylogenetic reconstruction of these sequences (Figure 9-4) indicates a recent duplication event. Two clades within the *Streptomyces* lineage, accounting for all of the PYK homologues found within the genus. These two clades include sequences from *Kitasatospora*, a closely related specie member of the *Streptomycineae* suborder, showing that PYK expansion is conserved beyond *Streptomyces*. Both clades have a strikingly similar topology, suggesting a steady evolutionary rate, maybe due to a conserved function. Furthermore, the identity between the members of the two clades is 68%, supporting a common enzymatic function. SCO2014 (*pyk1*), is the core PYK in *S. coelicolor* and shares a low but significant degree of synteny with the single PYK ortholog of *Catenulispora acidiphila*, (genome sequence accession: NC_013131; Copeland et al, 2009). This is the closest *Actinobacteria* with a single PYK. In contrast, SCO5423 (*pyk2*) is the expanded paralog, and shows no synteny with the *C. acidiphila* PYK. This gene was acquired after a duplication event occurred before the divergence of the genus.

With the aim of characterizing a potential mediation of the interaction between central and NP metabolism by the expanded glycolytic enzymes PGM and PYK, *in vivo* experiments were conducted. As a control expansion, the enolase enzyme (ENO) that takes the product of PGM and produces the substrate of PYK, was also selecte, members of the enolase enzyme family have been linked to phosphonate NP production in at least three *Streptomyces* strains in other words the expansion and metabolic fate of the enolase expansion can be explained for its recruitment in NP biosynthesis (Chapter II of this thesis), while the PGM and PYK cannot, providing a comparative framework for their characterization. Furthermore its phylogenetic analysis, mutant construction and some functional insights have been already presented in chapter II.

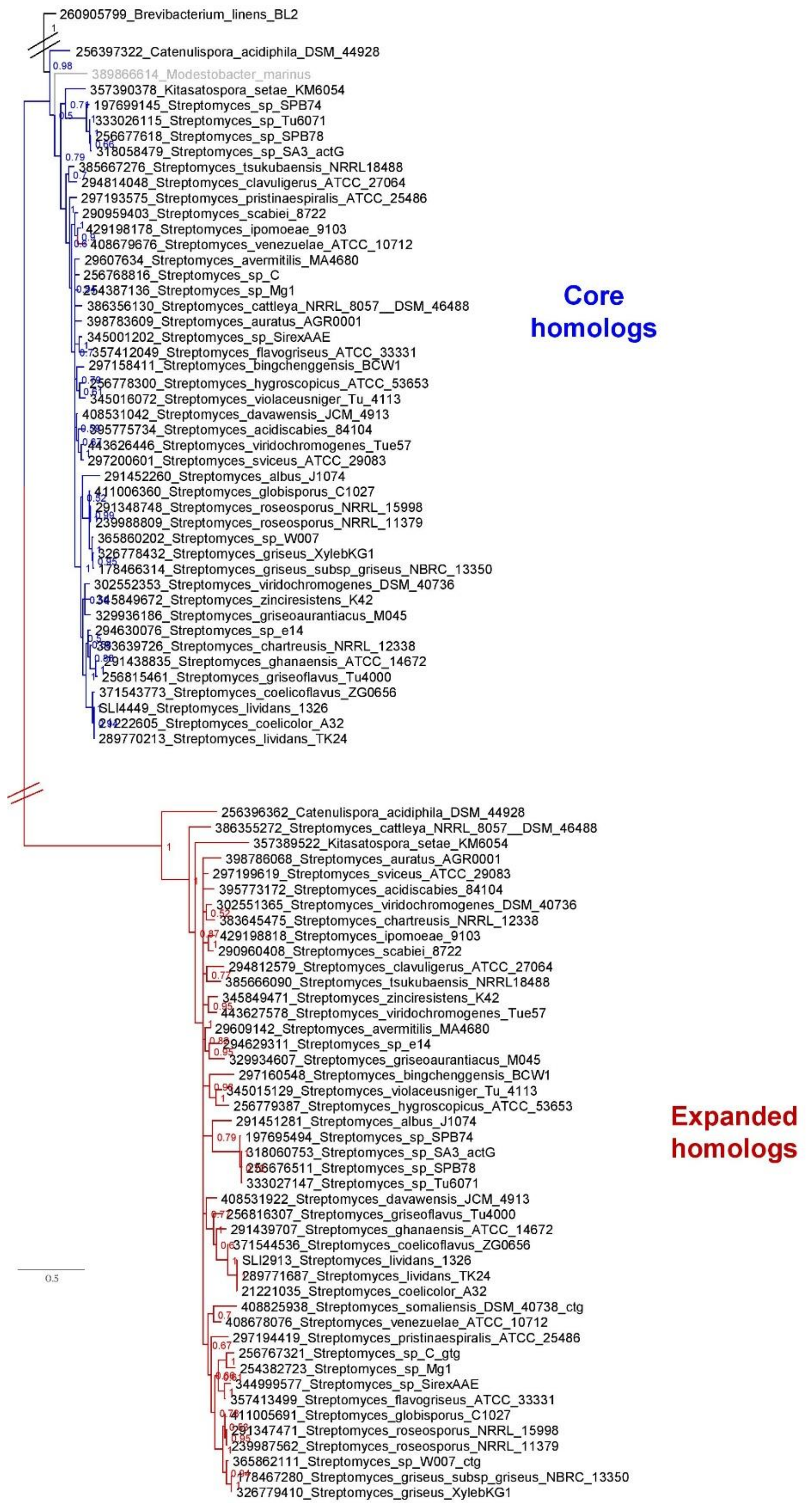

**Figure 9-3. Phylogenetic reconstruction of PGMs.** Within the *Streptomycineae* suborder, PGMs are grouped in two distant clades, which share a similar topology despite the sequence divergence between both groups of orthologues. Double diagonal lines indicate places were other taxa were present but were edited to fit this page. An unedited version of this tree is available as supplementary material (S11-PGM).

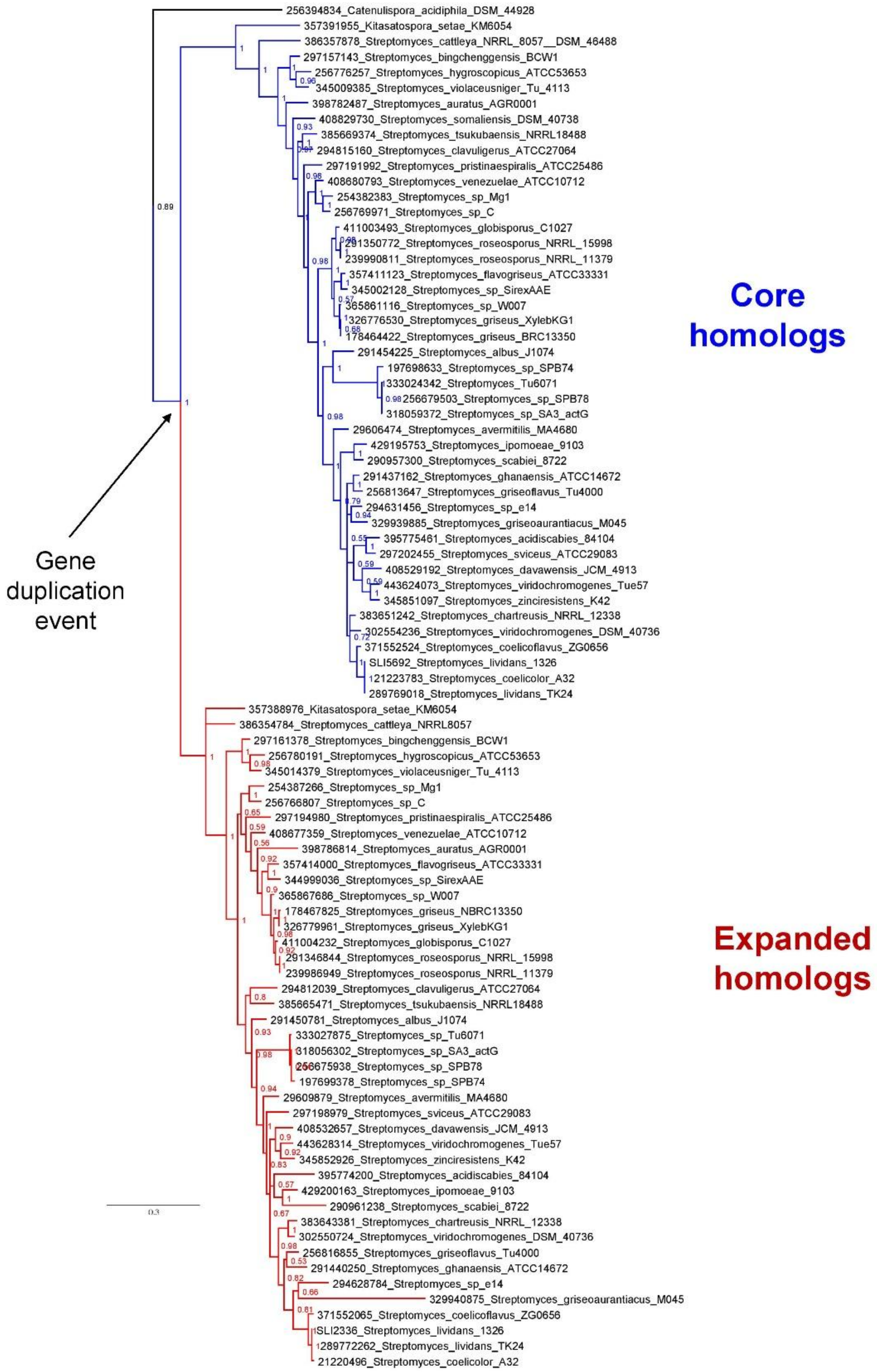

**Figure 9-4. Phylogenetic reconstruction of PYKs.** The duplication event that occurred in the last common ancestor of the *Streptomicinaceae* family is highlighted; the tree is rooted with C*atenulispora*. The closest relative of S*treptomyces* with only one PYK. *Kitasatospora setae*, a member of the *Streptomicinaceae* family, also has two pyruvate kinases. The tree shown here has been edited to fit this page, but an unedited version is available as supplementary material (S11) the closest relative of S*treptomyces* with only one pyruvate kinase, *Kitasatospora setae* a member of the *Streptomicinaceae* family also has two pyruvate kinases.

# 9.5 *In vivo* Characterization of Conserved PSCP Enzyme expansions.

## 9.5.1 Construction of knock out mutants and phenotypical analyses

To study the physiological role and impact of ancestral PSCP expansions on *Streptomyces* central metabolism, the six homologues of the glycolytic enzymes PGM, ENO, and PYK encoded in the *S. coelicolor* genome were characterized (Table 9-2). We first investigate the biochemical equivalence of both enzymes by in frame PCR-targeted gene replacement of their coding sequences using the Redirect system (Gust et al, 2003).

**Table 9-2.** Conserved glycolytic EEs in *S. coelicolor*

| Enzyme | E.C | *S. coelicolor* loci | EE Frequency | ID[1] |
|---|---|---|---|---|
| Phosphoglycerate mutase | 5.4.2.1 | SCO4209 (*pgm1*-core)<br>SCO2576 (*pgm2*-expansion) | 97% | 22% |
| Enolase | 4.2.1.11 | SCO3096 (*eno1*-core)<br>SCO7638 (*eno2*-expansion) | 47% | 57% |
| Pyruvate kinase | 2.7.1.40 | SCO2014 (*pyk1*-core)<br>SCO5423 (*pyk2*-expansion) | 97% | 68% |

[1] Amino acid sequence identity

Mutants of the six genes were obtained and confirmed by PCR amplification. Interestingly, the *pyk1* (core *pyk*, SCO2014::*acc(3)IV*) and *pgm1* (core *pgm*, SCO4209::*acc(3)IV*) mutants were difficult to obtain, and large numbers of colonies (*pyk1*: >50; *pgm1*: >100) had to be screened for double crossover ex-conjugants. In fact, *pgm1*::*acc(3)IV* was obtained only when the selection of double cross-over mutants was done in rich media (R2YE) supplemented with alanine, which was added to bypass the glycolytic production of pyruvate. In contrast, *pyk2* (SCO5423::*acc(3)IV*) *eno1* (SCO3096::*acc(3)IV*), *eno2* (SCO7638::*acc(3)IV*) and *pgm2* (SCO2576::*acc(3)IV*) mutants were obtained after the screening of few colonies (<10). This simple observation implies a differential role for each homologue.

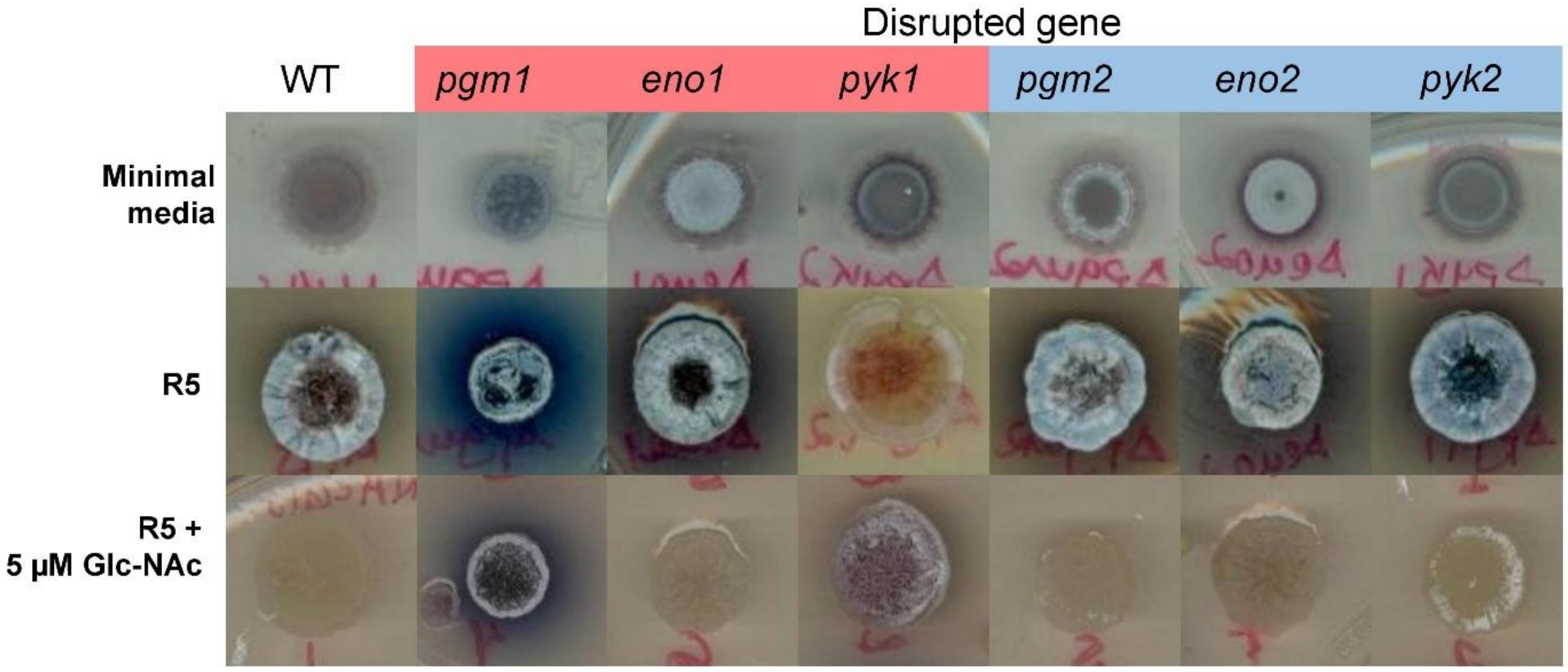

**Figure 9-5.** Phenotypes of PGM, PYK and ENO mutants in *S. coelicolor*. Mutants lacking the core homologues are highlighted in red, whereas mutants lacking the expanded homologues are in blue

To confirm the previous observation, the mutant strains were grown in different media and carbon sources. In summary, these experiments showed that (i) the lack of *pgm1* produces insensitivity to GlcNAc, and arrest of development when supplemented in rich media. This observation is in agreement with previous reports (Rigali et al, 2008). (ii) The lack of each enolase homologue accelerates development of aerial mycelium in minimal media when glucose is added as the single carbon source; and (iii) the lack of *pyk1* arrests the development of aerial mycelium in R5 (rich) media.

Phenotypic screening using a high-throughput platform (SFN2 micro plate; Biolog) provided further evidence of a differential metabolic role for the core and expanded homologues. In this system, the *S. coelicolor* production of pigmented NPs was observed (Table 9-3). This screening revealed a drastic loss in the production of pigments in the *pgm1* mutant, and differential patterns of pigment production in the different media, supporting a carbon source-dependent differential role for core and expanded glycolytic enzymes in *S. coelicolor,* given that PEP is the donor of the phosphate group in the PTS system, these differences may be related to carbon catabolite repression.

**Table 9-3. Phenotypes of PGM, PYK and ENO mutants in *S. coelicolor*.**

| Compound[1] | WT | *eno*1 | *eno*2 | *pyk*1 | *pyk*2 | *pgm*1 | *pgm*2 |
|---|---|---|---|---|---|---|---|
| Dextrin | 1 | 1 | 1 | 0 | 1 | 0 | 1 |
| Glycogen | 1 | 0 | 1 | 1 | 1 | 0 | 1 |
| Tween 40 | 1 | 0 | 1 | 1 | 0 | 0 | 1 |
| Tween 80 | 1 | 1 | 1 | 1 | 1 | 0 | 1 |
| N-Acetyl-D- Glucosamine | 0 | 2 | 2 | 0 | 1 | 0 | 1 |
| D-Arabitol | 0 | 0 | 1 | 0 | 0 | 0 | 1 |
| D-Cellobiose | 1 | 1 | 1 | 1 | 0 | 0 | 1 |
| D-Fructose | 1 | 1 | 1 | 2 | 2 | 0 | 1 |
| D-Galactose | 1 | 1 | 1 | 1 | 0 | 0 | 1 |
| Gentiobiose | 1 | 1 | 1 | 0 | 0 | 0 | 1 |
| α-D-Glucose | 1 | 1 | 1 | 1 | 1 | 0 | 1 |
| m-Inositol | 2 | 1 | 1 | 2 | 1 | 0 | 1 |
| α-D-Lactose | 1 | 1 | 1 | 2 | 1 | 0 | 1 |
| Lactulose | 1 | 1 | 1 | 1 | 0 | 0 | 0 |
| Maltose | 1 | 1 | 1 | 1 | 0 | 0 | 1 |
| D-Mannitol | 1 | 1 | 1 | 0 | 0 |  | 0 |
| D-Mannose | 1 | 2 | 1 | 1 | 1 | 0 | 1 |
| β-Methyl-D- Glucoside | 0 | 1 | 1 | 0 | 0 |  | 0 |
| L-Rhamnose | 0 | 1 | 1 | 0 | 0 | 0 | 0 |
| Turanose | 0 | 0 | 0 | 0 | 1 | 0 | 0 |
| Pyruvic Acid Methyl Ester | 1 | 1 | 1 | 1 | 1 | 0 | 1 |
| Acetic Acid | 0 | 1 | 0 | 0 | 0 | 0 | 0 |
| Cis-Aconitic Acid | 0 | 1 | 0 | 0 | 0 | 0 | 0 |
| Citric Acid | 1 | 1 | 1 | 1 | 0 | 0 | 1 |
| D-Gluconic Acid | 0 | 1 | 1 | 0 | 0 | 0 | 0 |
| D-Glucosaminic Acid | 0 | 1 | 0 | 0 | 0 | 0 | 0 |
| D-Glucuronic Acid | 1 | 1 | 0 | 1 | 1 | 0 | 0 |
| β- Hydroxybutyric Acid | 2 | 2 | 2 | 2 | 2 | 0 | 2 |
| Malonic Acid | 0 | 0 | 0 | 1 | 0 | 0 | 0 |
| Propionic Acid | 0 | 0 | 0 | 1 | 0 | 0 | 0 |
| Sebacic Acid | 0 | 0 | 0 | 1 | 1 | 0 | 0 |
| Bromosuccinic Acid | 0 | 1 | 1 | 1 | 0 | 0 | 0 |
| L-Alaninamide | 0 | 1 | 0 | 0 | 0 | 0 | 0 |
| L-Alanine | 1 | 2 | 1 | 1 | 2 | 0 | 1 |
| L-Alanyl- Glycine | 1 | 2 | 2 | 1 | 1 | 0 | 1 |
| L-Aspartic Acid | 1 | 0 | 0 | 0 | 0 | 0 | 1 |
| L-GlutamicAcid | 1 | 1 | 1 | 0 | 0 | 0 | 1 |
| Glycyl-L-Glutamic Acid | 1 | 1 | 1 | 1 | 1 | 0 | 0 |
| L-Histidine | 2 | 1 | 1 | 2 | 2 | 0 | 2 |
| Hydroxy-L- Proline | 1 | 1 | 1 | 1 | 1 | 0 | 1 |
| L-Phenylalanine | 1 | 1 | 1 | 1 | 1 | 0 | 0 |
| L-Proline | 1 | 2 | 2 | 1 | 2 | 0 | 2 |
| L-Pyroglutamic Acid | 1 | 1 | 0 | 2 | 1 | 0 | 0 |
| D-Serine | 0 | 0 | 0 | 1 | 0 | 0 | 0 |
| L-Serine | 1 | 1 | 1 | 1 | 1 | 0 | 1 |
| L-Threonine | 1 | 1 | 1 |  | 1 | 0 | 1 |
| γ-Aminobutyric Acid | 1 | 0 | 0 | 1 | 1 | 0 | 0 |
| Putrescine | 1 | 1 | 1 | 1 | 1 | 0 | 1 |
| Glycerol | 1 | 1 | 1 | 1 | 1 | 0 | 0 |

[1]96 different carbon and nitrogen sources were tested, 50 of them allowed the production of pigments in at least one strain. 0, indicates no production of pigments; 1, a mild production and 2, high production.

## 9.5.2 Transcriptional analysis

Given the differential phenotypes, one possibility is a differential transcriptional program for each homologue. In order to investigate the expression pattern of these genes RT-PCR of the *pgm*, *eno*, and *pyk* genes, in mutant and WT strains, was determined. These experiments were done in R2YE media, which has an identical composition than R5 but it is obtained following a different preparation method: yet, similar phenotypic results were obtained. Under these conditions, five of the six genes studied were constitutively expressed (*pgm*1 and 2, *eno*1 and 2, *pyk*1). The main observation derived from this experiment is that the expression of *pyk*1 is higher than *pyk*2.

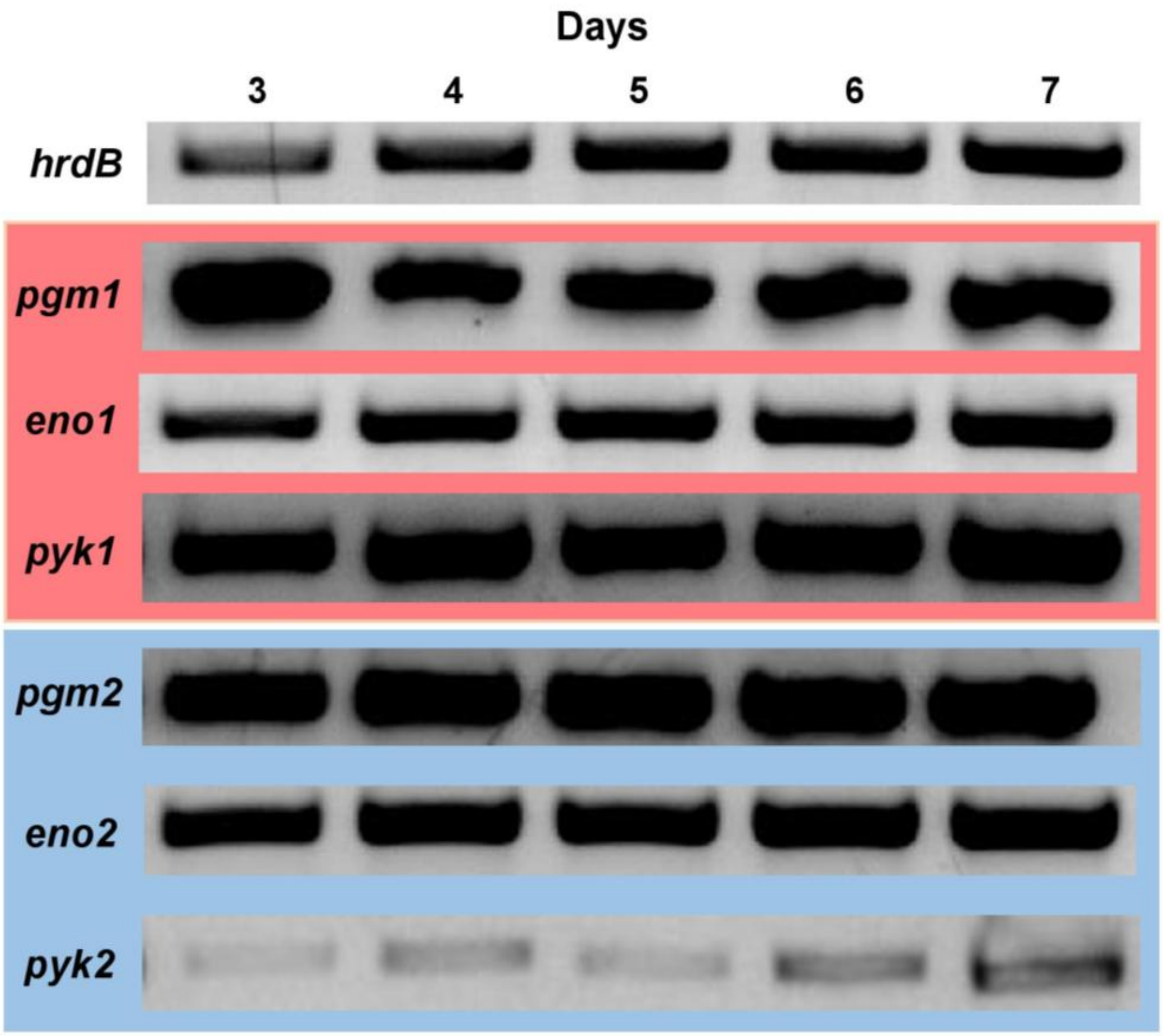

**Figure 9-6.** Expression profile of PGM, PYK and ENO genes in *S. coelicolor*. Total RNA was extracted from mycelium grown on solid R2YE media for 3 to 7 days.

### 9.5.3 Over-expression of PGM, PYK and ENO in *S. coelicolor*

To address the possible role of conserved glycolytic EE in the provision of precursors for NP biosynthesis, the six glycolytic genes were cloned in a tunable plasmid, pAV11b. This plasmid integrates into the chromosome as a single copy locus, specifically at the phage *phi* attachment site *att*. This allows a controlled expression of the cloned gene under the control of the tetracycline-inducible *tetris* cassette using the *tcp830* promoter (Khaleel et al, 2011). The integrity of the six cloned genes was confirmed by PCR amplification and DNA sequencing. Integration in *S. coelicolor* M145 was performed via conjugation and confirmed by PCR. Spores of each of the transformant strains were plated on R2YE media in a gradient of anhydro-tetracycline (ATC) to over-express the genes (Figure 9-7).

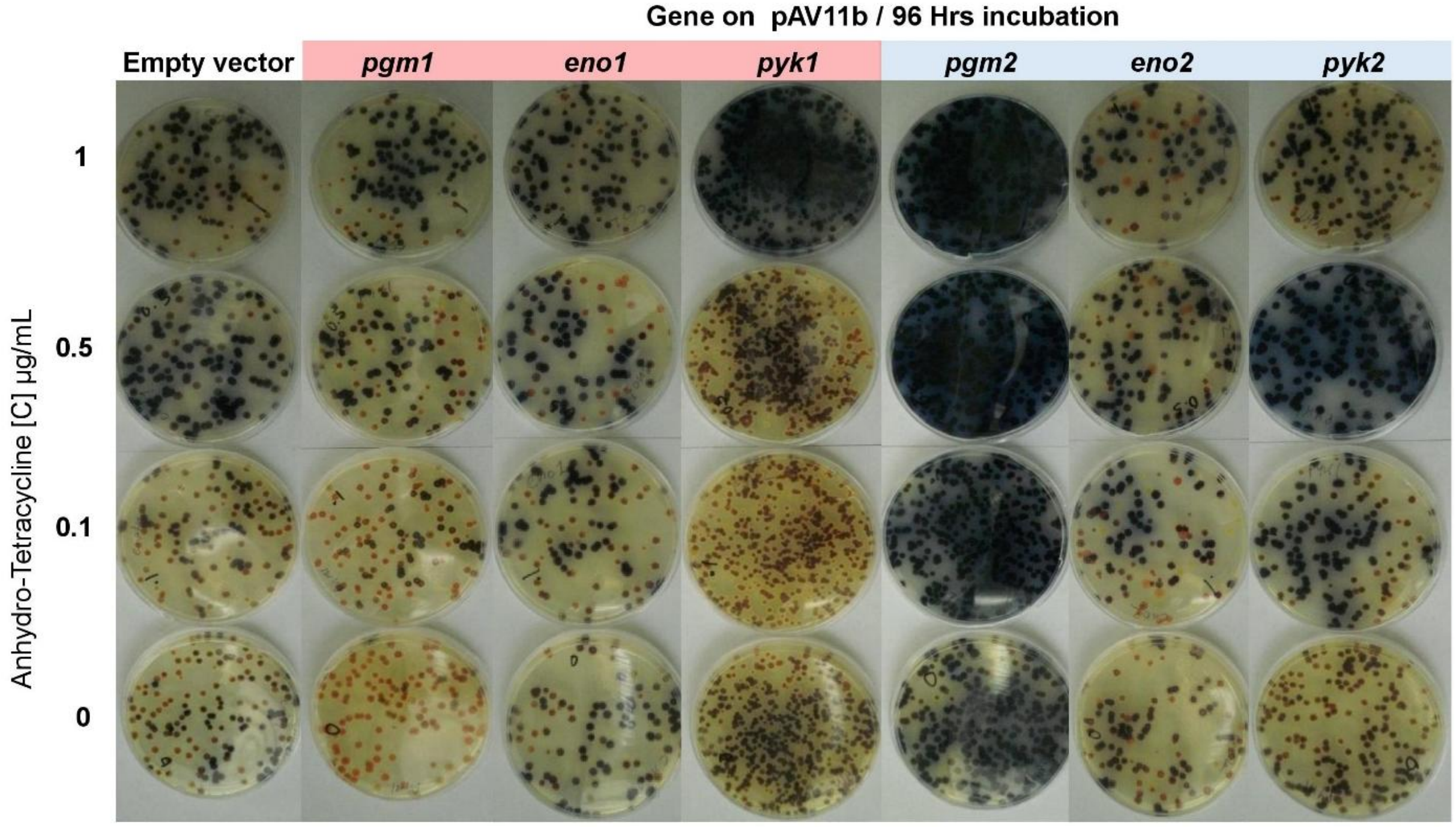

**Figure 9-7.** Over-expression assay of PGM, PYK and ENO in *S. coelicolor*. Transformant strains were plated on R2YE media with 0, 0.1, 0.5 and 1 µg of anhydro-tetracycline and incubated for 72 hours. Core homologues are highlighted in red and expanded homologues are in blue.

Different phenotypes between the strains were observed (Figure 9-7) even without induction by ATC. This can be explained by the fact that leaky expression from the *tcp80* promoter has been previously reported (Jyothikumar et al, 2012). Nonetheless, induction with ATC caused phenotypic differences depending on the gene that was over-expressed. Copious production of pigmented NP was associated by the over expression of both *pyk* genes, and *pgm2*. The leaky expression of pgm2 is enough to over produce pigments, a phenotype that increases sharply with the concentration of ATC. Pigments over-production was achieved with 1 µg/ml of ATC in the *pyk1* over-expressing strain, while only 0.5 µg/ml was required to achieve a high production of pigments in the strain over-expressing *pyk2*. A hypothesis derived from this observation, is that less Pyk2 is necessary to obtain the same phenotype, in such case, differential enzymatic efficiency, allosteric regulation or product inhibition, is needed to explain the differences. To address such hypothesis, further analysis of the *pyk* transcripts in the over expression strains and enzyme kinetics are required. All together, these results provide further support for a differential physiological role for both PYK paralogues in *Streptomyces*. Given that the glycolytic pathway is an important source of precursors for the synthesis of NPs, is tempting to speculate that a function for the expanded glycolytic enzymes is the deviation of precursors for the biosynthesis of NPs.

# 10 Discussion

## 10.1 General overview

The aim of this work was to contribute to the understanding of the evolutionary mechanisms behind the overwhelming diversity of NPs produced by *Actinobacteria*, with a particular emphasis on the genus *Streptomyces.* The main objective of this research was to exploit evolutionary concepts, such as the mechanistically diverse enzyme superfamilies, and the expansion and recruitment of enzyme families. These concepts were embraced using a series of methods, namely, comparative genomics, phylogenetic and gene context analyses, gene knockout and phenotyping experiments for the discovery of new biosynthetic pathways. The exploration of evolutionary concepts with bioinformatic methods led to the development of EvoMining, a phylogenomic approach which can efficiently predict Natural product biosynthetic pathways in an unbiased manner. In addition, during the development of EvoMining, the analysis of the patterns of enzyme expansions in central metabolism revealed what seems to be a conserved adaptation of the glycolytic pathway in *Streptomyces*, which was speculated to be linked to the ability of these microorganisms to produce NPs.

The model organism *S. lividans* was selected to test the performance of the EvoMining approach. The genome sequence of this organism was not only used for mining of NPs, but also for comparative genomic analyses with the closely related species *S. coelicolor*, as well as with the industrial strain TK24. These analyses led to the identification of the previously elusive plasmid, SLP3, together with a high number of metal homeostasis-related genes, all located within a genomic island lost during the obtaining of the strain TK24. Relevant for this project, it was found that this genomic island encodes for novel NP biosynthetic pathways, providing evidence of the loss of adaptive traits during the stabilization of environmental conditions caused by growth in the laboratory.

## 10.2 Natural products genome mining: A practical application of the theory of evolution

The development and use of the EvoMining approach on mining actinobacterial genomes for the identification of new NP biosynthetic pathways led to the following conclusions: (i) the enzyme repertoire of central metabolism is an important source of precursors for the evolution of NP chemical diversity; (ii) Expansion of central metabolic repertoire occurred in the most proficient producers of NPs, therefore is likely that this mechanism is required for the evolution of new biosynthetic pathways; and (iii) reconstruction of the evolutionary history of enzyme superfamilies, including identification of their divergent members, in combination with genome context analysis, can be exploited for the prediction of novel biosynthetic pathways in microbial genomes.

As proof-of-concept for the usefulness of EvoMining approach, a comprehensive analysis of the actinobacterial enolase enzyme family was performed. This analysis revealed that a natural metabolic fate for the divergent members of this enzyme family is the dehydratation (E1CB elimination) of an intermediary containing a C-P bond within the biosynthesis of phosphonate and phosphinate NPs. This is the same reaction chemistry used for the glycolytic enolase function. Structural modelling of enzyme-substrate interactions showed the presence of an almost intact active site, yet pointed out the residues in the active site that may have been selected during evolution of substrate specificity. These results actually suggested that the phosphonate enolases may still allow the binding and conversion of the original substrate, a notion that is in agreement with the concept of mechanistically diverse enzyme superfamilies.

The analysis presented in chapter II demonstrated that EvoMining can be applied systematically to identify novel biosynthetic gene clusters of NPs in an unbiased fashion, i.e. without using NRPS or PKS domains as queries for genome mining. This pipeline opens the possibility for the prediction of biosynthetic pathways of unknown classes of compounds. As such, EvoMining provides an alternative for genome mining using traditional approaches, even in low-quality genome sequences as demonstrated using the *S. sviceus* genome. To provide a glimpse of the potential of this approach, a preliminary (i.e not exhaustive) EvoMining analysis of the available actinobacterial genomes was performed.

Applying the same criteria and methods described for the proof-of-concept analysis using the enolase enzyme superfamily, seventy three new biosynthetic gene clusters, involving thirty recruited enzyme families, were identified. Amongst these, twenty predictions included unprecedented combinations of enzymes, which suggests that the products of these putative biosynthetic pathways would be novel metabolites. Only further characterization of these putative pathways can provide structural information and confirm the prediction of the structurally and biosynthetically novel molecules. This could only be achieved using molecular microbiology and analytical chemistry techniques that although feasible, are time-consuming and go beyond the scope of this thesis.

To fully demonstrate the potential of the EvoMining approach for the discovery of new biosynthetic pathways, the model strain *S. lividans* 66 was selected for EvoMining analysis. Its genome sequence was obtained, and its analysis led to the identification of three previously undescribed biosynthetic pathways. The three biosynthetic systems are predicted to have unprecedented biosynthetic strategies, such as the formation of peptide bonds in a tRNA-dependent manner by novel enzyme families, and the very first example of the incorporation of arsenate into a NP. Nonetheless, one of the most relevant results presented in this thesis is that these biosynthetic systems could not be identified by the current genome mining approaches, and were only revealed after the use of evolution-inspired mining strategies. This result supports the hypothesis that the application of evolutionary principles in NP Genome Mining provides a predictive power in an unbiased fashion.

## 10.3 Links between natural products, metal homeostasis, mobile genetic elements and adaptive evolution revealed in the *S. lividans 66* genome

The sequencing and assembly strategy used to obtain the complete genome sequence of *S. lividans* 66, involving most of the commercially available sequencing platforms, provides a useful guideline for the ever-growing number of microbiological laboratories engaged in genome sequencing. Of particular relevance for genome evolutionary comparative analyses is the possibility of performing *de novo* assemblies, leading to a single scaffold that can be used for comparisons. As demonstrated, the proposed hybrid assembly pipeline, including the long reads obtained with single-molecule real time sequencing technology, (SMRT; Pacific Biosystems) increases the quality of the assemblies and therefore the possibility of identifying genetic diversity amongst closely related strains that otherwise would have been lost.

In agreement with this result, during the preparation of this thesis, a second *Streptomyces* genome obtained solely with SMRT reads was published (Hoefler et al, 2013). This genome sequence was used to demonstrate that all multi-modular biosynthetic assembly lines (PKSs and NRPSs) present in the genome are truncated when using short read technologies such as 454 or Illumina, leading to highly fragmented genomes. This is an important observation when considering the fact that currently most *Streptomyces* genomes are sequenced for the discovery of novel NPs. Furthermore, the importance of subtle differences at both the gene-level and single-nucleotide polymorphisms can be associated to adaptive evolution, or from a more anthropocentric view, to genotypes related to desirable features within industrial settings.

Amongst the gene diversity uncovered by this study, particular attention was paid to SLIGI-1, a large genomic island which harbors the elusive plasmid SLP3. This discovery opens the door for detailed analysis of the elements mediating the transfer mechanism of SLP3. Indeed, current experimental efforts are showing that SLP3 can be integrated in the genetic repertoires of distant and closely related *Streptomyces* species. These efforts have been eased by the availability of a genome sequence of *S. lividans* 66. This sequence also provides a framework for the biotechnological analysis of *S. lividans* TK24, which is one of the most adopted hosts for heterologous production of recombinant proteins, where mobile elements could have interesting applications.

Several genes within SLIGI-1 and SLP3 were predicted to form part of metal tolerance systems. Direct evidence of the involvement of an important fraction of these genes in metal homeostasis was obtained by comparative phenotypic analysis and transcriptomic analysis of the copper response. The data provided reveals an unexpected enrichment of metal responsive genes, which in turn leads to the speculation that *S. lividans* was originally adapted to an environment where metal stress was an important challenge. Further gene-specific experiments to explore the genotype – phenotype link are envisaged to test this hypothesis.

Interestingly, as mentioned, SliGI-1 also harbors three NP biosynthetic systems: A lantipeptide biosynthetic system, the livipeptin biosynthetic gene cluster, and the arsonolividin biosynthetic gene cluster, both discovered using EvoMining. The location of these biosynthetic systems hints towards a direct involvement of their metabolic products in metal homeostasis. Furthermore, the fact that even in a thoroughly investigated bacterium, such as *S. lividans*, novel NPs biosynthetic systems remain to be discovered, highlights the need for alternative genome mining strategies that can direct the discovery of exciting new biosynthetic pathways. This not only emphasizes the large chemical potential of *Streptomyces*, but also brings about hopes and lessons for the discovery of novel antibiotics when pathogenic multidrug resistance is about to become a serious threat to human health.

Indeed, two lessons can be learnt from these discoveries. First, laboratory strains grown for long periods in axenic and artificial laboratory conditions may end up losing their mobile genetic elements which may contain these accessory metabolic features (Medema et al, 2010; Kinashi, 2011). Whether or not these elements are the main vehicle of horizontal transfer of NP biosynthetic gene clusters is yet to be established, however it is clear that the occurrence of these genes in plasmids is not uncommon. Second, beyond the anthropocentric use of NPs as antibiotics, these compounds do have diverse biological and physiological roles relevant in real environmental conditions (Price-Whelan et al, 2006; Yim et al, 2007). These notions suggest that NPs are adaptive traits whose evolutionary implications have been largely neglected. From a biotechnological point-of-view, they also provide meaningful biological settings that can direct efforts for the discovery of novel NPs with relevant biomolecular activities in the laboratory.

## 10.4 Adaptation of *Streptomyces* central metabolism for the production of natural products.

To investigate the role of the *Streptomyces* glycolytic node, consisting of PGM, ENO and PYK expansions, these genes were subject of *in silico* and in *vivo* analyses. The *In silico* analyses revealed the presence of conserved expansions in highly connected nodes of central metabolism in members of the genus *Streptomyces*. Of particular interest turned out to be the phosphoglyceromutase (PGM) and pyruvate kinase (PYK) expansions. Both glycolytic enzymes produce important precursors used for biomass and energy production, as well as for NP biosynthesis. The evolutionary history of the PGM and PYK enzyme families revealed that the expansion of the PGM and PYK occurred before radiation of the genus *Streptomyces*. The expanded PGMs (*pgm2*) are highly divergent, and as they are also conserved in NP producer strains beyond the *Streptomyces* genus, an event of horizontal gene transfer cannot be ruled out. The expansion of PYKs clearly arose via an event of gene duplication, as suggested by the fact that both paralogs are highly conserved at the sequence level. Irrespective of the sequence identity and the potential mechanisms leading to these expansions, the high conservation of the four genes in *Streptomyces* suggests a major physiological role.

To complement the analysis of PGM and PYK, the ENO gene was also included in this investigation, as this enzyme function connects the two previous enzymes. From the *in vivo* analyses few relevant phenotypic differences were observed when the enolase coding genes were knocked-out and/or over-expressed. These differences, however, did not go to the point to argue against extensive functional overlap between both homologues, at least at the biochemical level. This observation coincides with the lower conservation of this expansion, and served as a comparison point for the analysis of the highly conserved EEs, PGM and PYK.

Developmental arrest in rich media, together with reduced production of pigments in the presence of GlcNAc, was observed in the *pyk1* knockout mutant. This observation suggests a role for this homologue in carbon catabolic repression (CCR). An impairment in the GlcNAc acquisition system can be expected if the amount of PEP, the phosphate donor of the GlcNAc-PTS system, is depleted. This observation is supported by the transcriptional analysis, which showed higher expression levels of *pyk*1 than *pyk2* genes. Thus, when *pyk*1 is mutated, a stronger effect in central metabolic processes such as CCR, could be expected. All together, these results imply a non-overlapping physiological role of these paralogs, at least under certain conditions. A similar situation was found when the *pgm* mutants where analyzed, these mutants showed strikingly different phenotypes. While deletion of *pgm*1 produced insensitivity to the effect of GlcNAc, and led to the reduction of the biomass produced by the strain, the *pgm*2 mutant is very similar to the wild type under the same conditions.

An interesting result of this study was that over expression of *pyk*1 and *pyk*2 led to overproduction of pigmented NPs, however, this was obtained under different induction conditions. In contrast, over-expression of *pgm*2 showed a more direct relationship with overproduction of pigments, even using low induction conditions. This result gains relevance when the high conservation of the *pgm* expansion is considered. It is tempting to speculate that over-expression of *pgm*2 in other *Streptomyces* species may lead to the overproduction of other more interesting and valuable NPs.

# 11 Perspectives

## 11.1 Ongoing projects and follow-up ideas

### 11.1.1 Evolution inspired Genome mining tools

The EvoMining pipeline, and its associated databases, were written and collected for in-house use. Since then, this approach has been used as the basis for the development of a user friendly and more flexible EvoMining web-based tool. This platform is under development in collaboration with Christian Martinez, an experienced programmer working in our laboratory. An advantage of this new tool will be that up-to-date and improved databases can be easily generated, and therefore, it will be possible in the near future to extend the EvoMining analysis to other taxonomic groups beyond *Actinobacteria* and to other biological questions related to metabolism.

More specifically, to improve the PSCP database, a fundamental input of the pipeline, the use of the subsystems platform, which is a fast and comprehensive system that allows construction of metabolic models from genomic data (Henry et al, 2010), has been adopted. This web-tool is planned to be released for public use as an on-line analysis platform that helps in the discovery of novel NPs, but more importantly, to spread the evolutionary ideas and concepts behind the EvoMining approach. The release of this tool is expected to be done together with a bioinformatics-focused article.

### 11.1.2 New biosynthetic systems

An important product of this research project is the discovery of livipeptin, which belongs to a family of NPs of industrial and therapeutic interest. Therefore, a patent application is in preparation to protect the intellectual property over this discovery, and thus further details cannot be disclosed in the context of this thesis. Nevertheless, the biosynthetic mechanisms and the biological implications of livipeptin are expected to be highly interesting for the chemical biology community. Its investigation therefore should lead to a scientific publication regarding this subjects, and this is currently being prepared. This discovery will also open the door for further research on synthetic biology: The SLI0883-0884 system is a small peptide biosynthetic system. Indeed, the LFT enzyme, the core of the system, has already been proven to be able to incorporate promiscuously natural and unnatural amino acids (Taki et al, 2009). Therefore, the assembly of designed molecules is a feasible goal in the near future with a broad range of applications.

A number of NP biosynthetic predictions obtained with EvoMining have been recently subject of preliminary studies in our laboratory. Experimental characterization of some of these predictions, including the *psv* gene cluster found in *S. sviceus* and described as proof-of-concept in Chapter II, has actually started. Recently, we have succeeded in introducing exogenous DNA and producing knockout mutants in *S. sviceus* **(**Verduzco-Castro**,** PhD project**)**. This key technical advancement has opened the doors for the characterization of the *psv* biosynthetic pathway. Plasmids carrying a mutagenic cassette for gene knockout directed towards the PPM-PPD genes encoded in the *psv gene* cluster have been constructed. Soon, it will be possible to perform comparative metabolomic analyses of WT and mutant strains to identify the product(s) of the pathway.

Several important experimental advances made by Fernanda Iruegas-Bocardo and Luis Yanez-Guerra, master graduate students, have paved the way for elucidation of the biosynthetic pathway of the so-called arsonolividin metabolite. Currently, the experimental work has consisted of mutating key genes encoding regulators, transporters and biosynthetic enzymes, as well as subsequent phenotype characterization. This has led to identification of conditions for the induction of the expression of this gene cluster. Using this knowledge and strains, comparative metabolomics and the heterologous expression of key biosynthetic elements of the pathway have also been performed. Therefore, it is likely that the full elucidation of the pathway will be achieved in a relatively short period of time.

An important observation derived from the phylogenomic analysis of the arsonolividin biosynthetic system is that the AroA-PGM biosynthetic pair is present in other genomes, within gene clusters that include arsenic-related genes. Among them, a hybrid PKS NRPS system, suggesting that this biosynthetic system can be recruited to drive evolution of new NPs. To explore this idea, it will be interesting to experimentally analyze some of the divergent arsonolividin-like biosynthetic systems. Several of the genomes were this system was found belong to ant symbiotic *Streptomyces* species. This may imply a role for this potential new class of NPs on the relationship between insects and *Actinobacteria*, or perhaps it is linked to a particular environmental condition.

### 11.1.3 The enzyme expansions on the *Streptomyces* glycolytic node

Regarding the study of the glycolytic node, the results obtained in this thesis have been experimentally corroborated by the research group of Dr. Paul Hoskisson in Scotland, namely, by the work of his PhD student Jana Hiltner. Investigation of PGM and PYK by this group is part of a project aimed at studying a presumed link between central metabolism and NP biosynthesis from a physiological point of view. A natural collaboration has raised from this common interest and further experiments, including *in vitro* pyruvate kinase activity assays, are ongoing. The aim is now to deepen the understanding of the role of the pyruvate kinase in carbon catabolic repression, and the deviation of building blocks from central metabolism to NPs biosynthesis.

### 11.1.4 Adaptation of *Streptomyces lividans* to laboratory conditions

Based in the genetic differences identified between *S. lividans* 66 and the industrial strain TK24, a comparative analysis based in metabolic models from genomic data, is under development in collaboration with the group behind the subsystems technology (Henry et al, 2010). The main aim of this project is to understand the metabolic changes that occurred during the evolution of the strain TK24 during adaptation to industrial conditions. Several differences in terms of carbon source utilization, metal homeostasis, and development have been identified using phenotypic arrays and hypothesis-directed phenotyping. This experimental data is feeding the most complete metabolic models of a *Streptomyces* strain so far constructed. The comparison of closely related strains using genome-scale metabolic models has never been reported before.

### 11.1.5 Evolution of the enolase enzyme superfamily

Since the enolases identified using EvoMining have evolved from a glycolytic function, towards NP biosynthesis, it is likely that the recruited enolases are capable of catalyzing the glycolytic reaction promiscuously. This possibility opens the door for the functional study of the role of enzyme promiscuity and the evolution of novel enzymatic functions in the biosynthesis of NPs. This model may help understand how difficult is to reach a new enzyme function, and whether the novel functions were already present in the ancestral enzymes prior to the expansion events, or they arose after expansion. In such cases the question related to the original function promiscuously catalyzed and appearance of evolutionary intermediaries is an interesting one.

These latter ideas have been inspired by previous work on enzyme evolution done in our laboratory, using as model systems enzymes from central metabolism (Verdel-Aranda, in preparation; Noda-Garcia et al, 2013; Juárez-Vásquez, 2011, master thesis; Cruz-Morales, 2009, master thesis). To be able to exploit this model to test these ideas, the development of an *in vivo* positive selection system for the enolase functions in *S. sviceus*, *S. viridochromogenes*, or *S. coelicolor* is needed. Such selection system could be constructed based in the knockout mutants and phenotypic data obtained during this work. Meanwhile, a collection of glycolytic and recruited enolases (phosphinothricin and the *psv* gene cluster) from different *Streptomyces* species has been cloned into pAV11b, making it available for complementation and overexpression assays.

## 11.2 Long term goals: Integration of evolution, ecology, synthetic biology and bioprospection

In the final stretch of this research project I have become deeply interested in **how natural selection works on NPs and their biosynthetic pathways**. In most cases, these molecules have unknown functions, and little is known about **their role in natural conditions or their population level dynamics**. Several lines of evidence are needed to understand how the rare event of bioactivity arises in nature (see Firn 2010 for an inspiring collection of ideas regarding this subject).

One of the most intriguing questions regarding this subject is **how a biosynthetic pathway for a compound that performs a specific biological activity evolves**. When considering that NP biosynthetic pathways typically require the concerted action of several enzymes, and the benefit that the producing organism gain is only due to the final product of the pathway, positive selection on intermediary biosynthetic steps is needed to explain that these pathways evolved sequentially.

An alternative explanation is the presence of **biosynthetic subunits**, dedicated to the synthesis of moieties that have been selected for other roles, but which can be recruited for fast evolution of novel biosynthetic pathways. A first step to explain how these putative biosynthetic subunits were selected in nature prior to recruitments is their identification from within NP biosynthetic systems. These units could then be contrasted with the occurrence of homologous systems in central metabolism. An example of this phenomenon is pyrrole biosynthesis; pyrrolic groups are present in several NPs, but also in heme biosynthesis. From a biosynthetic and mechanistic point of view it seems apparent that these biosynthetic systems share several chemical and mechanistic features (Walsh et al 2006; Williamson et al 2006), and most likely a common evolutionary history as well.

Further functional and evolutionary insights will come from the understanding of the biological role of NPs in Nature. The main question to address here is: which are the driving forces behind the vast chemical diversity of NPs**.** This question can only be answered properly after studying the environmental conditions in which *Streptomyces* thrive, including their interactions with other inhabitants of their niches, the stability of their resources, the triggers of developmental changes and NP production, the pace of NP arising, and how antibiotic-utilizing microbes deal with resistance, among other subjects. In other words **to understand how natural selection works on NPs, an ecological perspective of NP metabolism is needed**.

These studies will naturally derive into biotechnological developments such as (i) discovery of novel biosynthetic pathways and the development of strategies for the activation of cryptic pathways based in ecological knowledge; (ii) the increase in the production of bioactive molecules and the rational design of industrial strains based in the modification and modelling of key metabolic features identified after systematic comparative genomic analysis; and ultimately (iii) **the design of *in vivo* selection systems which will make possible to obtain bioactive molecules "*a la carte*"** by the combination of biosynthetic subunits using synthetic biology and experimental evolution techniques.